\documentclass [letter, 12pt]{article}
\usepackage{a4wide, amsmath, amsfonts, amssymb, geometry}

\usepackage{bm}
\usepackage{fancyhdr, dsfont}
\usepackage[latin1]{inputenc}
\usepackage[pdftex]{graphicx} 
\usepackage{graphicx}
\usepackage{float}

\restylefloat{figure}
\usepackage{epstopdf}

\newcommand{\req}[1]{(\ref{#1})}

\def\vev#1{\langle #1 \rangle}
\def\lf{\left}
\def\ri{\right}
\def\bet{\beta}

\newcommand{\nwc}{\newcommand}
\nwc{\ba}  {\begin{array}}
\nwc{\ea}  {\end{array}}
\nwc{\bdm} {\begin{displaymath}}
\nwc{\edm} {\end{displaymath}}

\nwc{\bea} {\begin{equation}\ba{lcl}}
\nwc{\eea} {\ea\end{equation}}

\nwc{\bda} {\bdm\ba{lcl}} 
\nwc{\eda} {\ea\edm}

\nwc{\bc}  {\begin{center}}
\nwc{\ec}  {\end{center}}

\nwc{\ds}  {\displaystyle}
\nwc{\bmat}{\left(\ba}
\nwc{\emat}{\ea\right)}
\nwc{\nn}  {\nonumber}
\nwc{\nnn} {\nonumber \vspace{.2cm} \\ }
\nwc{\ra}  {\rightarrow}
\nwc{\lra} {\longrightarrow}

\nwc{\p} {\partial}

\def\slashchar#1{\setbox0=\hbox{$#1$}           
 \dimen0=\wd0                                 
 \setbox1=\hbox{/} \dimen1=\wd1               
 \ifdim\dimen0>\dimen1                        
    \rlap{\hbox to \dimen0{\hfil/\hfil}}      
    #1                                        
 \else                                        
    \rlap{\hbox to \dimen1{\hfil$#1$\hfil}}   
    /                                         
 \fi}

\newtheorem{pre-defi}[thm]{Definition}

\newtheorem{pre-rem}[thm]{Remark}

\newtheorem{pre-exple}[thm]{Example}

\newtheorem{pre-ue}[thm]{Exercise}

{\begin{sloppypar}\noindent{\bf Proof.}}%
{\hspace*{\fill}$\square$\bigskip\end{sloppypar}}

\def\beq{\begin{equation}}
\def\eeq{\end{equation}}

\def\eps{\varepsilon}

\newcommand{\vecb}{\left(\begin{array}{c}}
\newcommand{\vece}{\end{array}\right)}
\newcommand{\ccb}{\left(\begin{array}{cc}}
\newcommand{\cce}{\end{array}\right)}
\newcommand{\cccb}{\left(\begin{array}{ccc}}
\newcommand{\ccce}{\end{array}\right)}
\newcommand{\ccccb}{\left(\begin{array}{cccc}}
\newcommand{\cccce}{\end{array}\right)}
\newcommand{\cccccb}{\left(\begin{array}{ccccc}}
\newcommand{\ccccce}{\end{array}\right)}

\newcommand{\ve}{\vec}

\newcommand{\al}{\alpha}
\newcommand{\be}{\beta}
\newcommand{\ga}{\gamma}
\newcommand{\de}{\delta}
\newcommand{\ep}{\epsilon}
\newcommand{\vep}{\varepsilon}
\newcommand{\ze}{\zeta}
\newcommand{\io}{\iota}
\newcommand{\tet}{\theta}
\newcommand{\si}{\sigma}
\newcommand{\sib}{\bar\sigma}
\newcommand{\la}{\lambda}
\newcommand{\ka}{\kappa}
\newcommand{\om}{\omega}
\newcommand{\bi}{\bar\imath}
\newcommand{\bj}{\bar\jmath}
\newcommand{\Ga}{\Gamma}
\newcommand{\De}{\Delta}
\newcommand{\Si}{\Sigma}

\newcommand{\Om}{\Omega}

\newcommand{\mto}{\rightarrow}
\newcommand{\te}{\textrm}
\newcommand{\eq}{ \ \ = \ \ }

\newcommand{\co}{\ , \ \ \ \ \ \ }

\newcommand{\ee}{\mathrm{e}}
\newcommand{\half}{\tfrac{1}{2}}

\newcommand{\dal}{\dot{\alpha}}
\newcommand{\dbe}{\dot{\beta}}
\newcommand{\dga}{\dot{\gamma}}
\newcommand{\dde}{\dot{\delta}}
\newcommand{\dep}{\dot{\epsilon}}
\newcommand{\dze}{\dot{\zeta}}
\newcommand{\dka}{\dot{\kappa}}

\allowdisplaybreaks

\begin{document}

\title{\textbf{Higher~Point~Spin Field Correlators}\\ 
\textbf{in $\bm{D=4}$ Superstring Theory}\\[0.5cm]}
\author{D. H\"artl, O. Schlotterer, and\ S. Stieberger}
\date{}
\smallskip
\maketitle
\centerline{\it Max--Planck--Institut f\"ur Physik}
\centerline{\it Werner--Heisenberg--Institut}
\centerline{\it 80805 M\"unchen, Germany}

\medskip\bigskip\vskip2cm \abstract{\noindent Calculational tools are provided 
allowing to determine general tree--level
  scattering amplitudes for processes involving bosons and fermions in heterotic 
and superstring theories in
  four space--time dimensions.  We compute higher--point superstring correlators 
involving massless four--dimensional
  fermionic and spin fields.  In $D=4$ these correlators boil down to a product 
of two pure spin field correlators of
  left-- and right--handed spin fields.  This observation greatly simplifies the 
computation of such correlators.  The
  latter are basic ingredients to compute multi--fermion superstring amplitudes 
in $D=4$.  Their underlying fermionic
  structure and the fermionic couplings in the effective action are determined 
by these correlators.}

\vskip4cm
\begin{flushright}
{\small  MPP--2009--140}
\end{flushright}

\thispagestyle{empty}

\newpage
\setcounter{tocdepth}{2}
\tableofcontents

\numberwithin{equation}{section}

\newpage
\section{Introduction}
\label{sec:Introduction}

Multi--parton superstring amplitudes are of both considerable theoretical interest 
in the framework of a full--fledged
superstring theory \cite{ST,MHV,6G,Medina} 
and of phenomenological interest in describing the scattering 
processes underlying hadronic jet
production at high energy colliders  \cite{LHC1,LHC2}.
The key ingredient of these scattering 
amplitudes is the underlying superconformal
field theory (SCFT) governing the interactions of massless string states on the 
string world--sheet.

In four space--time dimensions this SCFT splits into an internal part and a 
space--time part. In the low--energy
effective action the latter determines the appropriate space--time Lorentz 
structure of the interactions, while the
internal part describes the internal degrees of freedom subject to the 
underlying compactification \cite{BD}.

In the (manifestly covariant) RNS fermionic string the space--time part of this SCFT 
comprises the $D=4$ matter fields $X^\mu,\psi^\mu,\ \mu=0,\ldots,3$, (covariant) spin fields $S_\al, S_{\dot\bet},\ \al,\dot\bet=1,2$, ghost-- and superghost system.
The  Neveu--Schwarz (NS) fermionic fields 
$\psi^{\mu}$ carry a space--time 
vector index $\mu$ and the  Ramond (R) spin fields $S_{\al}, S_{\dot\bet}$ 
carry spinor indices $\al,\dot\be$ under the Lorentz group $SO(1,3)$.
In the RNS formalism 
the fermionic coordinate fields $\psi^{\mu}$ of the two--dimensional world--sheet
theory are related to the bosonic 
coordinate fields $X^{\mu}$  through world--sheet supersymmetry.
On the other hand, the spin fields $S_{\al}, S_{\dot\bet}$ convert the 
fermionic boundary  conditions, i.e. intertwine the NS and R sectors. 
Their effect on the 
string world--sheet is the opening and closing of a branch cut \cite{FMS}. 

The fermion fields $\psi^\mu$ and $S_\al,S_{\dot\bet}$ enter the massless vertex operators of bosons and fermions, respectively. 
For concreteness, let us display the vertex operators of a gauge vector multiplet in $D=4$ type I superstring theory.
The gauge vector $A^a_\mu$  is created by the following vertex operator
\beq
V_{A^a}^{(-1)}(z,\xi,k) \ \ = \ \ g_A\ T^a\ \ee^{-\phi(z)}\ \xi^\mu\ 
\psi_\mu(z)\ \ee^{ik_\rho X^\rho(z)}\ ,
\eeq
with $\phi$ the scalar bosonizing the superghost system, $\xi^\mu$  the polarization vector of the gauge boson and 
some normalization $g_{A}$. Furthermore, $T^a$ are the Chan--Paton factors accounting
for the gauge degrees of freedom of the two open string ends.
On the other hand, the vertex operator of  the 
gauginos $\lambda^a,\overline\lambda^a$
of negative and positive helicity are given by   
\bea
\ds{V_{\lambda^{a,I}}^{(-1/2)}(z,u,k)} \,
&=& \, \ds{g_\lambda\ T^a\ \ee^{-\phi(z)/2}\ u^{\al}\  S_{\al}(z)\ 
\Si^I(z)\ \ee^{ik_\rho X^\rho(z)}\ ,}\\
\ds{V_{\overline\lambda^{a,I}}^{(-1/2)}(z,\bar u,k)}
 \, &=& \, \ds{g_\lambda\ T^a\ \ee^{-\phi(z)/2}\ \overline u_{\dbe} \ 
S^{\dbe}(z)\ \overline\Si^I(z)\
\ee^{ik_\rho X^\rho(z)}\ ,}
\eea
respectively. Above $u^\al,\overline u_{\dbe}$ are chiral spinors
satisfying the on--shell constraints $\slashchar{k}u=\slashchar{k}\overline u=0$
and $g_\lambda$ is a normalization constant.
The index $I$ labeling gaugino species may range from 1 to 4,
depending on the amount of supersymmetries,
while the associated  world--sheet fields $\Si^I$ of conformal dimension $3/8$
belong to the Ramond sector of internal SCFT \cite{BD}.

The world--sheet field theory is completely described by giving all 
correlation functions. Since $X^\mu,\psi^\mu$ are free fields and the spin fields 
$S_{\al}, S_{\dot\bet}$ are interacting fields the only 
non--trivial correlators are the $n+m$--point CFT correlators
\beq\label{BASIC}
\vev{\psi^{\mu_1}(z_1)\dots\psi^{\mu_n}(z_n)\,S_{\al_1}(x_1)\dots S_{\al_{r}}(x_{r})\,S_{\dbe_1}(y_1)\dots S_{\dbe_{s}}(y_{s})}
\eeq
involving $n$ NS fermionic fields and $m=r+s$ R spin fields $S_{\al}, S_{\dot\bet}$ 
with space--time  spinor indices $\al,\dot\bet$. 
Since the covariant spin field $S_\al$ is an interacting and double--valued 
field correlators of several spin fields cannot be computed by using
Wick's theorem and the underlying correlators as \req{BASIC} 
have to be determined by first principles.  Basically, these correlators
are completely specified by their properties under the $SO(1,3)$ current algebra 
of the RNS fermions and their singularity structure \cite{FMS,Cohn,KLLSW}.  
Indeed these  correlators can be constructed by analyzing 
their Lorentz and singularity structure. 
The latter is dictated by the relevant operator product expansion (OPE) of
the fields in the correlators. 

The calculation of fermionic string scattering amplitudes 
requires computing correlators of spin fields $S_\al, S_{\dot \be}$ and 
fermions $\psi^\mu$.  
Hence, correlators of the type \req{BASIC} are 
key ingredients entering the computation of 
multi--parton superstring amplitudes. Their underlying fermionic structure and the 
fermionic couplings in the low--energy effective action is revealed by the Lorentz 
structure of the CFT correlator~\req{BASIC}.

One of the main observation in this work is, that any correlator of the form 
\req{BASIC} may be first reduced to a $2n+m$--point correlator involving only spin fields
$S_\al$ and $S_{\dot\bet}$ by replacing each fermion $\psi^\mu$ by a pair of 
spin fields $S_\al, S_{\dot\bet}$. Moreover, in $D=4$ a correlator involving only spin fields $S_\al$ and $S_{\dot\bet}$  factorizes into 
products of two independent correlators of pure spin fields of one helicity
\bea\label{BASICs}
&&\vev{S_{\ka_1}(z_1)\dots S_{\ka_{n}}(z_{n})\,S_{\al_{1}}(x_{1})\dots S_{\al_{r}}(x_{r})}\ ,\\[3mm]
&&\vev{S_{\dka_1}(z_1)\dots S_{\dka_{n}}(z_{n})\,S_{\dbe_{1}}(y_{1})\dots S_{\dbe_{s}}(y_{s})}\ ,
\eea
respectively.
Hence, for any integers $n,m$ the correlator \req{BASIC} may be described by
products of the two pure spin field correlators \req{BASICs} involving $n+r$ and
$n+s$ spin fields of opposite helicity, respectively.

To compute fermionic processes with many external states one generically
needs CFT correlators \req{BASIC} for large integers $m$ and $n$.
The purpose of this work is to present the calculational tools and results 
necessary to compute correlators \req{BASIC} for any $m$ and $n$.
The latter enter the computation of general tree--level scattering amplitudes both in heterotic and superstring theory.

Covariant computation of fermion amplitudes in the RNS model 
has been advanced at tree--level in $D=10$ in \cite{GHMR,FMS,Cohn,Olaf} 
at the four--point level, 
while fermionic amplitudes in $D=10$
up to the six--point level are pioneered in \cite{KLLSW,KLSVWS}.
In $D=4$ superstring compactifications multi--parton amplitudes 
involving many bosons and some fermion fields
have been recently computed at disk tree--level \cite{MHV,6G,LHC2}, 
see also Refs. \cite{Barreiro:2005hv,DAN,Medina}.

In four--dimensional superstring theories, which preserve at least ${\cal N}=1$ spacetime supersymmetry, it is possible to evade the problem of an interacting CFT by using  the hybrid formalism  instead of the RNS approach to superstring theory 
\cite{Hybrid}. This formalism is based on some non--trivial field redefinitions, 
which replace the interacting RNS fields $\psi^\mu, S_\al$ and $S_{\dbe}$ by a new set of free world--sheet fields such that tree--amplitudes are completely fixed by appropriately summing their OPE singularities.

The organization of this work is as follows. In Section 2 we review the CFT 
of fermion and spin fields and list some basic CFT correlation functions involving
fermion and spin fields. In Section 3 we describe how a general correlator of the form
\req{BASIC} may be reduced to pure spin field correlators \req{BASICs}.
In Section 4 we give some group theoretical background allowing to classify and 
keep track of the Lorentz structure of the correlators \req{BASIC}. Moreover, we 
present a class of vanishing correlators, which give rise to non--trivial
consequences for the full string amplitudes in which those enter.
In Section 5 we determine the basic correlators \req{BASICs} for any 
numbers $r,s$ of spin fields. 
Equipped with these results in Section 6 we compute five-- through eight--point 
correlators \req{BASIC}, i.e. $n+m=5,\ldots,8$ and display their explicit expressions. The last two Sections 7 and 8 contain closed formulae for correlation functions with two spin fields and an arbitrary number of NS fermions. In Section 7 a basis of ordered products of sigma matrices is used for these correlators, while Section 8 
expresses them in terms of antisymmetrized products.
In Appendix A and B we present various $\si$--matrix identities in $D=4$.
The latter are needed to simplify correlators and to verify consistency checks of our results. Appendix C and D contain the proofs of the expressions for 
the two spin field correlators presented in Sections 7 and 8.

\section{Review of lower order correlators}
\label{sec:ReviewOfLowerOrderCorrelators}

In this Section we review the basic OPEs and some correlators of fermionic
 $\psi^\mu$ and spin fields~$S_\al$. 

The short distance behaviour of the vector fields $\psi^{\mu}$ from the NS sector and the left- and right-handed spin fields
$S_{\al},S_{\dbe}$ from the R sector is governed by the following OPEs 
\begin{subequations}
\label{rv,1}
\begin{align}
  \psi^{\mu}(z)\;\psi^{\nu}(w) \  &= \ \frac{\eta^{\mu \nu}}{z-w}\,+\,\dots\,, \label{rv,1a} \\
  S_{\al}(z)\;S_{\dbe}(w) \ &= \ \frac{1}{\sqrt{2}}\;(z-w)^{0}\,\si^{\mu}_{\al \dbe}\,\psi_{\mu}(w)\,+\,\dots\,, \label{rv,1b} \\
  S^{\dal}(z)\;S^{\be}(w) \ &= \ \frac{1}{\sqrt{2}}\;(z-w)^{0}\,\bar{\si}^{\mu \dal \be} \, \psi_{\mu}(w)\,+\,\dots\,, \label{rv,1c} \\
  S_{\al}(z)\;S_{\be}(w) \ &= \ -\,(z-w)^{-1/2}\,\vep_{\al \be}\,+\,\dots\,, \label{rv,1d} \\
  S_{\dal}(z)\;S_{\dbe}(w) \ &= \ +\,(z-w)^{-1/2}\,\vep_{\dal \dbe}\,+\,\dots\,,\label{rv,1e}
\end{align}
\end{subequations}
and:
\begin{subequations}
  \label{rv,2}
  \begin{align}
    \psi^{\mu}(z) \; S_{\al}(w) \ &= \ + \,\frac{1}{\sqrt{2}}\;(z-w)^{-1/2}\,\si^{\mu}_{\al \dbe}\,S^{\dbe}(w)\,+\,\dots\,, \label{rv,2a} \\
    \psi^{\mu}(z) \; S^{\dbe}(w) \ &= \ + \,\frac{1}{\sqrt{2}}\;(z-w)^{-1/2}\,\bar{\si}^{\mu \dbe \al}\,S_{\al}(w)\,+\,\dots \,.
    \label{rv,2b}
  \end{align}
\end{subequations}
The consistency of these OPEs is easily verified: Eqs. (\ref{rv,1a}) and (\ref{rv,1b}) require that the the OPEs of alike spin
fields must differ by a relative sign. If we take the sign convention as in (\ref{rv,1d}) and (\ref{rv,1e}) the OPEs
(\ref{rv,2}) are fixed.

One possible way to calculate correlation functions involving NS fermions and R spin fields is by applying all
possible OPEs (\ref{rv,1}) and (\ref{rv,2}). Then we can match the terms from the different
limits $z_i \to z_j$ to obtain the final result. In the case of the three-point function
\begin{align}
  \label{rv,3}
  \vev{\psi^\mu(z_{1}) \, S_{\al}(z_{2}) \, S_{\dbe}(z_{3})}\ \sim \ \left\{
  \begin{array}{lr}
    \ds{\frac{\si^\mu_{\al \dga} \, \vev{S^{\dga}(z_2)\,S_{\dbe}(z_3)}}{\sqrt{2}\,z_{12}^{1/2}} = 
    \frac{\si^\mu_{\al \dbe}}{\sqrt{2}\,(z_{12}\,z_{23})^{1/2}}} &\text{for} \ z_1 \to z_2\ , \\[5mm]
    \ds{\frac{\si^\nu_{\al \dbe} \, \vev{\psi^\mu(z_1)\,\psi_\nu(z_3)}}{\sqrt{2}} =
    \frac{\si^\mu_{\al \dbe}}{\sqrt{2} \, z_{13}}} &\text{for} \ z_{2} \to z_{3}\ ,
  \end{array} \right.
\end{align}
this method yields
\begin{equation}
  \label{rv,4}
  \vev{\psi^\mu(z_1)\,S_\al(z_2)\,S_{\dbe}(z_3)}\ =\ \frac{\si^\mu_{\al\dbe}}{\sqrt{2} \, (z_{12}\,z_{13})^{1/2}}\ ,
\end{equation}
with $z_{ij}:= z_i-z_j$. This way further correlators involving two spin fields have been calculated in \cite{MHV}:
\begin{subequations}
\begin{align}
  \vev{\psi^\mu(z_1)\,\psi^\nu(z_2)\,S_\al(z_3)\,S_\be(z_4)}\ &=\ \frac{- \, 1}{(z_{13}\,z_{14}\,z_{23}\,z_{24}\,z_{34})^{1/2}}
  \left(\eta^{\mu\nu}\,\vep_{\al\be} \,\frac{z_{13}\,z_{24}}{z_{12}}\,+\,(\si^\mu \, \sib^\nu \, \vep)_{\al\be} \, \frac{z_{34}}{2}\right)\,,  \label{rv,5a}\\
  \vev{\psi^\mu(z_1)\,\psi^\nu(z_2)\,S_{\dal}(z_3)\,S_{\dbe}(z_4)}\ &=\ \frac{+ \, 1}{(z_{13}\,z_{14}\,z_{23}\,z_{24}\,z_{34})^{1/2}}
  \left(\eta^{\mu\nu}\,\vep_{\dal\dbe} \,\frac{z_{13}\,z_{24}}{z_{12}}\,+\,(\vep \, \sib^\mu \, \si^\nu)_{\dal\dbe} \, \frac{z_{34}}{2}\right)\,.  \label{rv,5b}
\end{align}
\end{subequations}
In \cite{MHV} the four--point amplitude of one vector, two gauginos
and one scalar is derived. Its fermionic structure is determined by 
the correlators \req{rv,5}.
A more involved five--point amplitude involving three NS fermions and two R spin fields has been worked out in \cite{6G}:
\begin{align}
\label{rv,5}
&  \vev{\psi^\mu(z_1)\,\psi^\nu(z_2)\,\psi^\la(z_3)\,S_\al(z_4)\,S_{\dbe}(z_5)} \ = \
  \frac{1}{\sqrt{2}\,(z_{14}\,z_{15}\,z_{24}\,z_{25}\,z_{34}\,z_{35})^{1/2}} \notag \\
  &\hskip2cm
  \times\left(\frac{z_{45}}{2}\;(\si^\mu\,\bar{\si}^\nu\,\si^\la)_{\al\dbe}\,+\,
    \eta^{\mu\nu}\,\si^\la_{\al\dbe}\;\frac{z_{14}\,z_{25}}{z_{12}}\,-\,\eta^{\mu\la}\,\si^{\nu}_{\al \dbe}\;
    \frac{z_{14}\,z_{35}}{z_{13}}\,+\,\eta^{\nu\la}\,\si^\mu_{\al\dbe}\;\frac{z_{24}\,z_{35}}{z_{23}} \right)\ .
\end{align}
This correlator \req{rv,5} enters the computation of the six--point amplitude involving four scalars and two gauginos or chiral fermions \cite{6G}.
In addition to these cases with only two spin fields also some pure spin field correlators with four spinor indices are known \cite{MHV}:
\begin{subequations}
  \begin{align}
    \vev{S_\al(z_1)\,S_{\dbe}(z_2)\,S_\ga(z_3)\,S_{\dde}(z_4)} \ &= \ 
    -\frac{\vep_{\al\ga}\,\vep_{\dbe\dde}}{(z_{13}\,z_{24})^{1/2}}\ ,\label{4,1a} \\
    \vev{S_\al(z_1)\,S_\be(z_2)\,S_\ga(z_3)\,S_\de(z_4)} \ &= \  
    \frac{\vep_{\al\be}\,\vep_{\ga\de}\,z_{14}\ z_{23}\,-\,\vep_{\al\de}\,\vep_{\be\ga}\,z_{12}\,z_{34}}
    {(z_{12}\,z_{13}\,z_{14}\,z_{23}\,z_{24}\,z_{34})^{1/2}}\ , \label{4,1b} \\
    \vev{S_{\dal}(z_1)\,S_{\dbe}(z_2)\,S_{\dga}(z_3)\,S_{\dde}(z_4)} \ &= \ \frac{\vep_{\dal\dbe}\,\vep_{\dga\dde}\,z_{14}\,z_{23}\,-\,
      \vep_{\dal\dde}\,\vep_{\dbe\dga}\,z_{12}\,z_{34}}{(z_{12}\,z_{13}\,z_{14}\,z_{23}\,z_{24}\,z_{34})^{1/2}}\ . \label{4,1c}
  \end{align}
\end{subequations}
These correlators are basic ingredients of four--point amplitudes 
involving gauginos or chiral matter fermions \cite{MHV,LHC1}.
To check the individual limits $z_i \to z_j$ of the correlation functions \req{4,1a}--
\req{4,1c} the $z$--crossing identity
\begin{equation}
  \label{zcrossing}
  z_{ij}\,z_{kl} \ = \ z_{ik}\,z_{jl} \, + \, z_{il}\,z_{kj}
\end{equation}
proves to be useful. 

Some care is required to incorporate the complex phases which arise upon performing the OPEs. Since OPEs
are defined by the action of the involved fields on the vacuum state $|0 \rangle$, it is necessary to ``shift'' the
respective fields first to the right end of the correlation function before applying the OPE. The limit $z_2 \to z_3$ in
(\ref{rv,5a}), for instance, requires to commute $\psi^{\nu}(z_{2}) S_\al(z_3)$ past $S_\be(z_4)$. Due to the fractional
powers of $z-w$ in (\ref{rv,1d}) and (\ref{rv,2a}) both $\psi^{\nu}$ and $S_{\al}$ catch a phase of $i$ when they are
moved across $S_{\be}$. So an additional minus sign appears
\begin{equation}
  \label{wickcontract}
   \vev{\psi^\mu(z_1)\,\psi^\nu(z_2)\,S_\al(z_3)\,S_\be(z_4)}\ =\
  (-1)\,\frac{\si^\nu_{\al \dga}}{\sqrt{2}\,z_{23}^{1/2}}\,\vev{\psi^\mu(z_1)\, S^{\dga}(z_3) \, S_\be(z_4)} \ + \ {\cal O}(z_{23}^{1/2}) \,,
\end{equation}
which gives the correct $z_2\to z_3$ limit of (\ref{rv,5a}) using $-\si^{\nu} \bar{\si}^{\mu} = 2 \eta^{\mu \nu} +
\si^{\mu} \bar{\si}^{\nu}$. This relations shows that not all possible index terms are independent. The same thing
happens in the case of the correlations functions (\ref{4,1b}) and (\ref{4,1c}). Using the Fierz identity
\begin{equation}
  \label{fierzid}
  \vep_{\al\ga}\,\vep_{\be\de} \ = \ \vep_{\al\be}\,\vep_{\ga\de} \, + \, \vep_{\al\de}\,\vep_{\be\ga}
\end{equation}
one possible index configuration can be eliminated. In Section \ref{sec:GroupTheoreticalBackground} we systematically study
 how many independent index configurations exist for a particular correlator.

Determing all correlation functions by considering all possible OPEs only works consistently if all available
$\si$-- and Fierz identities are used to reduce the set of index terms to its minimal number. Finding these identities
for higher order correlators is quite involved. Hence, this method is rather inefficient for these cases. In fact, in the next Section we 
 propose a much more efficient way.

Besides, there is an efficient method to compute correlation functions with two 
fermion  fields $\psi$ at coinciding positions by making use of their Lorentz structure. 
The operators
\beq
J^{\mu \nu}(z) \ \ := \ \ \psi^{[\mu}(z)  \, \psi^{\nu]}(z)
\label{cur1}
\eeq
realize the $SO(1,3)$ current algebra at level $k=1$. 
Above the brackets $[\ldots]$ denote anti--symmetrization in the indices $\mu,\nu$.
Hence, their insertion into a correlator implements a Lorentz rotation. Any correlator including $J^{\mu \nu}$ can be reduced to its relatives with one current insertion less by means of the following prescription \cite{KZ} (see also \cite{FMS,Cohn,KLLSW})
\begin{align}
\vev{J^{\mu \nu}&(z) \, \psi_{\la_1}(z_1)\dots\psi_{\la_n}(z_n)\,S_{\al_1}(x_1)\dots S_{\al_{r}}(x_{r})\,S_{\dbe_1}(y_1)\dots S_{\dbe_{s}}(y_{s})} \notag \\
&= \ \ - \, \sum_{j=1}^n \, \frac{2 }{z-z_j} \; \de^{[ \mu}_{\la_j} \, \vev{ \psi_{\la_1}(z_1)\dots \psi^{\nu]}(z_j) \dots \psi_{\la_n}(z_n)\,S_{\al_1}(x_1)\dots S_{\al_{r}}(x_{r})\,S_{\dbe_1}(y_1)\dots S_{\dbe_{s}}(y_{s})} \notag \\
& \ \ \ \ - \, \sum_{j=1}^r \, \frac{1 }{2 \,(z-x_j)} \; \si^{\mu \nu} \, \! _{\al_j} \, \!^\ka \, \vev{ \psi_{\la_1}(z_1)\dots  \psi_{\la_n}(z_n)\,S_{\al_1}(x_1)\dots S_\ka(x_j) \dots S_{\al_{r}}(x_{r})\,S_{\dbe_1}(y_1)\dots S_{\dbe_{s}}(y_{s})} \notag \\
& \ \ \ \ + \, \sum_{j=1}^s \, \frac{1 }{2 \,(z-y_j)} \; \bar{\si}^{\mu \nu \dka} \, \!_{\dbe_j} \, \vev{ \psi_{\la_1}(z_1)\dots  \psi_{\la_n}(z_n)\,S_{\al_1}(x_1)\dots  S_{\al_{r}}(x_{r})\,S_{\dbe_1}(y_1)\dots S_{\dka}(y_j) \dots S_{\dbe_{s}}(y_{s})}
\label{cur}
\end{align}
as a result of  the $SO(1,3)$ action on the relevant fields: 
\begin{subequations}
\begin{align}
J^{\mu \nu}(z) \, \psi^{\la}(w) \ \ &= \ \ - \, \frac{2}{z-w} \; \eta^{\la [\mu} \, \psi^{\nu]}(w) \ + \ \ldots\,, \label{cur2} \\
J^{\mu \nu}(z) \, S_{\al}(w) \ \ &= \ \ - \, \frac{1}{2 \, (z -w)} \; \si^{\mu \nu} \, \!_{\al} \, \!^{\ka} \, S_{\ka}(w) \ + \ \ldots\,, \\ 
J^{\mu \nu}(z) \, S^{\al}(w) \ \ &= \ \ + \,\frac{1}{2 \, (z -w)} \; \si^{\mu \nu} \, \!_{\ka} \, \!^{\al} \, S^{\ka}(w) \ + \ \ldots\,, \label{cur3} \\
J^{\mu \nu}(z) \, S^{\dbe}(w) \ \ &= \ \ - \, \frac{1}{2 \, (z -w)} \; \bar{\si}^{\mu \nu \dbe} \, \!_{\dka} \,S^{\dka}(w) \ + \ \ldots\,, \\
J^{\mu \nu}(z) \, S_{\dbe}(w) \ \ &= \ \ + \,\frac{1}{2 \, (z -w)} \; \bar{\si}^{\mu \nu \dka} \, \!_{\dbe} \, S_{\dka}(w) \ + \ \ldots\,. \label{cur4}
\end{align}
\end{subequations}
Note that in case of several $J^{\mu \nu}$ insertions the central term of the current--current OPE arises:
\begin{align}
J^{\mu \nu}(z) \, J^{\la \rho}(w) \ \ = \ \ &\frac{1 }{(z-w)^2} \; \bigl( \eta^{\mu \rho} \, \eta^{\nu \la} \ - \ \eta^{\mu \la} \, \eta^{\nu \rho} \bigr) \notag \\ 
 + \ &\frac{1}{z-w} \; \bigl[ \eta^{\mu \la} \, J^{\nu \rho}(w) \ - \ \eta^{\mu \rho} \, J^{\nu \la}(w) \ - \ \eta^{\nu \la} \, J^{\mu \rho}(w)  \ + \ \eta^{\nu \rho} \, J^{\mu \la}(w) \bigr] \ + \ ... \ .
\label{cur5}
\end{align}
As a simple example of this method, let us compute the four point function (\ref{rv,5a}) at $z_1 = z_2$:
\begin{align}
&\vev{\psi^\mu(z_1)\,\psi^\nu(z_1)\,S_\al(z_3)\,S^\be(z_4)}\ \ = \ \ \vev{\psi^{[\mu}(z_1)\,\psi^{\nu]}(z_1)\,S_\al(z_3)\,S^\be(z_4)} \eq \vev{J^{\mu \nu}(z_1) \,S_\al(z_3)\,S_\be(z_4)} \notag \\
&\hskip1cm= \ \ - \, \frac{1}{z_{13}} \; \si^{\mu \nu} \, \! _{\al} \, \! ^{\ka} \, \langle S_\ka(z_3) \, S^{\be} (z_4) \rangle \ + \ \frac{1}{z_{14}} \; \si^{\mu \nu} \, \! _{\ka} \, \! ^{\be} \, \langle S_\al(z_3) \, S^{\ka} (z_4) \rangle \ = \ \frac{- \, z_{34}^{1/2}}{z_{13}\,z_{14}} \; \si^{\mu \nu} \! \,_{\al} \, \! ^{\be}\,. 
\end{align}
However, the goal of this article goes far beyond the application of Eq. (\ref{cur}). All the correlation functions will be given in full generality without any coinciding arguments. Of course, by a posteriori moving fermion positions together, one can obtain nice consistency checks for the results in the following Sections.

\section{From Ramond spin fields to Neveu--Schwarz fermions}
\label{sec:FromSpinFieldsToNSFermions}

After having collected existing results for some lower order correlation functions we now develop a new method to
systematically obtain correlators with arbitrarily many $\psi^{\mu}$ and $S_{\al},S_{\dbe}$ fields. First we show how NS
fermions can be reduced to a product of spin fields. Then it is demonstrated that in four space-time dimensions the
correlation function factorizes into a correlator of right-handed and a correlator of left-handed spin fields.

\subsection{Eliminating NS fermions}
\label{sec:NSFermionsInTermsOfSpinFields}

Let us first look at the $D$ dimensional generalization of the OPE (\ref{rv,1}) of two different spin fields. Spinor indices of $SO(1,D-1)$ will be denoted by $A,B$ and the corresponding gamma matrices by $\Ga^{\mu}_{AB}$. Since spin
fields $S_A(z_i)$ in $D$ space-time dimensions have conformal weight $D/16$, the OPE of $S_A$ and $S_B$ with appropriate relative chirality (alike in $D=4k+2$ and opposite in $D=4k$) is given by
\begin{equation}
  \label{Ddim}
  S_A(z) \,S_B(w) \eq \Ga^{\mu}_{AB} \, (z-w)^{1/2-D/8} \, \psi_{\mu}(w) \ + \ {\cal O}\bigl( (z-w)^{3/2-D/8} \bigr)\ .
\end{equation}
From the most singular term $\sim (z-w)^{1/2-D/8}$, one can read off the special property in $D=4$ dimensions -- there
is no singularity as $z \mto w$:
\begin{equation}
  \label{OPESSd}
  S_\al(z)\,S_{\dbe}(w)\ =\ \frac{1}{\sqrt{2}}\,\si^\mu_{\al\dbe}\, (z-w)^{0} \, \psi_\mu(w) \ + \  \mathcal{O}(z-w) \,.
\end{equation}
Setting $z=w$ leaves a non-trivial contribution on the right hand side:
\begin{equation}
  \label{OPESSd2}
  S_\al(z)\,S_{\dbe}(z)\ =\ \frac{1}{\sqrt{2}}\,\si^\mu_{\al\dbe}\,\psi_\mu(z)\ .
\end{equation}
Making use of $\si^\mu_{\ka\dka}\,\bar{\si}^{\nu\,\dka\ka}=-2\eta^{\mu \nu}$ this can be inverted:
\begin{equation}
  \label{psiss}
  \psi^\mu(z)\ =\ -\frac{1}{\sqrt{2}}\,\bar\si^{\mu\,\dka\ka}\,S_{\dka}(z)\,S_\ka(z)\ .
\end{equation}
Hence, it is possible to replace all NS fermions in the following correlator:
 \begin{align}
   &\vev{\psi^{\mu_1}(z_1)\dots\psi^{\mu_n}(z_n)\,S_{\al_1}(x_1)\dots S_{\al_{r}}(x_{r})\,S_{\dbe_1}(y_1)\dots S_{\dbe_{s}}(y_{s})} \ =\
   \prod_{i=1}^n \left(-\frac{\bar{\si}^{\mu_{i}\,\dka_i \ka_i}}{\sqrt{2}}\right)\notag\\
   &\,\times\,\vev{S_{\ka_1}(z_1)\dots S_{\ka_{n}}(z_{n})\,S_{\al_1}(x_1)\dots S_{\al_{r}}(x_{r})\,S_{\dka_1}(z_1)\dots S_{\dka_{n}}(z_{n})\,
     S_{\dbe_1}(y_1)\dots S_{\dbe_{s}}(y_{s})}\ .
\label{SSpsi2}
\end{align}
We see that an arbitrary correlation function can be written as a pure spin field correlator contracted by some $\si$
matrices. The next step is to systematically determine these correlators.

\subsection{Factorizing spin field correlators}
\label{sec:FactorizingSpinFieldCorrelators}

Looking at the simple result (\ref{4,1a}) for the four spin field correlation function $\langle S_\al S_{\dbe} S_\ga S_{\dde} \rangle$ one can identify it as
the product of the two point functions:
\begin{equation}
  \label{4Sprod}
  \vev{S_\al(z_1)\,S_\ga(z_3)} \ = \ -\frac{\;\vep_{\al\ga}\;}{z_{13}^{1/2}} \co \vev{S_{\dbe}(z_2)\,S_{\dde}(z_4)} \ = \ \frac{\;\vep_{\dbe\dde}\;}{z_{24}^{1/2}}\ .
\end{equation}
We prove now that this factorization property holds for an arbitrary number of spin fields. In order to do this it is
most convenient to treat them in bosonized form \cite{KLLSW}. The left- and right-handed spin fields in four dimensions
can be represented by two boson $H_{i=1,2}(z)$
\begin{align}
  \label{bosonization}
  S_{\al=1,2}(z) &\ \sim \ \ee^{\pm\frac{i}{2}[H_1(z)+H_2(z)]}\ =: \ \ee^{i\vec{p} \,\vec{H}(z)}\,,\notag\\
  S_{\dbe=1,2}(z) &\ \sim \ \ee^{\pm\frac{i}{2}[H_1(z)-H_2(z)]}\ =: \ \ee^{i\vec{q} \,\vec{H}(z)}\,,
\end{align}
with vector notation $\vec{H}(z)= \bigl(H_1(z),H_2(z) \bigr)$ for the bosons and weight vectors
$\vec{p}=\left(\pm\frac{1}{2},\pm\frac{1}{2}\right),\,\vec{q}=\left(\pm\frac{1}{2},\mp\frac{1}{2}\right)$. Note that the
weight vectors of distinct chiralities are orthogonal, $\vec{p}\, \vec{q}=0$. The two bosons fulfill the normalization
convention:
\begin{equation}
  \label{bosonnorm}
  \vev{H_i(z)\,H_j(w)}\ =\ \delta_{ij}\ \ln{(z-w)}\,.
\end{equation}
Cocycle factors which yield complex phases upon moving spin fields across each other are irrelevant for the following
discussion and are therefore neglected.\\
The OPEs (\ref{rv,1b})--(\ref{rv,1e}) as well as the four point functions (\ref{rv,5}) can be traced back to:
\begin{subequations}
\begin{align}
  \ee^{i\ve{p}  \, \ve{H}(z)}\,\ee^{i\ve{q} \, \ve{H}(w)} \ &\sim \ (z-w)^{\ve{p} \, \ve{q}}\;\ee^{i(\ve{p}+\ve{q}) \, \ve{H}(w)}\ +\ldots\ , \label{bos2a} \\
  \Bigl \langle \prod_{k=1}^n\ee^{i\ve{p}_k \ve{H}(z_k)}\Bigr\rangle \ &\sim \ \de\left(\sum_{k=1}^n\ve{p}_k\right)\ 
  \prod_{i,j=1\atop{i<j}}^nz_{ij}^{\ve{p}_i \ve{p}_j}\ .  \label{bos2b}
\end{align}
\end{subequations}
Hence the correlation function of $r$ left-handed and $s$ right-handed spin fields becomes:
\begin{align}
  \vev{S_{\al_1}&(z_1)\dots S_{\al_{r}}(z_{r})\,S_{\dbe_1}(w_1)\dots S_{\dbe_{s}}(w_{s})} \ =\ 
  \Bigl \langle
  \prod_{k=1}^r\ee^{i\ve{p}_k \ve{H}(z_k)}\ \prod_{l=1}^s\ee^{i\ve{q}_l \ve{H}(w_l)}\Bigr\rangle\notag\\
  &=\ \de\left(\sum_{k=1}^r\ve{p}_k+\sum_{l=1}^s\ve{q}_l\right)\,\prod_{i,j=1\atop{i<j}}^rz_{ij}^{\ve{p}_i\ve{p}_j}\,
 \ \prod_{\bi,\bj=1 \atop {\bi<\bj}}^{s} w_{\bi\bj}^{\ve{q}_{\bi}\ve{q}_{\bj}}\,
\ \underbrace{\prod_{m=1}^r\prod_{n=1}^s(z_m-w_n)^{\ve{p}_m\ve{q}_n}}_{=1}\notag\\
  &=\ \de\left(\sum_{k=1}^r\ve{p}_k\right)\,\prod_{i,j=1\atop{i<j}}^rz_{ij}^{\ve{p}_i \ve{p}_j}
 \ \de\left(\sum_{l=1}^s\ve{q}_l\right)\,\prod_{\bi,\bj=1\atop{\bi<\bj}}^sw_{\bi\bj}^{\ve{q}_{\bi} \ve{q}_{\bj}}=\ \Bigl\langle\prod_{k=1}^r\ee^{i\ve{p}_k \ve{H}(z_k)}\Bigr\rangle\ \Bigl\langle\prod_{l=1}^{s}\ee^{i\ve{q}_l \ve{H}(w_l)}\Bigr\rangle\notag\\
  &=\ \vev{S_{\al_1}(z_1)\dots S_{\al_r}(z_r)}\,\ \,\vev{S_{\dbe_1}(w_1)\dots S_{\dbe_{s}}(w_{s})}\,.
\label{factorize}
\end{align}
From the second to the third line we have used that $\ve{p}_m\ \ve{q}_n=0$ and the $\de$-function has been split into
the linearly independent $\ve{p}$ and $\ve{q}$ contributions. So we see that a general spin field correlation function
in four dimensions splits indeed into two correlators involving only left- and right-handed spin fields
respectively\footnote{ We want to stress that this result does not generalize to arbitrary dimensions. For $D=2k$ an
  even number of minus signs has to appear in the $k$ entries of $\vec{p}$, whereas $\vec{q}$ must contain an odd number
  of minus signs. Then for $k>2$ the crucial property $\vec{p}\, \vec{q}=0$ for weight vectors $\ve{p},\ve{q}$ of opposite chirality used in (\ref{factorize}) does not hold
  any longer.}.

Using the factorization property (\ref{factorize}) our previous result (\ref{SSpsi2}) becomes:  
\beq\label{noNS}
\boxed{
\ba{lcl}
&\ds{\vev{\psi^{\mu_1}(z_1)\dots\psi^{\mu_n}(z_n)\,S_{\al_{1}}(x_{1})\dots S_{\al_{r}}(x_{r})\,S_{\dbe_{1}}(y_{1})\dots S_{\dbe_{s}}(y_{s})}
  = \ \prod_{i=1}^n\left(-\frac{\bar{\si}^{\mu_{i}\dka_i\ka_i}}{\sqrt{2}}\right)} \\
  & \ds{\times \vev{S_{\ka_1}(z_1)\dots S_{\ka_{n}}(z_{n})\,S_{\al_{1}}(x_{1})\dots S_{\al_{r}}(x_{r})}
  \ \,\vev{S_{\dka_1}(z_1)\dots S_{\dka_{n}}(z_{n})\,S_{\dbe_{1}}(y_{1})\dots S_{\dbe_{s}}(y_{s})}\ .}
  \ea}
  \eeq
\vskip0.3cm

\noindent
This formula shows how correlators involving NS fermion  factorize into a product of correlators involving only left-- or
right-handed spin fields. Hence, if the latter correlators are known for an arbitrary number of spin fields it is
possible to calculate in principle any correlator.

\section{Group theoretical background}
\label{sec:GroupTheoreticalBackground}

Before we derive the formula for the correlator of an arbitrary number of alike spin fields let us have a look at the
$\psi$--$S$ correlators from a group theoretical point of view.

It is well known that the Lorentz algebra $so(1,3)$ decomposes into a direct sum of two $su(2)$ subalgebras, a left- and
a right-handed one. General representations of $SO(1,3)$ with spins $j_1,j_2$ with respect to the left- and right-handed
$SU(2)$ are denoted by $(\mathbf{j_1,j_2})$. The NS fields $\psi^\mu$ then transform as
$(\mathbf{\frac{1}{2}},\mathbf{\frac{1}{2}})$ under $SO(1,3)$, whereas the spin fields $S_\al$ and $S_{\dbe}$ transform
as $(\mathbf{\frac{1}{2}},\mathbf{0})$ and $(\mathbf{0},\mathbf{\frac{1}{2}})$ respectively.  Therefore an arbitrary
correlation function made up of these fields lies in the corresponding tensor product:
\begin{equation}
\label{tensorproduct}
  \vev{\psi^{\mu_1}\dots\psi^{\mu_n}\,S_{\al_1}\dots S_{\al_r}\,S_{\dbe_1}\dots S_{\dbe_s}} \ \in \
  (\mathbf{\half},\mathbf{\half})^{\otimes n}\,\otimes\,(\mathbf{\half},\mathbf{0})^{\otimes r}
  \,\otimes\,(\mathbf{0},\mathbf{\half})^{\otimes s}  \,.
\end{equation}

Looking back at the results of the correlators in Section \ref{sec:ReviewOfLowerOrderCorrelators} we see that they are
given as a sum over terms which consist of two parts: the index structure which carries all Lorentz and spinor indices
and the coefficients depending on the vertex operator positions $z_i$. It is clear that the coefficients are scalars
with respect to the four dimensional Lorentz group. Hence, the index terms have to be the \emph{Clebsch-Gordan
  coefficients} associated with the particular scalar representation.

The decomposition of the tensor product (\ref{tensorproduct}) does not help in finding the precise expressions for the
Clebsch-Gordan coefficients. In order to arrive at a minimal basis of index terms various identities (cf.\ appendix
\ref{sec:SigmaMatrixTechnology}) must be used which can for instance be derived from the Fierz identity (\ref{fierzid}). Yet it is
possible to find out the number of independent index terms from the appropriate tensor product. This is simply given by
the number of scalar representations $(\mathbf{0},\mathbf{0})$ within (\ref{tensorproduct}). Using our previous result
(\ref{noNS}) the number of scalar representations of an arbitrary correlation function is just the product of the number
of scalar representations of the appearing left- and right-handed spin field correlators.

In order to find the number of scalar representation of $\vev{S_{\al_1}(z_1)\dots S_{\al_N}(z_N)}$ one has to calculate
the tensor product $(\mathbf{\frac{1}{2}},\mathbf{0})^{\otimes N}=(\mathbf{\frac{1}{2}}^{\otimes N},\mathbf{0})$, where
as usual $\mathbf{\frac{i}{2}} \otimes \mathbf{\frac{1}{2}}=\mathbf{\frac{i+1}{2}} 
\oplus \mathbf{\frac{i-1}{2}}$. Finding the integer
coefficient $q(i,N)$ in front of the representation $(\mathbf{\frac{i}{2}},\mathbf{0})$ in the tensor product above is
in fact a common counting problem in combinatorics which is e.g.\ equivalent to a random walk with step size
$\frac{1}{2}$ on the positive real axis \cite{RW1}. The result is
\begin{equation}
  \label{12tensorproduct}
  (\mathbf{\tfrac{1}{2}},\mathbf{0})^{\otimes N}\ =\ \bigoplus_{i=0}^{N}\,q(i,N) \  (\mathbf{\tfrac{i}{2}},\mathbf{0})\,,
\end{equation}
where
\begin{equation}
  \label{catalantriangle}
  q(i,N) \ \equiv \ \frac{i+1}{N+1}\,\binom{N+1}{\frac{N-i}{2}}
\end{equation}
are the numbers appearing in the Catalan triangle. For $\frac{N-i}{2}\notin \mathbb{Z}$ the binomial coefficient is not
defined and in this case we set $q(i,N)$ to zero. The number of scalar representations is then given by
\begin{equation}
  \label{catalannumbers}
  q(0,N)\ =\ \frac{1}{N+1}\,\binom{N+1}{N/2}\,,
\end{equation}
which is only non-zero if $N$ is an even number. Then $q(0,N=2M)$ takes the well known form of the Catalan numbers:
\begin{equation}
  \label{catalaneven}
  q(0,2M)\ =\ \frac{1}{2M+1}\,\binom{2M+1}{M}\ =\ \frac{2M\,!}{M!\,(M+1)!}\,.
\end{equation}

An obvious consequence is that correlation functions consisting of an odd number of alike spin fields have to vanish due
to the fact that the corresponding tensor products yield no scalar representation. Using (\ref{noNS}) we also conclude
that the following correlators vanish ($n,r,s\in\mathbb{N}_0$):
\beq\label{vancor}
\boxed{\ba{lcl}
  \vev{\psi^{\mu_1}(z_1)\dots\psi^{\mu_{2n-1}}(z_{2n-1})\,S_{\al_1}(x_1)\dots S_{\al_{2r}}(x_{2r})\,S_{\dal_1}(y_1)\dots  S_{\dal_s}(y_s)}& =\ \ 0\ , \\[2mm]
  \vev{\psi^{\mu_1}(z_1)\dots\psi^{\mu_{2n-1}}(z_{2n-1})\,S_{\dal_1}(y_1)\dots S_{\dal_{2s}}(y_{2s})\,S_{\al_1}(x_1)\dots S_{\al_r}(x_r)}& =\ \ 0\ , \\[2mm]
  \vev{\psi^{\mu_1}(z_1)\dots\psi^{\mu_{2n}}(z_{2n})\,S_{\al_1}(x_1)\dots S_{\al_{2r-1}}(x_{2r-1})\,S_{\dal_1}(y_1)\dots  S_{\dal_s}(y_s)}& =\ \ 0\ , \\[2mm]
  \vev{\psi^{\mu_1}(z_1)\dots\psi^{\mu_{2n}}(z_{2n})\,S_{\dal_1}(y_1)\dots S_{\dal_{2s-1}}(y_{2s-1})\,S_{\al_1}(x_1)\dots S_{\al_r}(x_r)}& =\ \ 0\ .
\ea}
\eeq
\vskip0.3cm

\noindent
The vanishing of the correlators \req{vancor} has non--trivial consequences for the
full string amplitude, in which the latter enter. Hence some string amplitudes
involving bosons and fermions simply vanish as a result of  \req{vancor}.

Let us give two representative examples of tensor product decompositions and their help in determining the linear
independent set of Clebsch Gordan coefficients. We find for the following correlation functions
\begin{subequations}
\begin{align}
  \vev{S_\al\,S_\be\,S_\ga\,S_\de}\,&\in\,(\mathbf{\half,0})^{\otimes 4}\,=\,(\mathbf{2,0})\,\oplus\,3\ (\mathbf{1,0})\,
  \oplus\,\underline{2\ (\mathbf{0,0})}\label{example1}\,,\\
  \vev{\psi^\mu\,\psi^\nu\,\psi^\la\,S_\al\,S_{\dbe}}\,&\in\,(\mathbf{\half,\half})^{\otimes 3}\,\otimes\,(\mathbf{\half,0})\,
  \otimes\,(\mathbf{0,\half})\notag \\
  &\quad=\,(\mathbf{2,2})\,\oplus\,3\ (\mathbf{2,1})\,\oplus\,3\ (\mathbf{1,2})\,\oplus\,2\ (\mathbf{2,0})\,\oplus\,2\ (\mathbf{0,2})\notag \\
  &\qquad\oplus\,9\ (\mathbf{1,1})\,\oplus\,6\ (\mathbf{1,0})\,\oplus\,6\ (\mathbf{0,1})\,\oplus\,\underline{4\ (\mathbf{0,0})}\,. \label{example2}
\end{align}
\end{subequations}
We conclude that there are 2, respectively 4, independent index terms for these correlators, which agrees with
\eqref{4,1c} and \eqref{rv,5}. If we did not know about the Fierz identity (\ref{fierzid}), the number of scalars in
(\ref{example1}) would still tell us that there must be such a relation.

\section{Pure spin field correlators}
\label{sec:TheBasicBuildingBlockSpinFieldCorrelators}
 
In Section \ref{sec:FromSpinFieldsToNSFermions} we have seen that it is sufficient to know correlators with arbitrary
many alike spin fields in order to calculate any correlation functions \req{BASIC} involving fermions and spin fields. Deriving the
general formulas for the correlators of $N=2M$ alike spin fields is the aim of this Section. First we calculate the
correlation functions of four, six and eight left--handed spin fields. Guided by these results we then state the formula
for $2M$ left--handed spin fields and prove it by induction. This result is easily modified for the case of $2M$
right-handed spin fields.

The only possible index terms that can appear for the correlator of $2M$ alike spin fields are products of
$\vep$-tensors. In total there are $(2M-1)!!$ different index configurations, but only $\frac{2M!}{M!(M+1)!}$ are independent
according to (\ref{catalaneven}). When we calculate the correlation function by considering all OPEs we must therefore
eliminate the dependent terms from the set of index terms. This can be achieved by using generalizations of the Fierz
identity (\ref{fierzid}) and
\begin{equation}
  \label{antisymm}
 \de_{[\be}^{\al} \, \de^{\ga}_{\de} \de^{\ep}_{\zeta]} \eq 0 \ \ \ \Longleftrightarrow \ \ \ \vep_{[\al\underline{\be}}\,\vep_{\ga\underline{\de}}\,\vep_{\ep\underline{\ze}]} \eq 0\,,
\end{equation}
where we antisymmetrize over the underlined indices. Yet we will show that the results assume a nicer form if we use a
special non-minimal basis of $n!$ index terms. 

\subsection{Four--point, six--point and eight--point spin field correlators}

Let us start by reviewing the correlation function of four left-handed spin fields (\ref{4,1b}):
\bea\label{4s}
  \ds{\vev{S_\al(z_1)\,S_\be(z_2)\,S_\ga(z_3)\,S_\de(z_4)}}&=&\ds{
  \frac{1}{(z_{12}\,z_{13}\,z_{14}\,z_{23}\,z_{24}\,z_{34})^{1/2}}\ 
  \left(\vep_{\al\be}\,\vep_{\ga\de}\,z_{14}\,z_{23}\,-\,\vep_{\al\de}\,\vep_{\ga\be}\,z_{12}\,z_{43}\right)}\\[4mm]
  &=&\ds{\left(\frac{z_{12}\,z_{14}\,z_{23}\,z_{34}}{z_{13}\,z_{24}}\right)^{1/2}\ 
  \left(\frac{\vep_{\al\be}\,\vep_{\ga\de}}{z_{12}\,z_{34}}-\frac{\vep_{\al\de}\,\vep_{\ga\be}}{z_{14}\,z_{32}}\right)\ .}
  \eea
From the $3!!=3$ possible index terms we have eliminated the term $\vep_{\al\ga}\,\vep_{\be\de}$ using the Fierz
identity (\ref{fierzid}). The remaining two terms are independent which coincides with (\ref{catalaneven}) for $M=2$.

For the six point correlator $M=3$ there exist $5!!=15$ possible index terms, however, only five are independent. After taking into account all
possible OPEs one finds:
\goodbreak
\begin{align}
  \label{6smin}
  \vev{S_\al&(z_1)\,S_\be(z_2)\,S_\ga(z_3)\,S_\de(z_4)\,S_\ep(z_5)\,S_\ze(z_6)}\ =\ -\prod_{i<j}^6z_{ij}^{-1/2}
  \  \Big[ \vep_{\al\be}\,\vep_{\ga\de}\,\vep_{\ep\ze}\,z_{14}\,z_{15}\,z_{23}\,z_{26}\,z_{36}\,z_{45} \notag\\ 
  &+ \ \vep_{\al\be}\,\vep_{\ga\ze}\,\vep_{\ep\de}\,z_{14}\,z_{23}\,z_{56}\,(z_{15}\,z_{26}\,z_{34}-z_{12}\,z_{35}\,z_{46})
   \ + \ \vep_{\al\de}\,\vep_{\ga\ze}\,\vep_{\ep\be}\,z_{12}\,z_{13}\,z_{23}\,z_{45}\,z_{46}\,z_{56} \notag\\
  & +\ \vep_{\al\de}\,\vep_{\ga\be}\,\vep_{\ep\ze}\,z_{12}\,z_{36}\,z_{45}\,(z_{15}\,z_{26}\,z_{34}-z_{13}\,z_{24}\,z_{56})
   \ + \ \vep_{\al\ze}\,\vep_{\ga\be}\,\vep_{\ep\de}\,z_{12}\,z_{14}\,z_{24}\,z_{35}\,z_{36}\,z_{56} 
  \Big]\ .
\end{align}
The result assumes a more symmetric form if we introduce a sixth index term $\vep_{\al\ze}\,\vep_{\ga\de}\,\vep_{\ep\be}$:
\begin{align}
  \label{6snonmin}
  \vev{S_\al&(z_1)\,S_\be(z_2)\,S_\ga(z_3)\,S_\de(z_4)\,S_\ep(z_5)\,S_\ze(z_6)} \eq -\left(\frac{z_{12}\,z_{14}\,z_{16}\,z_{23}\,z_{25}\,z_{34}\,z_{36}\,z_{45}\,z_{56}}{z_{13}\,z_{15}\,z_{24}\,z_{26}\,z_{35}\,z_{46}}\right)^{1/2}\notag\\
  & \ \ \ \ \times\bigg(\frac{\vep_{\al\be}\,\vep_{\ga\de}\,\vep_{\ep\ze}}{z_{12}\,z_{34}\,z_{56}} \, - \, \frac{\vep_{\al\be}\,\vep_{\ga\ze}\,\vep_{\ep\de}}{z_{12}\,z_{36}\,z_{54}} \, + \, \frac{\vep_{\al\de}\,\vep_{\ga\ze}\,\vep_{\ep\be}}{z_{14}\,z_{36}\,z_{52}} \, - \, \frac{\vep_{\al\de}\,\vep_{\ga\be}\,\vep_{\ep\ze}}{z_{14}\,z_{32}\,z_{56}} \, + \, \frac{\vep_{\al\ze}\,\vep_{\ga\be}\,\vep_{\ep\de}}{z_{16}\,z_{32}\,z_{54}}
  -\frac{\vep_{\al\ze}\,\vep_{\ga\de}\,\vep_{\ep\be}}{z_{16}\,z_{34}\,z_{52}}\bigg)\notag\\
  &= \ \ -\prod_{i<j}^6z_{ij}^{-1/2}\ 
   \Big(\vep_{\al\be}\,\vep_{\ga\de}\,\vep_{\ep\ze}\,z_{14}\,z_{16}\,z_{23}\,z_{36}\,z_{25}\,z_{45}
   \, + \, \vep_{\al\be}\,\vep_{\ga\ze}\,\vep_{\ep\de}\,z_{14}\,z_{16}\,z_{23}\,z_{34}\,z_{25}\,z_{56}\notag\\[-.4cm]
  &\hspace{2.7cm}+ \, \vep_{\al\de}\,\vep_{\ga\be}\,\vep_{\ep\ze}\,z_{12}\,z_{16}\,z_{34}\,z_{36}\,z_{25}\,z_{45} \,
   - \, \vep_{\al\de}\,\vep_{\ga\ze}\,\vep_{\ep\be}\,z_{12}\,z_{16}\,z_{23}\,z_{34}\,z_{45}\,z_{56}\notag\\
  &\hspace{2.7cm}+ \, \vep_{\al\ze}\,\vep_{\ga\be}\,\vep_{\ep\de}\,z_{12}\,z_{14}\,z_{34}\,z_{36}\,z_{25}\,z_{56} \,
   + \, \vep_{\al\ze}\,\vep_{\ga\de}\,\vep_{\ep\be}\,z_{12}\,z_{14}\,z_{23}\,z_{36}\,z_{45}\,z_{56} \Big)\ .
\end{align}

In the case of the eight point function $M=4$ there are a total number of $7!!=105$ index terms from which only 14 are
independent. We give the result using a set of $4!=24$ dependent index terms:
\begin{align}
  \label{8s}
  \vev{S_\al(z_1)\,&S_\be(z_2)\,S_\ga(z_3)\,S_\de(z_4)\,S_\ep(z_5)\,S_\ze(z_6)\,S_\tet(z_7)\,S_\io(z_8)}\notag\\
  &=\left(\frac{z_{12}\,z_{14}\,z_{16}\,z_{18}\,z_{23}\,z_{25}\,z_{27}\,z_{34}\,z_{36}\,z_{38}\,z_{45}\,z_{47}\,z_{56}\,z_{58}\,z_{67}\,z_{78}}
    {z_{13}\,z_{15}\,z_{17}\,z_{24}\,z_{26}\,z_{28}\,z_{35}\,z_{37}\,z_{46}\,z_{48}\,z_{57}\,z_{68}}\right)^{1/2}\notag\\[.1cm]
  &\hspace{1cm}\times \bigg(\frac{\vep_{\al\be}\,\vep_{\ga \de}\,\vep_{\ep\ze}\,\vep_{\tet\io}}{z_{12}\,z_{34}\,z_{56}\,z_{78}}
   -\frac{\vep_{\al\be}\,\vep_{\ga\de}\,\vep_{\ep\io}\,\vep_{\tet\ze}}{z_{12}\,z_{34}\,z_{58}\,z_{76}}
   +\frac{\vep_{\al\be}\,\vep_{\ga\ze}\,\vep_{\ep\io}\,\vep_{\tet\de}}{z_{12}\,z_{36}\,z_{58}\,z_{74}}
   -\frac{\vep_{\al\be}\,\vep_{\ga\ze}\,\vep_{\ep\de}\,\vep_{\tet\io}}{z_{12}\,z_{36}\,z_{54}\,z_{78}}\notag\\[.2cm]
  &\hspace{1.3cm}+\frac{\vep_{\al\be}\,\vep_{\ga\io}\,\vep_{\ep\de}\,\vep_{\tet\ze}}{z_{12}\,z_{38}\,z_{54}\,z_{76}}
   -\frac{\vep_{\al\be}\,\vep_{\ga\io}\,\vep_{\ep\ze}\,\vep_{\tet\de}}{z_{12}\,z_{38}\,z_{56}\,z_{74}}
   -\frac{\vep_{\al\de}\,\vep_{\ga\be}\,\vep_{\ep\ze}\,\vep_{\tet\io}}{z_{14}\,z_{32}\,z_{56}\,z_{78}}
   +\frac{\vep_{\al\de}\,\vep_{\ga\be}\,\vep_{\ep\io}\,\vep_{\tet\ze}}{z_{14}\,z_{32}\,z_{58}\,z_{76}}\notag\\[.2cm]
  &\hspace{1.3cm}-\frac{\vep_{\al\de}\,\vep_{\ga\ze}\,\vep_{\ep\io}\,\vep_{\tet\be}}{z_{14}\,z_{36}\,z_{58}\,z_{72}}
   +\frac{\vep_{\al\de}\,\vep_{\ga\ze}\,\vep_{\ep\be}\,\vep_{\tet\io}}{z_{14}\,z_{36}\,z_{52}\,z_{78}}
   -\frac{\vep_{\al\de}\,\vep_{\ga\io}\,\vep_{\ep\be}\,\vep_{\tet\ze}}{z_{14}\,z_{38}\,z_{52}\,z_{76}}
   +\frac{\vep_{\al\de}\,\vep_{\ga\io}\,\vep_{\ep\ze}\,\vep_{\tet\be}}{z_{14}\,z_{38}\,z_{56}\,z_{72}}\notag\\[.2cm]
  &\hspace{1.3cm}+\frac{\vep_{\al\ze}\,\vep_{\ga\be}\,\vep_{\ep\de}\,\vep_{\tet\io}}{z_{16}\,z_{32}\,z_{54}\,z_{78}}
   -\frac{\vep_{\al\ze}\,\vep_{\ga\be}\,\vep_{\ep\io}\,\vep_{\tet\de}}{z_{16}\,z_{32}\,z_{58}\,z_{74}}
   +\frac{\vep_{\al\ze}\,\vep_{\ga\de}\,\vep_{\ep\io}\,\vep_{\tet\be}}{z_{16}\,z_{34}\,z_{58}\,z_{72}}
   -\frac{\vep_{\al\ze}\,\vep_{\ga\de}\,\vep_{\ep\be}\,\vep_{\tet\io}}{z_{16}\,z_{34}\,z_{52}\,z_{78}}\notag\\[.2cm]
  &\hspace{1.3cm}+\frac{\vep_{\al\ze}\,\vep_{\ga\io}\,\vep_{\ep\be}\,\vep_{\tet\de}}{z_{16}\,z_{38}\,z_{52}\,z_{74}}
   -\frac{\vep_{\al\ze}\,\vep_{\ga\io}\,\vep_{\ep\de}\,\vep_{\tet\be}}{z_{16}\,z_{38}\,z_{54}\,z_{72}}
   -\frac{\vep_{\al\io}\,\vep_{\ga\be}\,\vep_{\ep\de}\,\vep_{\tet\ze}}{z_{18}\,z_{32}\,z_{54}\,z_{76}}
   +\frac{\vep_{\al\io}\,\vep_{\ga\be}\,\vep_{\ep\ze}\,\vep_{\tet\de}}{z_{18}\,z_{32}\,z_{56}\,z_{74}}\notag\\[.2cm]
  &\hspace{1.3cm}-\frac{\vep_{\al\io}\,\vep_{\ga\de}\,\vep_{\ep\ze}\,\vep_{\tet\be}}{z_{18}\,z_{34}\,z_{56}\,z_{72}}
   +\frac{\vep_{\al\io}\,\vep_{\ga\de}\,\vep_{\ep\be}\,\vep_{\tet\ze}}{z_{18}\,z_{34}\,z_{52}\,z_{76}}
   -\frac{\vep_{\al\io}\,\vep_{\ga\ze}\,\vep_{\ep\be}\,\vep_{\tet\de}}{z_{18}\,z_{36}\,z_{52}\,z_{74}}
   +\frac{\vep_{\al\io}\,\vep_{\ga\ze}\,\vep_{\ep\de}\,\vep_{\tet\be}}{z_{18}\,z_{36}\,z_{54}\,z_{72}}\bigg)\ .
\end{align}

Comparing Eqs. (\ref{4s}), (\ref{6snonmin}) and (\ref{8s}) the following similarities are visible: in all cases the
pre--factor consists of all possible terms of the schematic form $(z_\text{odd even}\ z_\text{even odd})^{1/2}$ in the numerator
and $(z_\text{odd odd}\ z_\text{even even})^{1/2}$ in the denominator. Furthermore, the first index at every
$\vep$-tensor belongs to a spin field with argument $z_\text{odd}$ whereas the second index stems from a spin field with
argument $z_{\text{even}}$, and finally every $\vep$-tensor comes with the corresponding factor
$(z_\text{odd}-z_\text{even})^{-1}$. The overall sign can be traced back to $(-1)^M$ coming from the OPE (\ref{rv,1d})
whereas the relative signs between the index terms can be understood as the sign of the respective permutation of the
spinor indices.

\subsection[The generalization to $2M$ spin fields]{The generalization to $\bm{2M}$ spin fields}
\label{sec:TheGeneralizationTo2nSpinFields}

The results (\ref{4s}), (\ref{6snonmin}) and (\ref{8s}) suggest the following expression for the $2M$ point function of left-handed spin fields:\\
\beq\label{2ns}
\boxed{\ba{lcl}
&\ds{\vev{S_{\al_1}(z_1) S_{\al_2}(z_2)\dots S_{\al_{2M-1}}(z_{2M-1})\,S_{\al_{2M}}(z_{2M})}\ =\
  (-1)^M\,\bigg(\prod_{i \leq j}^M z_{2i-1,2j}\,\prod_{\bi<\bj}^{M} z_{2\bi,2\bj-1}\bigg)^{1/2} }\\
  &\hskip4cm \ds{\times \left(\prod_{k<l}^M z_{2k-1,2l-1}\,z_{2k,2l}\right)^{-1/2} \ \sum_{\rho \in S_M}\te{sgn}(\rho)\,
  \prod_{m=1}^M\frac{\vep_{\al_{2m-1}\al_{\rho(2m)}}}{z_{2m-1,\rho(2m)}}\ .}
\ea}
\eeq
\vskip0.3cm

\noindent
We  prove this expression by induction. For the base case $M=1$ this gives correctly the vev of the OPE
(\ref{rv,1d}). The inductive step makes use of the fact that the $2M-2$ correlator should appear from the $2M$
correlator if we replace two spin fields by the OPE in the corresponding limit $z_i \to z_j$. As every spin field can be
permuted to the very right in the correlator we study without loss of generality the case $z_{2M-1}\to z_{2M}$:
\begin{align}
  \label{inducstep}
  \vev{S_{\al_1}(z_1)\,&\dots\,S_{\al_{2M-2}}(z_{2M-2})\,S_{\al_{2M-1}}(z_{2M-1})\,S_{\al_{2M}}(z_{2M})}\Bigl.\Bigr|_{z_{2M-1}\ra z_{2M}} 
\notag\\[.2cm]
  &=\ - \,\frac{\vep_{\al_{2M-1}\al_{2M}}}{z_{2M-1,2M}^{1/2}}\, \vev{S_{\al_1}(z_1) \, \dots \, S_{\al_{2M-2}}(z_{2M-2})}\,+\,\mathcal{O}(z_{2M-1,2M})\notag\\
  &=\ - \,\frac{\vep_{\al_{2M-1}\al_{2M}}}{z_{2M-1,2M}}\,z_{2M-1,2M}^{1/2}\,(-1)^{M-1}\,\left(\prod_{i\leq j}^{M-1}z_{2i-1,2j}\,
  \prod_{\bi<\bj}^{M-1}z_{2\bi,2\bj-1}\right)^{1/2} \, \left(\prod_{k<l}^{M-1}z_{2k-1,2l-1}\,z_{2k,2l} \right)^{-1/2}\notag\\ 
  &\ \ \ \ \ \times \,\underbrace{\left(\frac{\prod_{p=1}^{M-1}z_{2p-1,2M}\,z_{2p,2M-1}}{\prod_{q=1}^{M-1}z_{2q-1,2M-1}\,z_{2q,2M}}
  \right)^{1/2}}_{=\,1\,+\,\mathcal{O}(z_{2M-1,2M})} \; \sum_{\rho \in S_{M-1}}\te{sgn}(\rho)\,\prod_{m=1}^{M-1}
  \frac{\vep_{\al_{2m-1}\al_{\rho(2m)}}}{z_{2m-1,\rho(2m)}}\,+\,\mathcal{O}(z_{2M-1,2M})\notag\\
  &= \ (-1)^{M} \, \left( \prod_{i \leq j}^{M}z_{2i-1,2j} \, \prod_{\bi<\bj}^{M}z_{2\bi,2\bj-1} \right)^{1/2} \, \left(
   \prod_{k<l}^{M}z_{2k-1,2l-1}\,z_{2k,2l} \right)^{-1/2}\notag\\
  &\ \ \ \ \ \times \sum_{\rho\in S_{M}}\te{sgn}(\rho)\,\prod_{m=1}^{M} \de_{\rho(2M),2M}\,\frac{\vep_{\al_{2m-1}\al_{\rho(2m)}}}{z_{2m-1,\rho(2m)}}\,+\,\mathcal{O}(z_{2M-1,2M})\,.
\end{align}
The most singular piece of (\ref{2ns}) in $z_{2M-1,2M}$ is the subset of $S_M$ permutations $\rho$ with $\rho(2M) =
2M$. This is precisely what we get by applying the OPE of $S_{\al_{2M-1}}(z_{2M-1}) S_{\al_{2M}}(z_{2M})$ and assuming
the claimed expression for $\vev{S_{\al_1}(z_1) \, \dots \, S_{\al_{2M-2}}(z_{2M-2})}$. This completes the proof of
(\ref{2ns}).

The correlator of $2M$ right-handed spin fields reads:
\beq\label{2nsd}
\boxed{\ba{lcl}
&\ds{\vev{S_{\dal_1}(z_1)S_{\dal_2}(z_2)\dots S_{\dal_{2M-1}}(z_{2M-1})\,S_{\dal_{2M}}(z_{2M})}\ =\
  \bigg(\prod_{i \leq j}^M z_{2i-1,2j}\,\prod_{\bi<\bj}^{M} z_{2\bi,2\bj-1}\bigg)^{1/2}}\\
  &\hskip4cm\ds{ \times \left(\prod_{k<l}^M z_{2k-1,2l-1}\,z_{2k,2l}\right)^{-1/2} \ \sum_{\rho \in S_M}\te{sgn}(\rho)\,
  \prod_{m=1}^M\frac{\vep_{\dal_{2m-1}\dal_{\rho(2m)}}}{z_{2m-1,\rho(2m)}}\ .}
  \ea}
  \eeq
\vskip0.3cm

\noindent
The factor $(-1)^M$ drops out because of the plus sign in the OPE (\ref{rv,1e}).

By plugging (\ref{2ns}) and (\ref{2nsd}) into (\ref{noNS}) we are now able to calculate any correlator involving
fermions $\psi^\mu$ and spin fields $S_\al,S_{\dbe}$.

\section{Explicit examples with more than four spin fields}
\label{sec:ExplicitExamplesWithSpinFields}
 
After presenting the setup to calculate various correlation function 
of the type \req{BASIC} in this Section we derive  all --so far unknown--
non--vanishing correlators of fermions and spin fields up to the eight--point level. 
Generically we give the result in a
non--minimal basis of index terms. When appropriate we also derive the result 
in a minimal basis.  All our correlators obey non--trivial consistency checks by means
of considering various OPEs between different pairs of fields and comparing the 
resulting correlators with lower point correlators. For this one needs various 
identities between $\si$--matrices. These relations are listed in Appendices A and B.

\subsection{Five--point functions}
\label{sec:FivePointFunctions}

For two cases $n=1,\ r=3,\ s=1$ and $n=1,\ r=1,\ s=3$ 
we obtain the non--minimal correlators
\bea\label{exp1a}
\ds{\langle \psi^{\mu}(z_{1}) \, S_{\al}(z_{2}) \, S_{\be}(z_{3}) \, S_{\ga}(z_{4}) \, S_{\dde}(z_{5}) \rangle} &=&\ds{\frac{-1}{\sqrt{2} \, ( z_{12} \, z_{13} \, z_{14} \, z_{15})^{1/2} \, (z_{23} \, z_{24} \, z_{34})^{1/2}}\ ,} \\[4mm]
&\times&\ds{ \Bigl( \si^{\mu}_{\ga \dde} \, \vep_{\al \be} \, z_{12} \, z_{13} \ - \ \si^{\mu}_{\be \dde} \, \vep_{\al \ga} \, z_{12} \, z_{14} \ + 
\ \si^{\mu}_{\al \dde} \, \vep_{\be \ga} \, z_{13} \, z_{14} \Bigr)\ ,}
\eea
\beq\label{exp1b}\ba{lcl}
\ds{\langle \psi^{\mu}(z_{1}) \, S_{\al}(z_{2}) \, S_{\dbe}(z_{3}) \, S_{\dga}(z_{4}) \, S_{\dde}(z_{5}) \rangle }&=& 
\ds{\frac{1}{\sqrt{2} \, ( z_{12} \, z_{13} \, z_{14} \, z_{15})^{1/2} \, (z_{34} \, z_{35} \, z_{45})^{1/2}} \ ,}\\[4mm]
&\times&\ds{\Bigl( \si^{\mu}_{\al \dbe} \, \vep_{\dga \dde} \, z_{14} \, z_{15} \ - \ \si^{\mu}_{\al \dga} \, \vep_{\dbe \dde} \, z_{13} \, z_{15} \ + \ \si^{\mu}_{\al \dde} \, \vep_{\dbe \dga} \, z_{13} \, z_{14} \Bigr)\ ,}
\ea
\eeq
which become in minimal representation
\bea\label{4,5a}
\ds{\langle \psi^{\mu}(z_{1}) \,S_{\al}(z_{2}) \, S_{\be}(z_{3}) \, S_{\ga}(z_{4}) \, S_{\dde}(z_{5}) \rangle } &\eq&\ds{ 
\frac{-1}{\sqrt{2} \, ( z_{12} \, z_{13} \, z_{14} \, z_{15})^{1/2} \, (z_{23} \, z_{24} \, z_{34})^{1/2}}  }\\[4mm]
&\times&\ds{\lf(\si^{\mu}_{\al \dde} \, \vep_{\be \ga} \, z_{14} \, z_{23} \ - \ \si^{\mu}_{\ga \dde} \, \vep_{\al \be} \, z_{12} \, z_{34}\ri)\ ,}
\eea
\bea\label{4,5b}
\ds{\langle \psi^{\mu}(z_{1}) \, S_{\al}(z_{2}) \, S_{\dbe}(z_{3}) \, S_{\dga}(z_{4}) \, S_{\dde}(z_{5}) \rangle}  &\eq&\ds{\frac{1}{\sqrt{2} \, ( z_{12} \, z_{13} \, z_{14} \, z_{15})^{1/2} \, (z_{34} \, z_{35} \, z_{45})^{1/2}}}\\[4mm]
&\times&\ds{\lf(\si^{\mu}_{\al \dbe} \, \vep_{\dga \dde} \, z_{15} \, z_{34} \ - \ \si^{\mu}_{\al \dde} \, \vep_{\dbe \dga} \, z_{13} \, z_{45}\ri) \ ,}
\eea
respectively.

\subsection{Six--point functions}
\label{sec:SixPointFunctions}

For the case $n=2,\ r=2,\ s=2$ we obtain
\begin{align}
 &\langle \psi^{\mu}(z_{1}) \,\psi^{\nu}(z_{2}) \, S_{\al}(z_{3}) \, S_{\dbe}(z_{4}) \, S_{\ga}(z_{5}) \, S_{\dde}(z_{6}) \rangle \eq \frac{- \, 1}{2 \, z_{12} \, ( z_{13} \, z_{14} \, z_{15} \, z_{16} \, z_{23} \, z_{24} \, z_{25} \, z_{26})^{1/2} \, (z_{35} \, z_{46} )^{1/2}} \notag \\
& \times \ \Biggl\{ \si^{\mu}_{\al \dde} \, \si^{\nu}_{\ga \dbe} \, z_{14} \, z_{15} \, z_{23} \, z_{26} \ - \ \si^{\mu}_{\al \dbe} \, \si^{\nu}_{\ga \dde} \, z_{15} \, z_{16} \, z_{23} \, z_{24} + \si^{\mu}_{\ga \dbe} \, \si^{\nu}_{\al \dde} \, z_{13} \, z_{16} \, z_{24} \, z_{25} \ - \ \si^{\mu}_{\ga \dde} \,  \si^{\nu}_{\al \dbe} \, z_{13} \, z_{14} \, z_{25} \, z_{26} \Biggr\}\ .\label{exp2}
\end{align}
The correlator \req{exp2} is needed to compute the five--point disk scattering of one gauge vector and four chiral
fermions \cite{LHC2}.  Actually, in this reference an other variant of \req{exp2} has been used:
\begin{align}
\label{spinii}
\vev{\psi^{\mu}(z_1)\,\psi&^{\nu}(z_2)\,S_\al(z_3)\,S_{\dot\bet}(z_4)\,S_\gamma(z_5)\,S_{\dot\delta}(z_6)} \eq 
\frac{-\,1}{z_{12}\,(z_{13}\,z_{14}\,z_{15}\,z_{16}\,z_{23}\,z_{24}\,z_{25}\,z_{26})^{1/2}\,(z_{35}\,z_{46})^{1/2}} \notag\\
&\times\,\Bigg\{\eta^{\mu\nu}\,\vep_{\al\ga}\,\vep_{\dbe\dde}\,z_{13}\,z_{15}\,z_{24}\,z_{26}
+\frac{1}{2}\left(\si^\mu_{\al\dbe}\,\si^\nu_{\ga\dde}\,z_{24}\,z_{36}-\si^\mu_{\al\dde}\,\si^\nu_{\ga\dbe}z_{26}\,z_{34}\right)z_{12}\,z_{15}\notag\\
&\ \ \ \ \ \ -\frac{1}{2}\left(\si^\mu_{\ga\dde}\,\si^\nu_{\al\dbe}\,z_{26}\,z_{45}+\si^\mu_{\ga\dbe}\,\si^\nu_{\al\dde}z_{24}\,z_{56} \right)z_{12}\,z_{13}\Bigg\}\,.
\end{align}

Next, we consider the case $n=2,\ r=4,\ s=0$, for which we derive:
\begin{align}
\langle \psi^{\mu}(z_{1}) \, &\psi^{\nu}(z_{2}) \, S_{\al}(z_{3}) \, S_{\be}(z_{4}) \, S_{\ga}(z_{5}) \, S_{\de}(z_{6}) \rangle \eq \frac{1}{z_{12} \, (z_{13} \, z_{14} \, z_{15} \, z_{16} \, z_{23} \, z_{24} \, z_{25} \, z_{26})^{1/2}} \, \notag \\
& \times \ \frac{1}{(z_{34} \, z_{35} \, z_{36} \, z_{45} \, z_{46} \, z_{56})^{1/2}} \ \times \ \Biggl\{ \eta^{\mu \nu} \, \vep_{\al \be} \, \vep_{\ga \de} \, z_{13} \, z_{16} \, z_{24} \, z_{25} \, z_{36} \, z_{45}  \Biggr. \notag \\
& \ \ \ \ \Biggl. - \ \eta^{\mu \nu} \, \vep_{\al \de} \, \vep_{\be \ga} \, z_{13} \, z_{14} \, z_{25} \, z_{26 } \, z_{34} \, z_{56} \ + \ \frac{1}{2} \; (\si^{\mu}  \, \bar{\si}^{\nu}\, \vep)_{\al \be} \, \vep_{\ga \de} \, z_{12} \, z_{15} \, z_{25} \, z_{34} \, z_{36} \, z_{46} \Biggr. \notag \\
& \ \ \ \ \Biggl. - \, \frac{1}{2} \; (\si^{\mu} \, \bar{\si}^{\nu} \, \vep)_{\al \ga} \, \vep_{\be \de} \, z_{12} \, z_{14} \, z_{24} \, z_{35} \, z_{36} \, z_{56} \ + \ \frac{1}{2} \; (\si^{\mu} \, \bar{\si}^{\nu}\, \vep )_{ \be \ga } \, \vep_{\al \de} \, z_{12} \, z_{13} \, z_{23} \, z_{45} \, z_{46} \, z_{56} \Biggr\}\ . \label{exp3}
\end{align}
In a slightly more symmetric representation with respect to the spin fields we find:
\begin{align}
\langle &\psi^{\mu}(z_{1})\,\psi^{\nu}(z_{2})\,S_{\al}(z_{3})\,S_{\be}(z_{4})\,S_{\ga}(z_{5})\,S_{\de}(z_{6})\rangle \notag \\
&= \ \ \frac{1}{(z_{13} \, z_{14} \, z_{15} \, z_{16} \, z_{23} \, z_{24} \, z_{25} \, z_{26})^{1/2} \, \  \, (z_{34} \, z_{35} \, z_{36} \, z_{45} \, z_{46} \, z_{56})^{1/2}} \, \notag \\
& \ \times \ \Biggl\{ \frac{\eta^{\mu \nu}}{z_{12}} \; (\vep_{\al \be} \, \vep_{\ga \de} \, z_{36} \, z_{45}) \, \  \, ( z_{13} \, z_{24} ) \, \  \, ( z_{15} \, z_{26}) \ + \ \frac{\eta^{\mu \nu}}{z_{12}} \; (\vep_{\al \de} \, \vep_{\ga \be} \, z_{34} \, z_{56}) \, \  \, ( z_{13} \, z_{26} ) \, \  \, ( z_{15} \, z_{24})  \Biggr. \notag \\
& \ \ \ \ \ \ \ \ \Biggl. + \ \left[ (\si^{\mu} \, \bar{\si}^{\nu} \, \vep)_{\al \be} \,  \frac{z_{34}}{2} \right] \, (\vep_{\ga \de} \, z_{36} \, z_{45}) \  (z_{15} \, z_{26}) \ + \  \left[ (\si^{\mu} \, \bar{\si}^{\nu} \, \vep)_{\al \de} \,   \frac{z_{36}}{2} \right] \, (\vep_{\ga \be} \, z_{34} \, z_{56}) \  (z_{15} \, z_{24}) \Biggr. \notag \\
& \ \ \ \ \ \ \ \ \Biggl. + \ \left[ (\si^{\mu} \, \bar{\si}^{\nu} \, \vep)_{\ga \de} \, \frac{z_{56}}{2} \right] \, (\vep_{\al \be} \, z_{36} \, z_{45}) \  (z_{13} \, z_{24}) \ + \ \left[ (\si^{\mu} \, \bar{\si}^{\nu} \, \vep)_{ \ga \be} \, \frac{z_{54}}{2} \right] \, (\vep_{\al \de} \, z_{34} \, z_{56}) \  (z_{13} \, z_{26}) \Biggr\}\ .
\label{exp4}
\end{align}
Moreover, \req{exp4} may be cast into manifestly $\psi^\mu(z_1) 
\leftrightarrow \psi^\nu(z_2)$ anti--symmetric form:
\begin{align}
\langle &\psi^{\mu}(z_{1}) \, \psi^{\nu}(z_{2}) \, S_{\al}(z_{3}) \, S_{\be}(z_{4}) \, S_{\ga}(z_{5}) \, S_{\de}(z_{6}) \rangle \notag \\
&= \ \ \frac{1}{2 \, (z_{13} \, z_{14} \, z_{15} \, z_{16} \, z_{23} \, z_{24} \, z_{25} \, z_{26})^{1/2} \, \  \, (z_{34} \, z_{35} \, z_{36} \, z_{45} \, z_{46} \, z_{56})^{1/2}} \, \notag \\
& \ \times \ \Biggl\{ \frac{\eta^{\mu \nu}}{z_{12}} \;  (\vep_{\al \be} \, \vep_{\ga \de} \, z_{36} \, z_{45} ) \, \  \,  ( z_{13} \, z_{24} \, z_{25} \, z_{16} \ + \ z_{23} \, z_{14} \, z_{15} \, z_{26}) \Biggr.  \notag \\
& \ \ \ \ \ \ \ \ \Biggl. + \ \frac{\eta^{\mu \nu}}{z_{12}} \; (\vep_{\al \de} \, \vep_{\ga \be} \, z_{34} \, z_{56}) \, \  \, ( z_{13} \, z_{14} \, z_{25} \, z_{26} \ + \ z_{23} \, z_{24} \, z_{15} \, z_{16})  \Biggr. \notag \\
& \ \ \ \ \ \ \ \ \ \ \ \ \ \ \Biggl. + \ \Bigl[ (\si^{\mu \nu} \, \vep)_{\al \be} \, z_{34} \Bigr] \, (\vep_{\ga \de} \, z_{36} \, z_{45}) \  \frac{z_{15} \, z_{26} \, + \, z_{16} \, z_{25}}{2} \Biggr. \notag \\
& \ \ \ \ \ \ \ \ \ \ \ \ \ \ \Biggl. + \  \Bigl[ (\si^{\mu \nu} \, \vep)_{\al \de} \, z_{36} \Bigr] \, (\vep_{\ga \be} \, z_{34} \, z_{56}) \  \frac{z_{15} \, z_{24} \, + \, z_{14} \, z_{25}}{2} \Biggr. \notag \\
& \ \ \ \ \ \ \ \ \ \ \ \ \ \ \Biggl. + \ \Bigl[ (\si^{\mu \nu} \, \vep)_{\ga \de} \,  z_{56}  \Bigr] \, (\vep_{\al \be} \, z_{36} \, z_{45}) \  \frac{z_{13} \, z_{24} \, + \, z_{14} \, z_{23}}{2} \Biggr. \notag \\
& \ \ \ \ \ \ \ \ \ \ \ \ \ \ \Biggl. + \ \Bigl[ (\si^{\mu \nu} \, \vep)_{ \ga \be} \, z_{54} \Bigr] \, (\vep_{\al \de} \, z_{34} \, z_{56}) \   \frac{z_{13} \, z_{26} \, + \, z_{16} \, z_{23}}{2} \Biggr\}\,.
\label{exp5}
\end{align}

\subsection{Seven--point functions}
\label{sec:SevenPointFunctions}

For the case $n=3,\ r=3,\ s=1$ we obtain:
\begin{align}
\langle &\psi^\mu(z_1) \, \psi^\nu(z_2) \, \psi^\la (z_3) \, S_\al (z_4) \, S_\be (z_5 ) \, S_\ga (z_6) \, S_{\dde}(z_7) \rangle \eq \frac{- \,1}{2 \sqrt{2} \, z_{23}} \notag \\
& \ \times \ \frac{1}{ (z_{14} \, z_{15} \, z_{16} \, z_{17} \, z_{24} \, z_{25} \, z_{26} \, z_{27} \, z_{34} \, z_{35} \, z_{36} \, z_{37} \, )^{1/2} \, (z_{45} \, z_{46} \, z_{56})^{1/2} } \notag \\
& \ \times \biggl\{ - \ \frac{2 \, \eta^{\mu \nu}}{z_{12}} \; \vep_{\al \be} \, \si^{\la}_{\ga \dde} \, z_{13} \, z_{14} \, z_{25} \, z_{26} \, z_{27} \, z_{34} \, z_{56} \ - \ \frac{2 \, \eta^{\mu \la}}{z_{13}} \; \vep_{\al \be} \, \si^{\nu}_{\ga \dde} \, z_{12} \, z_{14} \, z_{35} \, z_{36} \, z_{37} \, z_{24} \, z_{56} \biggr. \notag \\
& \ \ \ \ \ \ \ + \ \frac{2 \, \eta^{\mu \nu}}{z_{12}} \; \vep_{ \be \ga } \, \si^{\la}_{\al \dde} \, z_{23} \, z_{14} \, z_{15} \, z_{26} \, z_{27} \, z_{36} \, z_{45} \ - \ \frac{2 \, \eta^{\mu \la}}{z_{13}} \; \vep_{ \be \ga } \, \si^{\nu}_{\al \dde} \, z_{23} \, z_{14} \, z_{15} \, z_{26} \, z_{36} \, z_{37} \, z_{45} \notag \\
& \ \ \ \ \ \ \  + \ (\si^{\mu} \, \bar{\si}^{\nu} \, \vep)_{\al \be} \, \si^{\la}_{\ga \dde} \, z_{16} \, z_{26} \, z_{27} \, z_{34} \, z_{35} \, z_{45} \ + \ (\si^{\mu} \, \bar{\si}^{\la} \, \vep)_{\al \be} \, \si^{\nu}_{\ga \dde} \, z_{16} \, z_{36} \, z_{37} \, z_{24} \, z_{25} \, z_{45} \notag \\
& \ \ \ \ \ \ \  + \ (\si^{\mu} \, \bar{\si}^{\nu} \, \vep)_{ \be \ga } \, \si^{\la}_{\al \dde} \, z_{14} \, z_{24} \, z_{27} \, z_{35} \, z_{36} \, z_{56} \ + \ (\si^{\mu} \, \bar{\si}^{\la} \, \vep)_{ \be \ga } \, \si^{\nu}_{\al \dde} \, z_{14} \, z_{34} \, z_{37} \, z_{25} \, z_{26} \, z_{56} \notag \\
& \ \ \ \ \ \ \ \ \biggl. - \ (\si^{\mu} \, \bar{\si}^{\nu} \, \vep)_{\al \ga} \, \si^{\la}_{\be \dde} \, z_{15} \, z_{25} \, z_{27} \, z_{34} \, z_{36} \, z_{46} \ - \ (\si^{\mu} \, \bar{\si}^{\la} \, \vep)_{\al \ga} \, \si^{\nu}_{\be \dde} \, z_{15} \, z_{35} \, z_{37} \, z_{24} \, z_{26} \, z_{46} \biggr\} \label{exp6} \ .
\end{align}
The result \req{exp6} may be also cast into a form with most of the $z_{23}$ poles cancelled:
\goodbreak
\begin{align}
\langle &\psi^\mu(z_1) \, \psi^\nu(z_2) \, \psi^\la (z_3) \, S_\al (z_4) \, S_\be (z_5 ) \, S_\ga (z_6) \, S_{\dde}(z_7) \rangle  \notag \\
&= \ \ \frac{1}{ 2 \sqrt{2} \, (z_{14} \, z_{15} \, z_{16} \, z_{17} \, z_{24} \, z_{25} \, z_{26} \, z_{27} \, z_{34} \, z_{35} \, z_{36} \, z_{37} \, )^{1/2} \, (z_{45} \, z_{46} \, z_{56})^{1/2} } \notag \\
& \ \times \biggl\{ \frac{2 \, \eta^{\mu \nu}}{z_{12}} \; \si^{\la}_{\ga \dde} \, \vep_{\al \be} \, z_{14} \, z_{16} \, z_{25} \, z_{27} \, z_{34} \, z_{56} \ + \ \frac{2 \, \eta^{\mu \nu}}{z_{12}} \; \si^{\la}_{\al \dde} \, \vep_{\ga \be} \, z_{14} \, z_{16} \, z_{25} \, z_{27} \, z_{36} \, z_{45} \biggr. \notag \\
& \ \ \ \ \ \biggl. \ - \ \frac{2 \, \eta^{\mu \la}}{z_{13}} \; \si^{\nu}_{\ga \dde} \, \vep_{\al \be} \, z_{14} \, z_{16} \, z_{24} \, z_{35} \, z_{37} \, z_{56} \ - \ \frac{2 \, \eta^{\mu \la}}{z_{13}} \; \si^{\nu}_{\al \dde} \, \vep_{\ga \be} \, z_{14} \, z_{16} \, z_{26} \, z_{35} \, z_{37} \, z_{45} \biggr. \notag \\
& \ \ \ \ \ \biggl. \ + \ \frac{2 \, \eta^{\nu \la}}{z_{23}} \; \si^{\mu}_{\ga \dde} \, \vep_{\al \be} \, z_{14} \, z_{24} \, z_{26} \, z_{35} \, z_{37} \, z_{56} \ + \ \frac{2 \, \eta^{\nu \la}}{z_{23}} \; \si^{\mu}_{\al \dde} \, \vep_{\ga \be} \, z_{16} \, z_{24} \, z_{26} \, z_{35} \, z_{37} \, z_{45} \biggr. \notag \\
& \ \ \ \ \ \biggl. \ + \ (\si^{\mu} \, \bar{\si}^{\nu} \, \si^{\la})_{\ga \dde} \, \vep_{\al \be} \, z_{14} \, z_{25} \, z_{34} \,  z_{56} \, z_{67}  \ + \ (\si^{\mu} \, \bar{\si}^{\nu} \, \si^{\la})_{\al \dde} \, \vep_{\ga \be} \, z_{16} \, z_{25} \, z_{36} \, z_{45} \, z_{47}  \biggr. \notag \\
& \ \ \ \ \ \biggl. \ - \ (\si^{\mu} \, \bar{\si}^{\nu} \, \vep)_{\al \be} \, \si^{\la}_{\ga \dde} \, z_{23} \, z_{16} \, z_{47} \,  z_{45} \, z_{56}  \ + \ (\si^{\mu} \, \bar{\si}^{\nu} \, \vep)_{\ga \be} \, \si^{\la}_{\al \dde} \, z_{23} \, z_{14} \, z_{67} \,  z_{45} \, z_{56}  \biggr. \notag \\
& \ \ \ \ \ \biggl. \ + \ (\si^{\nu} \, \bar{\si}^{\la} \, \vep)_{\al \be} \, \si^{\mu}_{\ga \dde} \, z_{14} \, z_{26} \, z_{37} \,  z_{45} \, z_{56}  \ - \ (\si^{\nu} \, \bar{\si}^{\la} \, \vep)_{\ga \be} \, \si^{\mu}_{\al \dde} \, z_{16} \, z_{24} \, z_{37} \,  z_{45} \, z_{56}  \biggr\}\,.
\label{neu}
\end{align}

Next, for the case $n=1,\ r=5,\ s=1$ we obtain
\begin{align}
  \vev{\psi^\mu(z_1)\,&S_\al(z_2)\,S_\be(z_3)\,S_\ga(z_4)\,S_\de(z_5)\,S_\ep(z_6)\,S_{\dze}(z_7)}\notag\\
  &=\frac{1}{\sqrt{2}\,(z_{12}\,z_{13}\,z_{14}\,z_{15}\,z_{16}\,z_{17})^{1/2}(z_{23}\,z_{24}\,z_{25}\,z_{26}\,z_{34}\,z_{35}\,z_{36}\,z_{45}\,z_{46}\,z_{56})^{1/2}}\notag\\
  &\hspace{.6cm}\times\Big(\si^\mu_{\al\dze}\,\vep_{\be\ga}\,\vep_{\de\ep}\,z_{16}\,z_{23}\,z_{45}(z_{14}\,z_{25}\,z_{36}+z_{12}\,z_{34}\,z_{56})
   -\si^\mu_{\al\dze}\,\vep_{\be\ep}\,\vep_{\ga\de}\,z_{15}\,z_{16}\,z_{23}\,z_{24}\,z_{34}\,z_{56}\notag\\
  &\hspace{.8cm}+\si^\mu_{\ep\dze}\,\vep_{\al\be}\,\vep_{\ga\de}\,z_{12}\,z_{34}\,z_{56}(z_{14}\,z_{25}\,z_{36}+z_{16}\,z_{23}\,z_{45})
   -\si^\mu_{\ep\dze}\,\vep_{\al\de}\,\vep_{\be\ga}\,z_{12}\,z_{13}\,z_{23}\,z_{45}\,z_{46}\,z_{56}\notag\\
  &\hspace{.8cm}-\si^\mu_{\ga\dze}\,\vep_{\al\be}\,\vep_{\de\ep}\,z_{12}\,z_{16}\,z_{26}\,z_{34}\,z_{35}\,z_{45}\Big)\notag\\
  &=\frac{1}{\sqrt{2}\,(z_{12}\,z_{13}\,z_{14}\,z_{15}\,z_{16}\,z_{17})^{1/2}(z_{23}\,z_{24}\,z_{25}\,z_{26}\,z_{34}\,z_{35}\,z_{36}\,z_{45}\,z_{46}\,z_{56})^{1/2}}\notag\\
  &\hspace{.6cm}\times\Big(\si^\mu_{\al\dze}\,\vep_{\be\ga}\,\vep_{\de\ep}\,z_{14}\,z_{16}\,z_{23}\,z_{25}\,z_{36}\,z_{45}
   -\si^\mu_{\al\dze}\,\vep_{\be\ep}\,\vep_{\ga\de}\,z_{14}\,z_{16}\,z_{23}\,z_{25}\,z_{34}\,z_{56}\notag\\
  &\hspace{.8cm}-\si^\mu_{\ga\dze}\,\vep_{\al\be}\,\vep_{\de\ep}\,z_{12}\,z_{16}\,z_{25}\,z_{34}\,z_{36}\,z_{45}
   +\si^\mu_{\ga\dze}\,\vep_{\al\de}\,\vep_{\be\ep}\,z_{12}\,z_{16}\,z_{23}\,z_{34}\,z_{45}\,z_{56}\notag\\
  &\hspace{.8cm}+\si^\mu_{\ep\dze}\,\vep_{\al\be}\,\vep_{\ga\de}\,z_{12}\,z_{14}\,z_{25}\,z_{34}\,z_{36}\,z_{56}
   -\si^\mu_{\ep\dze}\,\vep_{\al\de}\,\vep_{\be\ga}\,z_{12}\,z_{14}\,z_{23}\,z_{36}\,z_{45}\,z_{56}\Big)\,,
\label{exp7}
\end{align}
where the first result is given in minimal and the second in non-minimal representation.

Finally, for the case $n=1,\ r=s=3$  we get  in minimal form:
\begin{align}
  \vev{\psi^\mu(z_1)&S_\al(z_2)\,S_\be(z_3)\,S_\ga(z_4)\,S_{\dde}(z_5)\,S_{\dep}(z_6)\,S_{\dze}(z_7)}\notag\\
  &=\frac{-1}{\sqrt{2}\,(z_{12}\,z_{13}\,z_{14}\,z_{15}\,z_{16}\,z_{17})^{1/2}(z_{23}\,z_{24}\,z_{34}\,z_{56}\,z_{57}\,z_{67})^{1/2}}
  \times\Big(\si^\mu_{\ga\dze}\,\vep_{\al\be}\,\vep_{\dde\dep}\,z_{12}\,z_{15}\,z_{34}\,z_{67}\notag\\
   &\quad-\si^\mu_{\ga\dde}\,\vep_{\al\be}\,\vep_{\dep\dze}\,z_{12}\,z_{17}\,z_{34}\,z_{56}+\si^\mu_{\al\dde}\,\vep_{\be\ga}\,\vep_{\dep\dze}\,z_{14}\,z_{17}\,z_{23}\,z_{56}
   -\si^\mu_{\al\dze}\,\vep_{\be\ga}\,\vep_{\dde\dep}\,z_{14}\,z_{15}\,z_{23}\,z_{67}\Big)\ .
\label{exp8}
\end{align}

\subsection{Eight--point functions}
\label{sec:EightPtFunctions}

For the case $n=4,\ r=s=2$ we obtain:
\begin{align}
\langle &\psi^{\mu}(z_{1}) \, \psi^{\nu}(z_{2}) \, \psi^{\la}(z_{3}) \, \psi^{\rho}(z_{4}) \, S_{\al}(z_{5}) \, S_{\dbe}(z_{6}) \, S_{\ga}(z_{7}) \, S_{\dde}(z_{8}) \rangle \notag \\
&= \ \ \frac{- \, 1}{4 \, \bigl( z_{15} \, z_{16} \, z_{17} \, z_{18} \, z_{25} \, z_{26} \, z_{27} \, z_{28} \, z_{35} \, z_{36} \, z_{37} \, z_{38} \, z_{45} \, z_{46} \, z_{47} \, z_{48} \bigr)^{1/2} \, \bigl( z_{57} \, z_{68} \bigr)^{1/2}} \notag \\
& \ \times \ \biggl\{ \frac{4 \, \eta^{\mu \nu} \, \eta^{\la \rho}}{z_{12} \, z_{34}} \; \vep_{\al \ga} \, \vep_{\dbe \dde} \, z_{15} \, z_{16} \, z_{27} \, z_{28} \, z_{35} \, z_{36} \, z_{47} \, z_{48} \biggr. \notag \\
& \ \ \ \ \biggl. - \ \frac{4 \, \eta^{\mu \la} \, \eta^{\nu \rho}}{z_{13} \, z_{24}} \; \vep_{\al \ga} \, \vep_{\dbe \dde} \, z_{15} \, z_{16} \, z_{25} \, z_{26} \, z_{37} \, z_{38} \, z_{47} \, z_{48}  \biggr. \notag \\
& \ \ \ \ \biggl. + \ \frac{4 \, \eta^{\mu \rho} \, \eta^{\nu \la}}{z_{14} \, z_{23}} \; \vep_{\al \ga} \, \vep_{\dbe \dde} \, z_{15} \, z_{16} \, z_{25} \, z_{26} \, z_{37} \, z_{38} \, z_{47} \, z_{48} \biggr. \notag \\
& \ \ \ \ \biggl. + \ \frac{2 \, \eta^{\mu \nu} }{z_{12}} \; \Bigl[ \vep_{\al \ga} \, (\vep \, \bar{\si}^{\la} \, \si^{\rho})_{\dbe \dde} \, z_{68} \, z_{15} \, z_{16} \, z_{27} \, z_{28} \, z_{35}  \, z_{47} \ + \  \vep_{\dbe \dde} \, (\si^{\la} \, \bar{\si}^{\rho} \, \vep)_{\al \ga} \, z_{57} \, z_{15} \, z_{16} \, z_{27} \, z_{28} \, z_{36} \, z_{48} \Bigr] \biggr. \notag \\
& \ \ \ \ \biggl. - \ \frac{2 \, \eta^{\mu \nu} }{z_{12}} \; \si^{\la}_{\al \dbe} \, \si^{\rho}_{\ga \dde} \, z_{57} \, z_{68} \, z_{34} \, z_{15} \, z_{17} \, z_{26} \, z_{28} \notag \\
& \ \ \ \ \biggl. + \ \frac{2 \, \eta^{\la \rho} }{z_{34}} \; \Bigl[ \vep_{\al \ga} \, (\vep \, \bar{\si}^{\mu} \, \si^{\nu})_{\dbe \dde} \, z_{68} \, z_{35} \, z_{36} \, z_{47} \, z_{48} \, z_{15}  \, z_{27} \ + \  \vep_{\dbe \dde} \, (\si^{\mu} \, \bar{\si}^{\nu} \, \vep)_{\al \ga} \, z_{57} \, z_{35} \, z_{36} \, z_{47} \, z_{48} \, z_{16} \, z_{28} \Bigr] \biggr. \notag \\
& \ \ \ \ \biggl. - \ \frac{2 \, \eta^{\la \rho} }{z_{34}} \; \si^{\mu}_{\al \dbe} \, \si^{\nu}_{\ga \dde} \, z_{57} \, z_{68} \, z_{12} \, z_{35} \, z_{37} \, z_{46} \, z_{48} \biggr. \notag \\
& \ \ \ \ \biggl. - \ \frac{2 \, \eta^{\mu \la} }{z_{13}} \; \Bigl[ \vep_{\al \ga} \, (\vep \, \bar{\si}^{\nu} \, \si^{\rho})_{\dbe \dde} \, z_{68} \, z_{15} \, z_{16} \, z_{37} \, z_{38} \, z_{25}  \, z_{47} \ + \  \vep_{\dbe \dde} \, (\si^{\nu} \, \bar{\si}^{\rho} \, \vep)_{\al \ga} \, z_{57} \, z_{15} \, z_{16} \, z_{37} \, z_{38} \, z_{26} \, z_{48} \Bigr] \biggr. \notag \\
& \ \ \ \ \biggl. + \ \frac{2 \, \eta^{\mu \la} }{z_{13}} \; \si^{\nu}_{\al \dbe} \, \si^{\rho}_{\ga \dde} \, z_{57} \, z_{68} \, z_{24} \, z_{15} \, z_{17} \, z_{36} \, z_{38} \biggr. \notag \\
& \ \ \ \ \biggl. - \ \frac{2 \, \eta^{\nu \rho} }{z_{24}} \; \Bigl[ \vep_{\al \ga} \, (\vep \, \bar{\si}^{\mu} \, \si^{\la})_{\dbe \dde} \, z_{68} \, z_{25} \, z_{26} \, z_{47} \, z_{48} \, z_{15}  \, z_{37} \ + \  \vep_{\dbe \dde} \, (\si^{\mu} \, \bar{\si}^{\la} \, \vep)_{\al \ga} \, z_{57} \, z_{25} \, z_{26} \, z_{47} \, z_{48} \, z_{16} \, z_{38} \Bigr] \biggr. \notag \\
& \ \ \ \ \biggl. + \ \frac{2 \, \eta^{\nu \rho} }{z_{24}} \; \si^{\mu}_{\al \dbe} \, \si^{\la}_{\ga \dde} \, z_{57} \, z_{68} \, z_{13} \, z_{25} \, z_{27} \, z_{46} \, z_{48} \biggr. \notag \\
& \ \ \ \ \biggl. + \ \frac{2 \, \eta^{\mu \rho} }{z_{14}} \; \Bigl[ \vep_{\al \ga} \, (\vep \, \bar{\si}^{\nu} \, \si^{\la})_{\dbe \dde} \, z_{68} \, z_{15} \, z_{16} \, z_{47} \, z_{48} \, z_{25}  \, z_{37} \ + \  \vep_{\dbe \dde} \, (\si^{\nu} \, \bar{\si}^{\la} \, \vep)_{\al \ga} \, z_{57} \, z_{15} \, z_{16} \, z_{47} \, z_{48} \, z_{26} \, z_{38} \Bigr] \biggr. \notag \\
& \ \ \ \ \biggl. - \ \frac{2 \, \eta^{\mu \rho} }{z_{14}} \; \si^{\nu}_{\al \dbe} \, \si^{\la}_{\ga \dde} \, z_{57} \, z_{68} \, z_{23} \, z_{15} \, z_{17} \, z_{46} \, z_{48} \biggr. \notag \\
& \ \ \ \ \biggl. + \ \frac{2 \, \eta^{\nu \la} }{z_{23}} \; \Bigl[ \vep_{\al \ga} \, (\vep \, \bar{\si}^{\mu} \, \si^{\rho})_{\dbe \dde} \, z_{68} \, z_{25} \, z_{26} \, z_{37} \, z_{38} \, z_{15}  \, z_{47} \ + \  \vep_{\dbe \dde} \, (\si^{\mu} \, \bar{\si}^{\rho} \, \vep)_{\al \ga} \, z_{57} \, z_{25} \, z_{26} \, z_{37} \, z_{38} \, z_{16} \, z_{48} \Bigr] \biggr. \notag \\
& \ \ \ \ \biggl. - \ \frac{2 \, \eta^{\nu \la} }{z_{23}} \; \si^{\mu}_{\al \dbe} \, \si^{\rho}_{\ga \dde} \, z_{57} \, z_{68} \, z_{14} \, z_{25} \, z_{27} \, z_{36} \, z_{38} \biggr. \notag \\
& \ \ \ \ \biggl. + \ 2 \, \eta^{\nu \rho} \, \vep_{\al \ga} \, (\vep \, \bar{\si}^{\mu} \,
  \si^{\la})_{\dbe \dde} \, z_{57} \, z_{68} \, z_{15} \, z_{23} \, z_{48} \, z_{67} \ - \ 2 \, \eta^{\nu \la} \,
  \vep_{\al \ga} \, (\vep\,\sib^{\mu} \, \si^{\rho})_{\dbe \dde}  \, z_{57} \, z_{68} \, z_{15} \, z_{23} \,
  z_{48} \, z_{67} \biggr. \notag \\
& \ \ \ \ \biggl. + \ 2 \, \eta^{\mu \la} \, \si^{\nu}_{\al \dbe} \, \si^{\rho}_{\ga \dde} \, z_{57} \, z_{68} \, z_{15} \, z_{23} \, z_{48} \, z_{67} \ - \ 2 \, \eta^{\mu \rho} \, \si^{\nu}_{\al \dbe} \, \si^{\la}_{\ga \dde} \, z_{57} \, z_{68} \, z_{15} \, z_{23} \, z_{48} \, z_{67} \biggr. \notag \\
& \ \ \ \ \biggl. + \ \vep_{\al \ga} \, (\vep \, \bar{\si}^{\mu} \, \si^{\nu} \, \bar{\si}^{\la} \, \si^{\rho})_{\dbe
  \dde} \, z_{68}^2 \, z_{15} \, z_{27} \, z_{35} \, z_{47} \ + \ \vep_{\dbe \dde} \, (\si^{\mu} \, \bar{\si}^{\nu} \, \si^{\la} \, \bar{\si}^{\rho} \, \vep)_{\al \ga} \, z_{57}^2 \, z_{16} \, z_{28} \, z_{36} \, z_{48} \biggr. \notag \\
& \ \ \ \ + \ \si^{\rho}_{\ga\dde}\,(\si^{\mu}\,\bar{\si}^{\nu}\,\si^{\la})_{\al\dbe}\,\bigl[z_{15}\,z_{28}\,z_{36}\,z_{47}-z_{14}\,z_{28}\,z_{37}\,z_{56}\bigr]z_{57}\,z_{68}\ + \ \si^{\nu}_{\ga\dde}\,(\si^{\la}\,\bar{\si}^{\rho}\,\si^{\mu})_{\al\dbe}\,z_{15}\,z_{23}\,z_{48}\,z_{67}\,z_{57}\,z_{68}\notag \biggr.\\
& \ \ \ \ - \ \si^{\mu}_{\ga\dbe}\,(\si^{\nu}\,\bar{\si}^{\la}\,\si^{\rho})_{\al\dde}\,\bigl[z_{15}\,z_{28}\,z_{36}\,z_{47}+z_{15}\,z_{23}\,z_{46}\,z_{78}\bigr]z_{57}\,z_{68}\ - \ \si^{\la}_{\ga\dbe}\,(\si^{\mu}\,\bar{\si}^{\nu}\,\si^{\rho})_{\al\dde}\,z_{16}\,z_{23}\,z_{47}\,z_{58}\,z_{57}\,z_{68}\notag \biggr.\\
& \ \ \ \ + \
\si^{\mu}_{\al\dbe}\,(\si^{\nu}\,\bar{\si}^{\la}\,\si^{\rho})_{\ga\dde}\,\bigl[z_{15}\,z_{23}\,z_{46}\,z_{78}-z_{14}\,z_{25}\,z_{36}\,z_{78}+z_{16}\,z_{25}\,z_{37}\,z_{48}\bigr]z_{57}\,z_{68}\notag
\biggr
.\\
& \ \ \ \ - \ \si^{\rho}_{\ga\dbe}\,(\si^{\mu}\,\bar{\si}^{\nu}\,\si^{\la})_{\al\dde}\,z_{16}\,z_{25}\,z_{37}\,z_{48}\,z_{57}\,z_{68}\biggr\}\ .
\label{exp9}
\end{align}

On the other hand, for the case $n=4,\ r=4,\ s=0$ we compute:
\begin{align}
\langle &\psi^{\mu}(z_{1}) \, \psi^{\nu}(z_{2}) \, \psi^{\la}(z_{3}) \, \psi^{\rho}(z_{4}) \, S_{\al}(z_{5}) \, S_{\be}(z_{6}) \, S_{\ga}(z_{7}) \, S_{\de}(z_{8}) \rangle \notag \\
&= \ \ \frac{1}{ \bigl(z_{15} \, z_{16} \, z_{17} \, z_{18} \, z_{25} \, z_{26} \, z_{27} \, z_{28} \, z_{35} \, z_{36} \, z_{37} \, z_{38} \, z_{45} \, z_{46} \, z_{47} \, z_{48} \bigr)^{1/2} \, \bigl(z_{56} \, z_{57} \, z_{58} \, z_{67} \, z_{68} \, z_{78} \bigr)^{1/2} } \notag \\
& \ \ \times \ \biggl\{ \, \frac{\eta^{\mu \nu} \, \eta^{\la \rho}}{z_{12} \, z_{34}} \; \bigl( \vep_{\al \be} \, \vep_{\ga \de} \, z_{58} \, z_{67} \ + \ \vep_{\al \de} \, \vep_{\ga \be} \, z_{56} \, z_{78} \bigr) \, \  \, (z_{15} \, z_{26} \, z_{17} \, z_{28}) \, \  \, ( z_{35} \, z_{46} \, z_{37} \, z_{48}) \biggr. \notag \\
& \ \ \ \ \ \ - \ \frac{\eta^{\mu \la} \, \eta^{\nu \rho}}{z_{13} \, z_{24}} \; \bigl( \vep_{\al \be} \, \vep_{\ga \de} \, z_{58} \, z_{67} \ + \ \vep_{\al \de} \, \vep_{\ga \be} \, z_{56} \, z_{78} \bigr)  \, \  \, (z_{15} \, z_{36} \, z_{17} \, z_{38}) \, \  \, ( z_{25} \, z_{46} \, z_{27} \, z_{48}) \Biggr. \notag \notag \\
& \ \ \ \ \ \ + \ \frac{\eta^{\mu \rho} \, \eta^{\nu \la}}{z_{14} \, z_{23}} \; \bigl( \vep_{\al \be} \, \vep_{\ga \de} \, z_{58} \, z_{67} \ + \ \vep_{\al \de} \, \vep_{\ga \be} \, z_{56} \, z_{78} \bigr) \, \  \, (z_{15} \, z_{46} \, z_{17} \, z_{48}) \, \  \, ( z_{25} \, z_{36} \, z_{27} \, z_{38}) \notag \\
& \ \ \ \ \ \ + \ \frac{\eta^{\mu \nu}}{z_{12}} \; \left[ (\si^{\la} \, \bar{\si}^{\rho} \, \vep)_{\al \be} \,  \frac{z_{56}}{2} \right] \, \bigl(\vep_{\ga \de} \, z_{58} \, z_{67} \bigr) \, \  \, (z_{37} \, z_{48}) \, \  \, (z_{15} \, z_{26} \, z_{17} \, z_{28}) \notag \\
& \ \ \ \ \ \ + \ \frac{\eta^{\mu \nu}}{z_{12}} \; \left[ (\si^{\la} \, \bar{\si}^{\rho} \, \vep)_{\al \de} \,  \frac{z_{58}}{2} \right] \, \bigl(\vep_{\ga \be} \, z_{56} \, z_{78} \bigr) \, \  \, (z_{37} \, z_{46}) \, \  \, (z_{15} \, z_{26} \, z_{17} \, z_{28}) \notag \\
& \ \ \ \ \ \ + \ \frac{\eta^{\mu \nu}}{z_{12}} \; \left[ (\si^{\la} \, \bar{\si}^{\rho} \, \vep)_{\ga \de} \,  \frac{z_{78}}{2} \right] \, \bigl(\vep_{\al \be} \, z_{58} \, z_{67} \bigr) \, \  \, (z_{35} \, z_{46}) \, \  \, (z_{15} \, z_{26} \, z_{17} \, z_{28}) \notag \\
& \ \ \ \ \ \ + \ \frac{\eta^{\mu \nu}}{z_{12}} \; \left[ (\si^{\la} \, \bar{\si}^{\rho} \, \vep)_{\ga \be} \,  \frac{z_{76}}{2} \right] \, \bigl(\vep_{\al \de} \, z_{56} \, z_{78} \bigr) \, \  \, (z_{35} \, z_{48}) \, \  \, (z_{15} \, z_{26} \, z_{17} \, z_{28}) \notag \\
& \ \ \ \ \ \ + \ \frac{\eta^{\la \rho}}{z_{34}} \; \left[ (\si^{\mu} \, \bar{\si}^{\nu} \, \vep)_{\al \be} \,  \frac{z_{56}}{2} \right] \, \bigl(\vep_{\ga \de} \, z_{58} \, z_{67} \bigr) \, \  \, (z_{17} \, z_{28}) \, \  \, (z_{35} \, z_{46} \, z_{37} \, z_{48}) \notag \\
& \ \ \ \ \ \ + \ \frac{\eta^{\la \rho}}{z_{34}} \; \left[ (\si^{\mu} \, \bar{\si}^{\nu} \, \vep)_{\al \de} \,  \frac{z_{58}}{2} \right] \, \bigl(\vep_{\ga \be} \, z_{56} \, z_{78} \bigr) \, \  \, (z_{17} \, z_{26}) \, \  \, (z_{35} \, z_{46} \, z_{37} \, z_{48}) \notag \\
& \ \ \ \ \ \ + \ \frac{\eta^{\la \rho}}{z_{34}} \; \left[ (\si^{\mu} \, \bar{\si}^{\nu} \, \vep)_{\ga \de} \,  \frac{z_{78}}{2} \right] \, \bigl(\vep_{\al \be} \, z_{58} \, z_{67} \bigr) \, \  \, (z_{15} \, z_{26}) \, \  \, (z_{35} \, z_{46} \, z_{37} \, z_{48}) \notag \\
& \ \ \ \ \ \ + \ \frac{\eta^{\la \rho}}{z_{34}} \; \left[ (\si^{\mu} \, \bar{\si}^{\nu} \, \vep)_{\ga \be} \,  \frac{z_{76}}{2} \right] \, \bigl(\vep_{\al \de} \, z_{56} \, z_{78} \bigr) \, \  \, (z_{15} \, z_{28}) \, \  \, (z_{35} \, z_{46} \, z_{37} \, z_{48}) \notag \\
& \ \ \ \ \ \ - \ \frac{\eta^{\mu \la}}{z_{13}} \; \left[ (\si^{\nu} \, \bar{\si}^{\rho} \, \vep)_{\al \be} \,  \frac{z_{56}}{2} \right] \, \bigl(\vep_{\ga \de} \, z_{58} \, z_{67} \bigr) \, \  \, (z_{27} \, z_{48}) \, \  \, (z_{15} \, z_{36} \, z_{17} \, z_{38}) \notag \\
& \ \ \ \ \ \ - \ \frac{\eta^{\mu \la}}{z_{13}} \; \left[ (\si^{\nu} \, \bar{\si}^{\rho} \, \vep)_{\al \de} \,  \frac{z_{58}}{2} \right] \, \bigl(\vep_{\ga \be} \, z_{56} \, z_{78} \bigr) \, \  \, (z_{27} \, z_{46}) \, \  \, (z_{15} \, z_{36} \, z_{17} \, z_{38}) \notag \\
& \ \ \ \ \ \ - \ \frac{\eta^{\mu \la}}{z_{13}} \; \left[ (\si^{\nu} \, \bar{\si}^{\rho} \, \vep)_{\ga \de} \,  \frac{z_{78}}{2} \right] \, \bigl(\vep_{\al \be} \, z_{58} \, z_{67} \bigr) \, \  \, (z_{25} \, z_{46}) \, \  \, (z_{15} \, z_{36} \, z_{17} \, z_{38}) \notag \\
& \ \ \ \ \ \ - \ \frac{\eta^{\mu \la}}{z_{13}} \; \left[ (\si^{\nu} \, \bar{\si}^{\rho} \, \vep)_{\ga \be} \,  \frac{z_{76}}{2} \right] \, \bigl(\vep_{\al \de} \, z_{56} \, z_{78} \bigr) \, \  \, (z_{25} \, z_{48}) \, \  \, (z_{15} \, z_{36} \, z_{17} \, z_{38}) \notag \\
& \ \ \ \ \ \ - \ \frac{\eta^{\nu \rho}}{z_{24}} \; \left[ (\si^{\mu} \, \bar{\si}^{\la} \, \vep)_{\al \be} \,  \frac{z_{56}}{2} \right] \, \bigl(\vep_{\ga \de} \, z_{58} \, z_{67} \bigr) \, \  \, (z_{17} \, z_{38}) \, \  \, (z_{25} \, z_{46} \, z_{27} \, z_{48}) \notag \\
& \ \ \ \ \ \ - \ \frac{\eta^{\nu \rho}}{z_{24}} \; \left[ (\si^{\mu} \, \bar{\si}^{\la} \, \vep)_{\al \de} \,  \frac{z_{58}}{2} \right] \, \bigl(\vep_{\ga \be} \, z_{56} \, z_{78} \bigr) \, \  \, (z_{17} \, z_{36}) \, \  \, (z_{25} \, z_{46} \, z_{27} \, z_{48}) \notag \\
& \ \ \ \ \ \ - \ \frac{\eta^{\nu \rho}}{z_{24}} \; \left[ (\si^{\mu} \, \bar{\si}^{\la} \, \vep)_{\ga \de} \,  \frac{z_{78}}{2} \right] \, \bigl(\vep_{\al \be} \, z_{58} \, z_{67} \bigr) \, \  \, (z_{15} \, z_{36}) \, \  \, (z_{25} \, z_{46} \, z_{27} \, z_{48}) \notag \\
& \ \ \ \ \ \ - \ \frac{\eta^{\nu \rho}}{z_{24}} \; \left[ (\si^{\mu} \, \bar{\si}^{\la} \, \vep)_{\ga \be} \,  \frac{z_{76}}{2} \right] \, \bigl(\vep_{\al \de} \, z_{56} \, z_{78} \bigr) \, \  \, (z_{15} \, z_{38}) \, \  \, (z_{25} \, z_{46} \, z_{27} \, z_{48}) \notag \\
& \ \ \ \ \ \ + \ \frac{\eta^{\mu \rho}}{z_{14}} \; \left[ (\si^{\nu} \, \bar{\si}^{\la} \, \vep)_{\al \be} \,  \frac{z_{56}}{2} \right] \, \bigl(\vep_{\ga \de} \, z_{58} \, z_{67}\bigr) \, \  \, (z_{27} \, z_{38}) \, \  \, (z_{15} \, z_{46} \, z_{17} \, z_{48}) \notag \\
& \ \ \ \ \ \ + \ \frac{\eta^{\mu \rho}}{z_{14}} \; \left[ (\si^{\nu} \, \bar{\si}^{\la} \, \vep)_{\al \de} \,  \frac{z_{58}}{2} \right] \, \bigl(\vep_{\ga \be} \, z_{56} \, z_{78}\bigr) \, \  \, (z_{27} \, z_{36}) \, \  \, (z_{15} \, z_{46} \, z_{17} \, z_{48}) \notag \\
& \ \ \ \ \ \ + \ \frac{\eta^{\mu \rho}}{z_{14}} \; \left[ (\si^{\nu} \, \bar{\si}^{\la} \, \vep)_{\ga \de} \,  \frac{z_{78}}{2} \right] \, \bigl(\vep_{\al \be} \, z_{58} \, z_{67}\bigr) \, \  \, (z_{25} \, z_{36}) \, \  \, (z_{15} \, z_{46} \, z_{17} \, z_{48}) \notag \\
& \ \ \ \ \ \ + \ \frac{\eta^{\mu \rho}}{z_{14}} \; \left[ (\si^{\nu} \, \bar{\si}^{\la} \, \vep)_{\ga \be} \,  \frac{z_{76}}{2} \right] \, \bigl(\vep_{\al \de} \, z_{56} \, z_{78}\bigr) \, \  \, (z_{25} \, z_{38}) \, \  \, (z_{15} \, z_{46} \, z_{17} \, z_{48}) \notag \\
& \ \ \ \ \ \ + \ \frac{\eta^{\nu \la}}{z_{23}} \; \left[ (\si^{\mu} \, \bar{\si}^{\rho} \, \vep)_{\al \be} \,  \frac{z_{56}}{2} \right] \, \bigl(\vep_{\ga \de} \, z_{58} \, z_{67}\bigr) \, \  \, (z_{17} \, z_{48}) \, \  \, (z_{25} \, z_{36} \, z_{27} \, z_{38})  \notag \\
& \ \ \ \ \ \ + \ \frac{\eta^{\nu \la}}{z_{23}} \; \left[ (\si^{\mu} \, \bar{\si}^{\rho} \, \vep)_{\al \de} \,  \frac{z_{58}}{2} \right] \, \bigl(\vep_{\ga \be} \, z_{56} \, z_{78}\bigr) \, \  \, (z_{17} \, z_{46}) \, \  \, (z_{25} \, z_{36} \, z_{27} \, z_{38})  \notag \\
& \ \ \ \ \ \ + \ \frac{\eta^{\nu \la}}{z_{23}} \; \left[ (\si^{\mu} \, \bar{\si}^{\rho} \, \vep)_{\ga \de} \,  \frac{z_{78}}{2} \right] \, \bigl(\vep_{\al \be} \, z_{58} \, z_{67}\bigr) \, \  \, (z_{15} \, z_{46}) \, \  \, (z_{25} \, z_{36} \, z_{27} \, z_{38}) \notag \\
& \ \ \ \ \ \ + \ \frac{\eta^{\nu \la}}{z_{23}} \; \left[ (\si^{\mu} \, \bar{\si}^{\rho} \, \vep)_{\ga \be} \,  \frac{z_{76}}{2} \right] \, \bigl(\vep_{\al \de} \, z_{56} \, z_{78}\bigr) \, \  \, (z_{15} \, z_{48}) \, \  \, (z_{25} \, z_{36} \, z_{27} \, z_{38})   \notag \\
& \ \ \ \ \ \ + \ (\si^{\mu} \, \bar{\si}^{\nu} \, \si^{\la} \, \bar{\si}^{\rho} \, \vep)_{\al \be} \; \left(\frac{z_{56}}{2} \right)^{2} \; \bigl(\vep_{\ga \de} \, z_{58} \, z_{67} \bigr) \, \  \, (z_{17} \, z_{28} \, z_{37} \, z_{48}) \notag \\
& \ \ \ \ \ \ + \ (\si^{\mu} \, \bar{\si}^{\nu} \, \si^{\la} \, \bar{\si}^{\rho} \, \vep)_{\al \de} \; \left(\frac{z_{58}}{2} \right)^{2} \; \bigl(\vep_{\ga \be} \, z_{56} \, z_{78} \bigr) \, \  \, (z_{17} \, z_{26} \, z_{37} \, z_{46}) \notag \\
& \ \ \ \ \ \ + \ (\si^{\mu} \, \bar{\si}^{\nu} \, \si^{\la} \, \bar{\si}^{\rho} \, \vep)_{\ga \de} \; \left(\frac{z_{78}}{2} \right)^{2} \; \bigl(\vep_{\al \be} \, z_{58} \, z_{67} \bigr) \, \  \, (z_{15} \, z_{26} \, z_{35} \, z_{46}) \notag \\
& \ \ \ \ \ \ + \ (\si^{\mu} \, \bar{\si}^{\nu} \, \si^{\la} \, \bar{\si}^{\rho} \, \vep)_{\ga \be} \; \left(\frac{z_{76}}{2} \right)^{2} \; \bigl(\vep_{\al \de} \, z_{56} \, z_{78} \bigr) \, \  \, (z_{15} \, z_{28} \, z_{35} \, z_{48}) \notag \\
& \ \ \ \ \ \ + \  \left[ (\si^{\mu} \, \bar{\si}^{\nu} \, \vep)_{\al \be} \; \frac{z_{56}}{2} \right] \, z_{58} \, z_{67} \, \left[ (\si^{\la} \, \bar{\si}^{\rho} \, \vep)_{\ga \de} \; \frac{z_{78}}{2} \right] \, (z_{17} \, z_{28} \, z_{35} \, z_{46} \ - \ z_{17}\, z_{25} \, z_{38}  \, z_{46}) \notag \\
& \ \ \ \ \ \ + \ \left[ (\si^{\mu} \, \bar{\si}^{\nu} \, \vep)_{\al \de} \; \frac{z_{58}}{2} \right] \, z_{56} \, z_{78} \, \left[ (\si^{\la} \, \bar{\si}^{\rho} \, \vep)_{\ga \be} \; \frac{z_{76}}{2} \right] \, (z_{17} \, z_{26} \, z_{35} \, z_{48} \ - \ z_{17}\, z_{25} \, z_{36}  \, z_{48}) \notag \\
& \ \ \ \ \ \ + \ \left[ (\si^{\mu} \, \bar{\si}^{\nu} \, \vep)_{\ga \de} \; \frac{z_{78}}{2} \right] \, z_{58} \, z_{67} \, \left[ (\si^{\la} \, \bar{\si}^{\rho} \, \vep)_{\al \be} \; \frac{z_{56}}{2} \right] \, (z_{15} \, z_{26} \, z_{37} \, z_{48} \ - \ z_{15}\, z_{27} \, z_{36}  \, z_{48}) \notag \\
& \ \ \ \ \ \ +  \ \left[ (\si^{\mu} \, \bar{\si}^{\nu} \, \vep)_{\ga \be} \; \frac{z_{76}}{2} \right] \, z_{56} \, z_{78} \, \left[ (\si^{\la} \, \bar{\si}^{\rho} \, \vep)_{\al \de} \; \frac{z_{58}}{2} \right] \, (z_{15} \, z_{28} \, z_{37} \, z_{46} \ - \ z_{15}\, z_{27} \, z_{38}  \, z_{46})  \notag \\
& \ \ \ \ \ \ + \ \left[ (\si^{\nu} \, \bar{\si}^{\la} \, \vep)_{\al \be} \; \frac{z_{56}}{2} \right] \, z_{58} \, z_{67} \, \left[ (\si^{\mu} \, \bar{\si}^{\rho} \, \vep)_{\ga \de} \; \frac{z_{78}}{2} \right] \, (z_{15} \, z_{27} \, z_{38} \, z_{46} ) \notag \\
& \ \ \ \ \ \ + \ \left[ (\si^{\nu} \, \bar{\si}^{\la} \, \vep)_{\al \de} \; \frac{z_{58}}{2} \right] \, z_{56} \, z_{78} \, \left[ (\si^{\mu} \, \bar{\si}^{\rho} \, \vep)_{\ga \be} \; \frac{z_{76}}{2} \right] \, (z_{15} \, z_{27} \, z_{36} \, z_{48} ) \notag \\
& \ \ \ \ \ \ + \ \left[ (\si^{\nu} \, \bar{\si}^{\la} \, \vep)_{\ga \de} \; \frac{z_{78}}{2} \right] \, z_{58} \, z_{67} \, \left[ (\si^{\mu} \, \bar{\si}^{\rho} \, \vep)_{\al \be} \; \frac{z_{56}}{2} \right] \, (z_{17} \, z_{25} \, z_{36} \, z_{48} ) \notag \\
& \ \ \ \ \ \ \ \biggl. + \ \left[ (\si^{\nu} \, \bar{\si}^{\la} \, \vep)_{\ga \be} \; \frac{z_{76}}{2} \right] \, z_{56} \, z_{78} \, \left[ (\si^{\mu} \, \bar{\si}^{\rho} \, \vep)_{\al \de} \; \frac{z_{58}}{2} \right] \, (z_{17} \, z_{25} \, z_{38} \, z_{46} )  \biggr\}\ . 
\label{exp10}
\end{align}
The correlators \req{exp9} and \req{exp10} are key ingredients for computing
the disk amplitude involving two gauge vectors and two Ramond closed string
states in superstring theory.

Furthermore, for the case $n=2,\ r=4,\ s=2$ we obtain:
\begin{align}
\langle \psi^{\mu}&(z_{1}) \, \psi^{\nu}(z_{2}) \, S_{\al}(z_{3}) \, S_{\be}(z_{4}) \, S_{\ga}(z_{5}) \, S_{\de}(z_{6}) \, S_{\dep}(z_7) \, S_{\dze}(z_8) \rangle \eq \frac{1}{2 \, \bigl( z_{34} \, z_{35} \, z_{36} \, z_{45} \, z_{46} \, z_{56} \bigr)^{1/2}} \notag \\
& \times \ \frac{1}{\bigl( z_{13} \, z_{14} \, z_{15} \, z_{16} \, z_{17} \, z_{18} \bigr)^{1/2} \, \bigl( z_{23} \, z_{24} \, z_{25} \, z_{26} \, z_{27} \, z_{28} \bigr)^{1/2} \, z_{78}^{1/2} } \notag \\
& \times \ \biggl\{ \frac{2 \, \eta^{\mu \nu}}{z_{12}} \; \bigl( \vep_{\al \be} \, \vep_{\ga \de} \, z_{36} \, z_{45} \ + \ \vep_{\al \de} \, \vep_{\ga \be} \, z_{34} \, z_{56} \bigr) \, \vep_{\dep \dze} \, z_{13} \, z_{24} \, z_{15} \, z_{26} \  (z_{17} \, z_{28}) \biggr. \notag \\
& \ \ \ \ \ \ \biggl. + \ \Bigl[ \bigl( \vep \, \bar{\si}^{\mu} \, \si^{\nu} \bigr)_{\dep \dze} \, z_{78} \Bigr] \; \bigl( \vep_{\al \be} \, \vep_{\ga \de} \, z_{36} \, z_{45} \ + \ \vep_{\al \de} \, \vep_{\ga \be} \, z_{34} \, z_{56} \bigr) \,  z_{13} \, z_{24} \, z_{15} \, z_{26}  \biggr. \notag \\
& \ \ \ \ \ \ \biggl. + \ \Bigl[ \si^{\mu}_{\al \dep} \, \si^{\nu}_{\be \dze} \, z_{18} \, z_{27} \ - \ \si^{\mu}_{\al \dze} \, \si^{\nu}_{\be \dep} \, z_{17} \, z_{28} \Bigr] \,  \bigl( \vep_{\ga \de} \, z_{36} \, z_{45} \bigr) \, z_{15} \, z_{26} \, z_{34} \biggr. \notag \\
& \ \ \ \ \ \ \biggl. + \ \Bigl[ \si^{\mu}_{\ga \dep} \, \si^{\nu}_{\be \dze} \, z_{18} \, z_{27} \ - \ \si^{\mu}_{\ga \dze} \, \si^{\nu}_{\be \dep} \, z_{17} \, z_{28} \Bigr] \,  \bigl( \vep_{\al \de} \, z_{34} \, z_{56} \bigr) \, z_{13} \, z_{26} \, z_{54} \biggr. \notag \\
& \ \ \ \ \ \ \biggl. + \ \Bigl[ \si^{\mu}_{\al \dep} \, \si^{\nu}_{\de \dze} \, z_{18} \, z_{27} \ - \ \si^{\mu}_{\al \dze} \, \si^{\nu}_{\de \dep} \, z_{17} \, z_{28} \Bigr] \,  \bigl( \vep_{\ga \be} \, z_{34} \, z_{56} \bigr) \, z_{15} \, z_{24} \, z_{36} \biggr. \notag \\
& \ \ \ \ \ \ \biggl. + \ \Bigl[ \si^{\mu}_{\ga \dep} \, \si^{\nu}_{\de \dze} \, z_{18} \, z_{27} \ - \ \si^{\mu}_{\ga \dze} \, \si^{\nu}_{\de \dep} \, z_{17} \, z_{28} \Bigr] \,  \bigl( \vep_{\al \be} \, z_{36} \, z_{45} \bigr) \, z_{13} \, z_{24} \, z_{56} \biggr\}\ .
\label{exp11}
\end{align}

Finally for $n=2,\ r=6,\ s=0$ we get:
\begin{align}
  \vev{\psi^{\mu}(z_1)\,\psi^\nu(z_2)\,&S_\al(z_3)\,S_\be(z_4)\,S_\ga(z_5)\,S_\de(z_6)\,S_\ep(z_7)\,S_\ze(z_8)}\ =\
  \frac{z_{12}^{1/2}}{2} \,\prod_{i<j}^8z_{ij}^{-1/2}\notag\\
  &\times\Big(- \; \frac{2 \, \eta^{\mu\nu}}{z_{12}} \; \vep_{\al\be}\,\vep_{\ga\de}\,\vep_{\ep\ze}\,z_{14}\,z_{16}\,z_{18}\,z_{23}\,z_{25}\,z_{27}\,z_{36}\,z_{38}\,z_{45}\,z_{47}\,z_{58}\,z_{67}\notag\\
  &\hspace{.8cm}-\; \frac{2 \, \eta^{\mu\nu}}{z_{12}} \; \vep_{\al\be}\,\vep_{\ga\ze}\,\vep_{\ep\de}\,z_{14}\,z_{16}\,z_{18}\,z_{23}\,z_{25}\,z_{27}\,z_{36}\,z_{38}\,z_{45}\,z_{47}\,z_{56}\,z_{78}\notag\\
  &\hspace{.8cm}+ \; \frac{2 \, \eta^{\mu\nu}}{z_{12}} \; \vep_{\al\de}\,\vep_{\be\ga}\,\vep_{\ep\ze}\,z_{14}\,z_{16}\,z_{18}\,z_{23}\,z_{25}\,z_{27}\,z_{34}\,z_{38}\,z_{47}\,z_{56}\,z_{58}\,z_{67}\notag\\
  &\hspace{.8cm}- \; \frac{2 \, \eta^{\mu\nu}}{z_{12}} \; \vep_{\al\de}\,\vep_{\be\ep}\,\vep_{\ga\ze}\,z_{14}\,z_{16}\,z_{18}\,z_{23}\,z_{25}\,z_{27}\,z_{34}\,z_{38}\,z_{45}\,z_{56}\,z_{67}\,z_{78}\notag\\
  &\hspace{.8cm}+ \; \frac{2 \, \eta^{\mu\nu}}{z_{12}} \; \vep_{\al\ze}\,\vep_{\be\ep}\,\vep_{\ga\de}\,z_{14}\,z_{16}\,z_{18}\,z_{23}\,z_{25}\,z_{27}\,z_{34}\,z_{36}\,z_{45}\,z_{58}\,z_{67}\,z_{78}\notag\\
  &\hspace{.8cm}- \; \frac{2 \, \eta^{\mu\nu}}{z_{12}} \; \vep_{\al\ze}\,\vep_{\be\ga}\,\vep_{\de\ep}\,z_{14}\,z_{16}\,z_{18}\,z_{23}\,z_{25}\,z_{27}\,z_{34}\,z_{36}\,z_{47}\,z_{56}\,z_{58}\,z_{78}\notag\\
  &\hspace{.8cm}- \, (\si^{\mu} \, \bar{\si}^{\nu} \vep)_{\be\al} \,\vep_{\ga\de}\,\vep_{\ep\ze}\,z_{16}\,z_{18}\,z_{25}\,z_{27}\,z_{34}\,z_{36}\,z_{38}\,z_{45}\,z_{47}\,z_{58}\,z_{67}\notag\\
  &\hspace{.8cm}+ \,  (\si^{\mu} \, \bar{\si}^{\nu} \vep)_{\be\al} \,\vep_{\ga\ze}\,\vep_{\de\ep}\,z_{16}\,z_{18}\,z_{25}\,z_{27}\,z_{34}\,z_{36}\,z_{38}\,z_{45}\,z_{47}\,z_{56}\,z_{78}\notag\\
  &\hspace{.8cm}+ \,  (\si^{\mu} \, \bar{\si}^{\nu} \vep)_{\be\ga} \,\vep_{\al\de}\,\vep_{\ep\ze}\,z_{16}\,z_{18}\,z_{23}\,z_{27}\,z_{34}\,z_{38}\,z_{45}\,z_{47}\,z_{56}\,z_{58}\,z_{67}\notag\\
  &\hspace{.8cm}- \,  (\si^{\mu} \, \bar{\si}^{\nu} \vep)_{\be\ga} \,\vep_{\al\ze}\,\vep_{\de\ep}\,z_{16}\,z_{18}\,z_{23}\,z_{27}\,z_{34}\,z_{36}\,z_{45}\,z_{47}\,z_{56}\,z_{58}\,z_{78}\notag\\ 
  &\hspace{.8cm}- \,  (\si^{\mu} \, \bar{\si}^{\nu} \vep)_{\be\ep} \,\vep_{\al\de}\,\vep_{\ga\ze}\,z_{16}\,z_{18}\,z_{23}\,z_{25}\,z_{34}\,z_{38}\,z_{45}\,z_{47}\,z_{56}\,z_{67}\,z_{78}\notag\\
  &\hspace{.8cm}+ \,  (\si^{\mu} \, \bar{\si}^{\nu} \vep)_{\be\ep} \,\vep_{\al\ze}\,\vep_{\ga\de}\,z_{16}\,z_{18}\,z_{23}\,z_{25}\,z_{34}\,z_{36}\,z_{45}\,z_{47}\,z_{58}\,z_{67}\,z_{78}\notag\\
  &\hspace{.8cm}+ \,  (\si^{\mu} \, \bar{\si}^{\nu} \vep)_{\de\al} \,\vep_{\be\ga}\,\vep_{\ep\ze}\,z_{14}\,z_{18}\,z_{25}\,z_{27}\,z_{34}\,z_{36}\,z_{38}\,z_{47}\,z_{56}\,z_{58}\,z_{67}\notag\\
  &\hspace{.8cm}- \,  (\si^{\mu} \, \bar{\si}^{\nu} \vep)_{\de\al} \,\vep_{\be\ep}\,\vep_{\ga\ze}\,z_{14}\,z_{18}\,z_{25}\,z_{27}\,z_{34}\,z_{36}\,z_{38}\,z_{45}\,z_{56}\,z_{67}\,z_{78}\notag\\
  &\hspace{.8cm}- \,  (\si^{\mu} \, \bar{\si}^{\nu} \vep)_{\de\ga} \,\vep_{\al\be}\,\vep_{\ep\ze}\,z_{14}\,z_{18}\,z_{23}\,z_{27}\,z_{36}\,z_{38}\,z_{45}\,z_{47}\,z_{56}\,z_{58}\,z_{67}\notag\\
  &\hspace{.8cm}+ \, (\si^{\mu} \, \bar{\si}^{\nu} \vep)_{\de\ga} \,\vep_{\al\ze}\,\vep_{\be\ep}\,z_{14}\,z_{18}\,z_{23}\,z_{27}\,z_{34}\,z_{36}\,z_{45}\,z_{56}\,z_{58}\,z_{67}\,z_{78}\notag\\
  &\hspace{.8cm}+ \,  (\si^{\mu} \, \bar{\si}^{\nu} \vep)_{\de\ep} \,\vep_{\al\be}\,\vep_{\ga\ze}\,z_{14}\,z_{18}\,z_{23}\,z_{25}\,z_{36}\,z_{38}\,z_{45}\,z_{47}\,z_{56}\,z_{67}\,z_{78}\notag\\
  &\hspace{.8cm}- \,  (\si^{\mu} \, \bar{\si}^{\nu} \vep)_{\de\ep} \,\vep_{\al\ze}\,\vep_{\be\ga}\,z_{14}\,z_{18}\,z_{23}\,z_{25}\,z_{34}\,z_{36}\,z_{47}\,z_{56}\,z_{58}\,z_{67}\,z_{78}\notag\\
  &\hspace{.8cm}- \,  (\si^{\mu} \, \bar{\si}^{\nu} \vep)_{\ze\al} \,\vep_{\be\ga}\,\vep_{\de\ep}\,z_{14}\,z_{16}\,z_{25}\,z_{27}\,z_{34}\,z_{36}\,z_{38}\,z_{47}\,z_{56}\,z_{58}\,z_{78}\notag\\
  &\hspace{.8cm}+ \,  (\si^{\mu} \, \bar{\si}^{\nu} \vep)_{\ze\al} \,\vep_{\be\ep}\,\vep_{\ga\de}\,z_{14}\,z_{16}\,z_{25}\,z_{27}\,z_{34}\,z_{36}\,z_{38}\,z_{45}\,z_{58}\,z_{67}\,z_{78}\notag\\
  &\hspace{.8cm}+ \, (\si^{\mu} \, \bar{\si}^{\nu} \vep)_{\ze\ga} \,\vep_{\al\be}\,\vep_{\de\ep}\,z_{14}\,z_{16}\,z_{23}\,z_{27}\,z_{36}\,z_{38}\,z_{45}\,z_{47}\,z_{56}\,z_{58}\,z_{78}\notag\\
  &\hspace{.8cm}- \, (\si^{\mu} \, \bar{\si}^{\nu} \vep)_{\ze\ga} \,\vep_{\al\de}\,\vep_{\be\ep}\,z_{14}\,z_{16}\,z_{23}\,z_{27}\,z_{34}\,z_{38}\,z_{45}\,z_{56}\,z_{58}\,z_{67}\,z_{78}\notag\\
  &\hspace{.8cm}- \, (\si^{\mu} \, \bar{\si}^{\nu} \vep)_{\ze\ep} \,\vep_{\al\be}\,\vep_{\ga\de}\,z_{14}\,z_{16}\,z_{23}\,z_{25}\,z_{36}\,z_{38}\,z_{45}\,z_{47}\,z_{58}\,z_{67}\,z_{78}\notag\\
  &\hspace{.8cm}+ \, (\si^{\mu} \, \bar{\si}^{\nu} \vep)_{\ze\ep} \,\vep_{\al\de}\,\vep_{\be\ga}\,z_{14}\,z_{16}\,z_{23}\,z_{25}\,z_{34}\,z_{38}\,z_{47}\,z_{56}\,z_{58}\,z_{67}\,z_{78}\Big)\ .
\label{exp12}
\end{align}

\section{Correlation functions with two spin fields}
\label{sec:TheSixAndSevenPointCorrelator}

Parton amplitudes are important for collider phenomenology since multijet production
is dominated by tree-level QCD scattering.
Therefore these parton amplitudes that are  generic to any string compactification are especially important, as they  may give rise to
universal string signals independent on any compactification details.
Amplitudes involving an arbitrary number of gluons $g$ (or gauginos $\chi$)
but only two quarks $\psi$ or
squarks $\phi$ are of this kind \cite{LHC2}. 
More precisely, the following  $N$--point amplitudes:
\beq\ba{lcl}
{\cal M}(g^{a_1}\ldots g^{a_N})\ ,&&
{\cal M}(\psi^{a_1}\overline\psi^{a_2}g^{a_3}\ldots g^{a_N})\ ,\\
{\cal M}(\chi^{a_1}\overline\chi^{a_2}g^{a_3}\ldots g^{a_N})\ ,&&
{\cal M}(\phi^{a_1}\overline\phi^{a_2}g^{a_3}\ldots g^{a_N})\ ,
\ea\label{UNIVERSAL}
\eeq
are completely universal to any string compactification, i.e. they
do not depend on the compactification details as geometry and  topology. Using space--time supersymmetry we may also replace gluons $g$ by gauginos $\chi$. Hence the amplitudes \req{UNIVERSAL} belong 
to an entire classes of string amplitudes, for which only the string Regge modes but
not the KK/winding modes contribute.

In this Section we derive the necessary correlators to compute the 
amplitudes \req{UNIVERSAL}.
While the amplitudes with only gluons and scalars only require correlators of  at most $2N-2$ NS fermions, the amplitudes with two fermions requires correlators 
involving two R spin fields and at most  $2N-5$ NS fermions $(n=N-2)$:
\beq\label{STARTover}
\langle \psi^{\mu_{1}}(z_{1}) \, \psi^{\mu_{2}}(z_{2}) \, ... \, \psi^{\mu_{2n-1}}(z_{2n-1}) \, S_{\al}(z_{A}) \, S_{\dbe}(z_{B}) \rangle\,.
\eeq

The factorization method presented in Section \ref{sec:FromSpinFieldsToNSFermions} 
in principle tells us how to compute any desired
correlation function involving arbitrary numbers of $\psi^{\mu}$, $S_{\al}$ and $S_{\dbe}$ fields. However, this method
has some disadvantages. The result obtained using the factorization method uses a non-minimal set of index terms. For a
large number of fields the number of index terms grows exponentially and hence sophisticated relations between these are
needed to reduce the result to a more compact form. Furthermore the computational effort increases rapidly with the
number of NS fermions because they contribute both to the left- and to the right handed spin field correlator.

\subsection{The six-- and seven--point function with two spin fields}
\label{sec:TheSixAndSevenPointFunctionWithTwoSpinFields}

For the subclass of correlators $\vev{\psi^{\mu_1} \dots \psi^{\mu_n} S_{\al} S_{\be , \dbe}}$ with precisely two spin
fields it is possible to derive short expressions by matching various $z_i \mto z_j$ limits. We apply this old-fashioned
method to the correlators
\begin{subequations}
\begin{align}
  \om_{(3)}^{\mu \nu \la \rho} \, _{\al} \,^{\be} (z_{1},...,z_{6}) \ \ &:= \ \ \langle \psi^{\mu} (z_{1}) \,\psi^{\nu}(z_{2}) \,
  \psi^{\la}(z_{3}) \, \psi^{\rho}(z_{4}) \, S_{\al}(z_{5}) \, S^{\be}(z_{6}) \rangle \,, \label{rv,9} \\
  \Om_{(3)}^{\mu \nu \la \rho \tau} \, _{\al \dbe} (z_{1},...,z_{7}) \ \ &:= \ \ \langle \psi^{\mu} (z_{1}) \,
  \psi^{\nu}(z_{2}) \, \psi^{\la}(z_{3}) \, \psi^{\rho}(z_{4}) \, \psi^{\tau}(z_{5}) \, S_{\al}(z_{6}) \, S_{\dbe}(z_{7})\rangle \label{rv,14}
\end{align}
\end{subequations}
in the first Subsection. By comparison with the lower order analogues (\ref{rv,5a}) and (\ref{rv,5}), a general formula
for the respective correlators with arbitrary number of NS fields is guessed. The prove by induction is postponed to the
Appendix \ref{sec:TheProofByInduction}.

Let us start with the correlator $\om_{(3)}^{\mu \nu \la \rho} \, _{\al} \,^{\be}$. If the spin field arguments $z_{5}$
and $z_{6}$ approach each other, we are left with a four NS field correlator:
\begin{equation}
 \langle \psi^{\mu} (z_{1}) \, \psi^{\nu}(z_{2}) \, \psi^{\la}(z_{3}) \, \psi^{\rho}(z_{4}) \rangle \eq \frac{\eta^{\mu \nu} \, \eta^{\la
    \rho}}{z_{12} \, z_{34}} \ - \ \frac{\eta^{\mu \la} \, \eta^{\nu \rho}}{z_{13} \, z_{24}} \ + \ \frac{\eta^{\mu
    \rho} \, \eta^{\nu \la}}{z_{14} \, z_{23}} \, .
\label{rv,10}
\end{equation}
The most constraining limits, however, are those which explicitly include the $\psi$-$S$ interaction. The five--point
function (\ref{rv,5}) allows to evaluate $\om_{(3)}$ in the $z_{i=1,2,3,4} \mto z_{6}$ regime. This leads to products of
up to 4 $\si$ matrices. We restrict the occurring terms to ordered $\si$ products, i.e.\ $\si^{\mu}$ is placed left of
$\si^{\nu},\si^{\la}, \si^{\rho}$ and so on. So we arrive at the unique result:
\begin{align}
\om_{(3)}^{\mu \nu \la \rho} \,_{\al} \,^{\be} &(z_{1},...,z_{6}) \ \ = \ \ \frac{- \, 1}{z_{56}^{1/2} \, (z_{15} \, z_{16} \, z_{25} \, z_{26} \, z_{35} \, z_{36} \, z_{45} \, z_{46})^{1/2}} \notag \\
&\times \Biggl\{ \eta^{\mu \nu} \, \eta^{\la \rho} \, \de_{\al}^{\be} \; \frac{z_{15} \, z_{26} \, z_{35} \, z_{46}}{z_{12} \, z_{34}} \ - \ \eta^{\mu \la} \, \eta^{\nu \rho} \, \de_{\al}^{\be} \; \frac{z_{15} \, z_{36} \, z_{25} \, z_{46}}{z_{13} \, z_{24}} \Biggr. \notag\\
& \ \ \ \ + \ \eta^{\mu \rho} \, \eta^{\nu \la} \, \de_{\al}^{\be} \; \frac{z_{15} \, z_{46} \, z_{25} \, z_{36}}{z_{14} \, z_{23}} \ + \ \left( \frac{z_{56}}{2} \right)^{2} \, (\si^{\mu} \, \bar{\si}^{\nu} \, \si^{\la} \, \bar{\si}^{\rho})_{\al} \,^{\be} \notag \\
& \ \ \ \ + \ \frac{z_{56}}{2} \; \left( + \ \eta^{\mu \nu} \, (\si^{\la} \, \bar{\si}^{\rho})_{\al} \,^{\be} \, \frac{z_{15} \, z_{26}}{z_{12}} \ + \ \eta^{\la \rho} \, (\si^{\mu} \, \bar{\si}^{\nu})_{\al} \,^{\be} \, \frac{z_{35} \, z_{46}}{z_{34}} \right. \notag \\
& \ \ \ \ \ \ \ \ \ \ \ \ \ \ \ \ \! - \ \eta^{\mu \la} \, (\si^{\nu} \, \bar{\si}^{\rho})_{\al} \,^{\be} \, \frac{z_{15} \, z_{36}}{z_{13}} \ - \ \eta^{\nu \rho} \, (\si^{\mu} \, \bar{\si}^{\la})_{\al} \,^{\be} \, \frac{z_{25} \, z_{46}}{z_{24}} \notag \\
& \ \ \ \ \ \ \ \ \ \ \ \ \ \ \ \ \Biggl. \left. + \ \eta^{\mu \rho} \, (\si^{\nu} \, \bar{\si}^{\la})_{\al} \,^{\be} \, \frac{z_{15} \, z_{46}}{z_{14}} \ + \ \eta^{\nu \la} \, (\si^{\mu} \, \bar{\si}^{\rho})_{\al} \,^{\be} \, \frac{z_{25} \, z_{36}}{z_{23}} \right) \Biggr\} \ .
\label{rv,13}
\end{align}
The analogue with dotted indices is given by:
\begin{align}
\bar{\om}_{(3)}^{\mu \nu \la \rho} &^{\dal} \, _{\dbe} (z_{1},...,z_{6}) \ \ := \ \ \langle \psi^{\mu} (z_{1}) \, \psi^{\nu}(z_{2}) \, \psi^{\la}(z_{3}) \, \psi^{\rho}(z_{4}) \, S^{\dal}(z_{5}) \, S_{\dbe}(z_{6}) \rangle \notag \\
&= \ \ \frac{1}{z_{56}^{1/2} \, (z_{15} \, z_{16} \, z_{25} \, z_{26} \, z_{35} \, z_{36} \, z_{45} \, z_{46})^{1/2}} \notag \\
& \ \ \ \ \times \Biggl\{ \eta^{\mu \nu} \, \eta^{\la \rho} \, \de_{\dal}^{\dbe} \; \frac{z_{15} \, z_{26} \, z_{35} \, z_{46}}{z_{12} \, z_{34}} \ - \ \eta^{\mu \la} \, \eta^{\nu \rho} \, \de_{\dal}^{\dbe} \; \frac{z_{15} \, z_{36} \, z_{25} \, z_{46}}{z_{13} \, z_{24}} \Biggr. \notag\\
& \ \ \ \ \ \ \ \ + \ \eta^{\mu \rho} \, \eta^{\nu \la} \, \de_{\dal}^{\dbe} \; \frac{z_{15} \, z_{46} \, z_{25} \, z_{36}}{z_{14} \, z_{23}} \ + \ \left( \frac{z_{56}}{2} \right)^{2} \, (\bar{\si}^{\mu} \, \si^{\nu} \, \bar{\si}^{\la} \, \si^{\rho})^{\dal} \, _{\dbe} \notag \\
& \ \ \ \ \ \ \ \ + \ \frac{z_{56}}{2} \; \left( + \ \eta^{\mu \nu} \, (\bar{\si}^{\la} \, \si^{\rho})^{\dal} \, _{\dbe} \, \frac{z_{15} \, z_{26}}{z_{12}} \ + \ \eta^{\la \rho} \, (\bar{\si}^{\mu} \, \si^{\nu})^{\dal} \, _{\dbe} \, \frac{z_{35} \, z_{46}}{z_{34}} \right. \notag \\
& \ \ \ \ \ \ \ \ \ \ \ \ \ \ \ \ \ \ \ \ \! - \ \eta^{\mu \la} \, (\bar{\si}^{\nu} \, \si^{\rho})^{\dal} \, _{\dbe} \, \frac{z_{15} \, z_{36}}{z_{13}} \ - \ \eta^{\nu \rho} \, (\bar{\si}^{\mu} \, \si^{\la})^{\dal} \, _{\dbe} \, \frac{z_{25} \, z_{46}}{z_{24}} \notag \\
& \ \ \ \ \ \ \ \ \ \ \ \ \ \ \ \ \ \ \ \ \Biggl. \left. + \ \eta^{\mu \rho} \, (\bar{\si}^{\nu} \, \si^{\la})^{\dal} \, _{\dbe} \, \frac{z_{15} \, z_{46}}{z_{14}} \ + \ \eta^{\nu \la} \, (\bar{\si}^{\mu} \, \si^{\rho})^{\dal} \, _{\dbe} \, \frac{z_{25} \, z_{36}}{z_{23}} \right) \Biggr\} \ .
\label{rv,13a}
\end{align}
With the correlators (\ref{rv,13}), (\ref{rv,13a}) at hand, one is ready to compute the seven point function $\Om_{(3)}$.

For the seven point function the limits $z_{1} \mto z_{i=2,3,4,5}$ are easily evaluated using the five point correlator
(\ref{rv,5}). We also have to perform the limits $z_{1} \mto z_{6}$ and $z_{1} \mto z_{7}$ which require the six point
correlators $\om_{(3)}$ and $\bar{\om}_{(3)}$. Using only ordered $\si$ chains as before we find the following result:
\begin{align}
\Om_{(3)}^{\mu \nu \la \rho \tau}&\,_{\al \dbe}(z_{1},...,z_{7}) \ \ = \ \ \frac{1}{\sqrt{2} \, (z_{16} \, z_{17} \, z_{26} \, z_{27} \, z_{36} \, z_{37} \, z_{46} \, z_{47} \, z_{56} \, z_{57} )^{1/2}} \notag \\
&\times \Biggl\{ + \ \si^{\mu}_{\al \dbe} \, \left( \eta^{\nu \la} \, \eta^{\rho \tau} \; \frac{z_{26} \, z_{37} \, z_{46} \, z_{57}}{z_{23} \, z_{45}} \ + \ \eta^{\nu \tau} \, \eta^{\la \rho} \; \frac{z_{26} \, z_{57} \, z_{36} \, z_{47}}{z_{25} \, z_{34}} \ - \ \eta^{\nu \rho} \, \eta^{\la \tau} \; \frac{z_{26} \, z_{47} \, z_{36} \, z_{57}}{z_{24} \, z_{35}} \right) \Biggr. \notag \\
& \ \ \ \ \ - \ \si^{\nu}_{\al \dbe} \, \left( \eta^{\mu \la} \, \eta^{\rho \tau} \; \frac{z_{16} \, z_{37} \, z_{46} \, z_{57}}{z_{13} \, z_{45}} \ + \ \eta^{\mu \tau} \, \eta^{\la \rho} \; \frac{z_{16} \, z_{57} \, z_{36} \, z_{47}}{z_{15} \, z_{34}} \ - \ \eta^{\mu \rho} \, \eta^{\la \tau} \; \frac{z_{16} \, z_{47} \, z_{36} \, z_{57}}{z_{14} \, z_{35}} \right) \notag \\
& \ \ \ \ \ + \ \si^{\la}_{\al \dbe} \, \left( \eta^{\mu \nu} \, \eta^{\rho \tau} \; \frac{z_{16} \, z_{27} \, z_{46} \, z_{57}}{z_{12} \, z_{45}} \ + \ \eta^{\mu \tau} \, \eta^{\nu \rho} \; \frac{z_{16} \, z_{57} \, z_{26} \, z_{47}}{z_{15} \, z_{24}} \ - \ \eta^{\mu \rho} \, \eta^{\nu \tau} \; \frac{z_{16} \, z_{47} \, z_{26} \, z_{57}}{z_{14} \, z_{25}} \right) \notag \\
& \ \ \ \ \ - \ \si^{\rho}_{\al \dbe} \, \left( \eta^{\mu \nu} \, \eta^{\la \tau} \; \frac{z_{16} \, z_{27} \, z_{36} \, z_{57}}{z_{12} \, z_{35}} \ + \ \eta^{\mu \tau} \, \eta^{\nu \la} \; \frac{z_{16} \, z_{57} \, z_{26} \, z_{37}}{z_{15} \, z_{23}} \ - \ \eta^{\mu \la} \, \eta^{\nu \tau} \; \frac{z_{16} \, z_{37} \, z_{26} \, z_{57}}{z_{13} \, z_{25}} \right) \notag \\
& \ \ \ \ \ + \ \si^{\tau}_{\al \dbe} \, \left( \eta^{\mu \nu} \, \eta^{\la \rho} \; \frac{z_{16} \, z_{27} \, z_{36} \, z_{47}}{z_{12} \, z_{34}} \ + \ \eta^{\mu \rho} \, \eta^{\nu \la} \; \frac{z_{16} \, z_{47} \, z_{26} \, z_{37}}{z_{14} \, z_{23}} \ - \ \eta^{\mu \la} \, \eta^{\nu \rho} \; \frac{z_{16} \, z_{37} \, z_{26} \, z_{47}}{z_{13} \, z_{24}} \right) \notag \\
& \ \ \ \ \ \ \ \ + \ \frac{z_{67}}{2} \, \biggl( + \ \eta^{\mu \nu} \, (\si^{\la} \, \bar{\si}^{\rho} \, \si^{\tau})_{\al \dbe} \; \frac{z_{16} \, z_{27}}{z_{12}} \ - \ \eta^{\mu \la} \, (\si^{\nu} \, \bar{\si}^{\rho} \, \si^{\tau})_{\al \dbe} \; \frac{z_{16} \, z_{37}}{z_{13}} \biggr. \notag \\
& \ \ \ \ \ \ \ \ \ \ \ \ \ \ \ \ \ \ \ + \eta^{\mu \rho} \, (\si^{\nu} \, \bar{\si}^{\la} \, \si^{\tau})_{\al \dbe} \; \frac{z_{16} \, z_{47}}{z_{14}} \ - \ \eta^{\mu \tau} \, (\si^{\nu} \, \bar{\si}^{\la} \, \si^{\rho})_{\al \dbe} \; \frac{z_{16} \, z_{57}}{z_{15}} \notag \\
& \ \ \ \ \ \ \ \ \ \ \ \ \ \ \ \ \ \ \ + \eta^{\nu \la} \, (\si^{\mu} \, \bar{\si}^{\rho} \, \si^{\tau})_{\al \dbe} \; \frac{z_{26} \, z_{37}}{z_{23}} \ - \ \eta^{\nu \rho} \, (\si^{\mu} \, \bar{\si}^{\la} \, \si^{\tau})_{\al \dbe} \; \frac{z_{26} \, z_{47}}{z_{24}} \notag \\
& \ \ \ \ \ \ \ \ \ \ \ \ \ \ \ \ \ \ \ + \eta^{\nu \tau} \, (\si^{\mu} \, \bar{\si}^{\la} \, \si^{\rho})_{\al \dbe} \; \frac{z_{26} \, z_{57}}{z_{25}} \ + \ \eta^{\la \rho} \, (\si^{\mu} \, \bar{\si}^{\nu} \, \si^{\tau})_{\al \dbe} \; \frac{z_{36} \, z_{47}}{z_{34}} \notag \\
& \ \ \ \ \ \ \ \ \ \ \ \ \ \ \ \ \ \ \ \biggl. \, - \ \eta^{\la \tau} \, (\si^{\mu} \, \bar{\si}^{\nu} \, \si^{\rho})_{\al \dbe} \; \frac{z_{36} \, z_{57}}{z_{35}} \ + \ \eta^{\rho \tau} \, (\si^{\mu} \, \bar{\si}^{\nu} \, \si^{\la})_{\al \dbe} \; \frac{z_{46} \, z_{57}}{z_{45}} \biggr) \notag \\
& \ \ \ \ \ \ \ \ \Biggl. + \ \left( \frac{z_{67}}{2} \right)^{2} \, (\si^{\mu} \, \bar{\si}^{\nu} \, \si^{\la} \, \bar{\si}^{\rho} \, \si^{\tau})_{\al \dbe} \Biggr\}\,.
\label{rv,16}
\end{align}

\subsection[The $n$--point correlators with two spin fields]{The $\bm{n}$--point correlators with two spin fields}
\label{sec:TheNPointGeneralization}

The correlators $\om_{(3)}$, $\bar{\om}_{(3)}$ and $\Om_{(3)}$ in their representations (\ref{rv,13}), (\ref{rv,13a})
and (\ref{rv,16}) have striking similarities in their structure. Let us denote the arguments of the spin fields as
$S_\al(z_{A})$, $S_{\dbe}(z_{B})$ and a generic NS field by $\psi^{\mu_{i}}(z_{i})$ in the following list of observations:
\begin{itemize}
\item Increasing numbers of $\si$ matrices are multiplied by increasing powers of $\frac{z_{AB}}{2}$. More precisely,
  each term $\si^{\mu_{i}} \bar{\si}^{\mu_{j}}$ comes with a factor $\frac{z_{AB}}{2}$.
\item The prefactor contains a $-1/2$ power of each $\psi \leftrightarrow S$ contraction, i.e. the correlators are
  proportional to $\prod_{i=1}^{N} (z_{iA} z_{iB})^{-1/2}$ where $N$ is the number of $\psi$'s involved.
\item Each Minkowski metric $\eta^{\mu_{i} \mu_{j}}$ with $i<j$ appears in combination with $\frac{z_{iA}z_{jB}}{z_{ij}}$.
\item The sign of each term depends on the ordering of the Lorentz indices, whether they appear as an odd or an even
  permutation of ${\mu_{1} \mu_{2} ... \mu_{N}}$.
\end{itemize}
These properties lead us to claim the following expression for $2n+1$ point function $\Om_{(n)}$:

\beq\label{npt,1} \hskip-0.25cm\boxed{
\ba{ll}
\ds{\Om_{(n)}^{\mu_{1} ... \mu_{2n-1}}\,_{\al \dbe}(z_{i}) }&:= 
\ds{\langle \psi^{\mu_{1}}(z_{1}) \, \psi^{\mu_{2}}(z_{2}) \, ... \, \psi^{\mu_{2n-1}}(z_{2n-1}) \, S_{\al}(z_{A}) \, S_{\dbe}(z_{B}) \rangle }\\[4mm]
& =\ds{
\frac{1}{\sqrt{2}} \prod_{i=1}^{2n-1} (z_{iA} \, z_{iB})^{-1/2} \, \sum_{\ell = 0}^{n-1} \, \biggl( \frac{z_{AB}}{2} \biggr)^{\ell} \! \! \! \! \sum_{\rho \in S_{2n-1}/{\cal P}_{n,\ell}} \! \! \!  \te{sgn}(\rho) \, \bigl(\si^{\mu_{\rho(1)}} \, \bar{\si}^{\mu_{\rho(2)}} \, ... \, \bar{\si}^{\mu_{\rho(2\ell)}} \, \si^{\mu_{\rho(2\ell+1)}} \bigr)_{\al \dbe}} \\[4mm]
&\ds{ \times \prod_{j=1}^{n-\ell-1} \frac{\eta^{\mu_{\rho(2\ell+2j)} \mu_{\rho(2\ell+2j+1)}}}{z_{\rho(2\ell+2j)} \, - \, z_{\rho(2\ell+2j+1)} } \; \bigl( z_{\rho(2\ell+2j)} \, - \, z_{A} \bigr) \, \bigl( z_{\rho(2\ell+2j+1)} \, - \, z_{B} \bigr)\ .}
\ea}
\eeq
The most important auxiliary correlators to verify the expression (\ref{npt,1}) are the $2n$ point correlation functions
$\om_{(n)}$ and $\bar{\om}_{(n)}$:
\begin{align}
\om_{(n)}&^{\mu_{1} ... \mu_{2n-2}}\,_{\al} \, ^{ \be}(z_{i}) \ \ := \ \ \langle \psi^{\mu_{1}}(z_{1}) \, \psi^{\mu_{2}}(z_{2}) \, ... \, \psi^{\mu_{2n-2}}(z_{2n-2}) \, S_{\al}(z_{A}) \, S^{\be}(z_{B}) \rangle \notag \\
&= \ \ \frac{- \, 1}{z_{AB}^{1/2}} \; \prod_{i=1}^{2n-2} (z_{iA} \, z_{iB})^{-1/2} \, \sum_{\ell = 0}^{n-1} \, \biggl( \frac{z_{AB}}{2} \biggr)^{\ell} \! \! \! \! \sum_{\rho \in S_{2n-2}/{\cal Q}_{n,\ell}} \! \! \!  \te{sgn}(\rho) \, \bigl(\si^{\mu_{\rho(1)}} \, \bar{\si}^{\mu_{\rho(2)}} \, ... \, \si^{\mu_{\rho(2\ell-1)}} \, \bar{\si}^{\mu_{\rho(2\ell)}} \bigr)_{\al} \, ^{\be} \notag \\
& \ \ \ \ \  \ \ \times \prod_{j=1}^{n-\ell-1} \frac{\eta^{\mu_{\rho(2\ell+2j-1)} \mu_{\rho(2\ell+2j)}}}{z_{\rho(2\ell+2j-1)} \, - \, z_{\rho(2\ell+2j)} } \; \bigl( z_{\rho(2\ell+2j-1)} \, - \, z_{A} \bigr) \, \bigl( z_{\rho(2\ell+2j)} \, - \, z_{B} \bigr)\,,
\label{npt,2} \\
\bar{\om}_{(n)}&^{\mu_{1} ... \mu_{2n-2} \dal} \, _{ \dbe}(z_{i}) \ \ := \ \ \langle \psi^{\mu_{1}}(z_{1}) \, \psi^{\mu_{2}}(z_{2}) \, ... \, \psi^{\mu_{2n-2}}(z_{2n-2}) \, S^{\dal}(z_{A}) \,  S_{\dbe}(z_{B}) \rangle \notag \\
&= \ \ \frac{+ \, 1}{z_{AB}^{1/2}} \; \prod_{i=1}^{2n-2} (z_{iA} \, z_{iB})^{-1/2} \, \sum_{\ell = 0}^{n-1} \, \biggl( \frac{z_{AB}}{2} \biggr)^{\ell} \! \! \! \! \sum_{\rho \in S_{2n-2}/{\cal Q}_{n,\ell}} \! \! \!  \te{sgn}(\rho) \, \bigl(\bar{\si}^{\mu_{\rho(1)}} \, \si^{\mu_{\rho(2)}} \, ... \, \bar{\si}^{\mu_{\rho(2\ell-1)}} \, \si^{\mu_{\rho(2\ell)}} \bigr)^{\dal} \,_{\dbe} \notag \\
& \ \ \ \ \  \ \ \times \prod_{j=1}^{n-\ell-1} \frac{\eta^{\mu_{\rho(2\ell+2j-1)} \mu_{\rho(2\ell+2j)}}}{z_{\rho(2\ell+2j-1)} \, - \, z_{\rho(2\ell+2j)} } \; \bigl( z_{\rho(2\ell+2j-1)} \, - \, z_{A} \bigr) \, \bigl( z_{\rho(2\ell+2j)} \, - \, z_{B} \bigr)\,.
\label{npt,3}
\end{align}
The proof of (\ref{npt,1}) together with (\ref{npt,2}) and (\ref{npt,3}) can be found in the appendix \ref{sec:TheProofByInduction}.

The summation ranges $\rho \in S_{2n-1}/{\cal P}_{n,\ell}$ and $\rho \in S_{2n-2}/{\cal Q}_{n,\ell}$ certainly require some explanation: We only sum over permutations $\rho$ of $(1,2,...,2n-1)$ or $(1,2,...,2n-2)$ which satisfy the following constraints:
\begin{itemize}
\item We only keep ordered $\si$ products: The indices $\mu_{\rho(i)}$ attached to a chain of $\si$ matrices are increasingly ordered, e.g. whenever the product $\si^{\mu_{\rho(i)}} \bar{\si}^{\mu_{\rho(j)}} \si^{\mu_{\rho(k)}}$ appears, we have $\rho(i) < \rho(j) < \rho(k)$.
\item On each metric $\eta^{\mu_{\rho(i)} \mu_{\rho(j)}}$ the first index is the ``lower'' one, i.e.\ $\rho(i) < \rho(j)$.
\item Products of several $\eta$'s are not double counted. So once we get $\eta^{\mu_{\rho(i)} \mu_{\rho(j)}}
  \eta^{\mu_{\rho(k)} \mu_{\rho(l)}}$, the term $\eta^{\mu_{\rho(k)} \mu_{\rho(l)}} \eta^{\mu_{\rho(i)} \mu_{\rho(j)}}$
  does not appear.
\end{itemize}
These restrictions to the occurring $S_{2n-1}$ (or $S_{2n-2}$-) elements are abbreviated by a quotient ${\cal
  P}_{n,\ell}$ and ${\cal Q}_{n,\ell}$. Let us give a more formal definition:
\begin{subequations}
\begin{align}
S_{2n-1}/ {\cal P}_{n,\ell} \ \ := \ \ \Bigl\{ &\rho \in S_{2n-1} : \ \rho(1) < \rho(2) < ... < \rho(2\ell + 1) \ , \Bigr. \notag \\
&\rho(2\ell + 2j) < \rho(2\ell + 2j + 1) \ \forall \ j = 1,2,...,n-\ell - 1 \ , \notag \\
\Bigl. &\rho(2\ell + 3) < \rho(2\ell + 5) < ... < \rho(2n -1) \Bigr\} 
\label{npt,4}\,, \\
S_{2n-2}/ {\cal Q}_{n,\ell} \ \ := \ \ \Bigl\{ &\rho \in S_{2n-2} : \ \rho(1) < \rho(2) < ... < \rho(2\ell) \ , \Bigr. \notag \\
&\rho(2\ell + 2j-1) < \rho(2\ell + 2j) \ \forall \ j = 1,2,...,n-\ell - 1 \ , \notag \\
\Bigl. &\rho(2\ell + 2) < \rho(2\ell + 4) < ... < \rho(2n -2) \Bigr\} \,.
\label{npt,5}
\end{align}
\end{subequations}
So the groups of subpermutations removed from the original $S_{2n-1}$, $S_{2n}$ are as follows:
\begin{subequations}
\begin{align}
{\cal P}_{n,\ell} \ \ &\leftrightarrow \ \ \left\{ \begin{array}{rl} S_{2\ell+1} &: \ \te{permute the $(2\ell + 1)$ matrices $\si^{\mu_i}$} \\ S_{n-\ell - 1} &: \ \te{permute the $n-\ell-1$ Minkowski metrics} \\ (S_{2})^{n-\ell-1} &: \ \te{exchange the indices of one of the $\eta$'s} \end{array} \right. \label{npt,6a} \\
{\cal Q}_{n,\ell} \ \ &\leftrightarrow \ \ \left\{ \begin{array}{rl} S_{2\ell} &: \ \te{permute the $(2\ell)$ matrices $\si^{\mu_i}$} \\ S_{n-\ell - 1} &: \ \te{permute the $n-\ell-1$ Minkowski metrics} \\ (S_{2})^{n-\ell-1} &: \ \te{exchange the indices of one of the $\eta$'s\ .} \end{array} \right.
\label{npt,6}
\end{align}
\end{subequations}
Since the full permutation group $S_{N}$ has $N!$ elements, one can conclude from (\ref{npt,6a}) and (\ref{npt,6}) how
many terms remain in the sums over $\rho$ in (\ref{npt,1}), (\ref{npt,2}) and (\ref{npt,3}):
\begin{subequations}
\begin{align}
\bigl| S_{2n-1} / {\cal P}_{n,\ell} \bigr| \ \ &= \ \ \frac{(2n\, - \, 1)!}{(2\ell \, + \, 1)! \, (n \, - \, \ell \, - \, 1)! \, 2^{n-\ell-1}}\,, \label{npt,7a} \\
\bigl| S_{2n-2} / {\cal Q}_{n,\ell} \bigr| \ \ &= \ \ \frac{(2n\, - \, 2)!}{(2\ell )! \, (n \, - \, \ell \, - \, 1)! \, 2^{n-\ell-1}}\,.
\label{npt,7}
\end{align}
\end{subequations}
We have already stated the cases $n=1,2,3$ for $\Om_{(n)}$, $\om_{(n)}$, $\bar{\om}_{(n)}$ in terms of ordered $\si$
products in Sections \ref{sec:ReviewOfLowerOrderCorrelators} and \ref{sec:TheSixAndSevenPointCorrelator}. So one can
check equations (\ref{npt,7a}), (\ref{npt,7}) by counting the number of Clebsch Gordan coefficients with fixed number of
$\si$'s in these cases. The following table gives an overview of all the previously presented examples:
\begin{equation}
\begin{array}{c|c|l|c|c|c|l|c|c|c}
n &\ell & & \bigl| S_{2n-2} / {\cal Q}_{n,\ell} \bigr| &\te{term} &\te{equation} & & \bigl| S_{2n-1} / {\cal P}_{n,\ell} \bigr| &\te{term} &\te{equation} \\\hline
1 &0 & & \frac{0!}{0! 0! 2^{0}} = 1 &\de^{\be}_{\al} &(\ref{rv,1b}) & &\frac{1!}{1! 0! 2^{0}} = 1 &\si^{\mu}_{\al \dbe} &(\ref{rv,4}) \\\hline
2 &0 & &\frac{2!}{0! 1! 2^{1}} = 1 &\eta^{\mu \nu} &(\ref{rv,5a}) & &\frac{3!}{1! 1! 2^{1}} = 3 &\eta^{\mu \nu} \si^{\la} &(\ref{rv,5}) \\
&1 & &\frac{2!}{2! 0! 2^{0}} = 1 &\si^{\mu} \bar{\si}^{\nu} & & &\frac{3!}{3! 0! 2^{0}} = 1 &\si^{\mu} \bar{\si}^{\nu} \si^{\la} & \\\hline
3 &0 &  & \frac{4!}{0! 2! 2^{2}} = 3 &\eta^{\mu \nu} \eta^{\la \rho} &(\ref{rv,13}) & &\frac{5!}{1!2!2^{2}} = 15 &\eta^{\mu \nu} \eta^{\la \rho} \si^{\tau} &(\ref{rv,16}) \\
&1 & & \frac{4!}{2! 1! 2^{1}} = 6 &\eta^{\mu \nu} \si^{\la } \bar{\si}^{\rho} & & &\frac{5!}{3! 1! 2^{1}} = 10 &\eta^{\mu \nu} \si^{\la} \bar{\si}^{\rho} \si^{\tau} & \\
&2 & & \frac{4!}{4! 0! 2^{0}} = 1 &\si^{\mu} \bar{\si}^{ \nu} \si^{\la } \bar{\si}^{\rho} & & &\frac{5!}{5! 0! 2^{0}} =
1 &\si^{\mu} \bar{\si}^{ \nu} \si^{\la} \bar{\si}^{\rho} \si^{\tau} &
\end{array}
\nonumber
\end{equation}
Obviously, (\ref{npt,7a}) and (\ref{npt,7}) yield exactly the number of index structures which we have used in the known correlators. However, this does not mean that we are using minimal sets. The spin field correlators discussed in Section \ref{sec:TheBasicBuildingBlockSpinFieldCorrelators} are excellent examples that reducing the index terms to a minimum might spoil certain symmetries or simplifications in the $z$ dependences.

The same is true for the correlators (\ref{npt,1}), (\ref{npt,2}) and (\ref{npt,3}): Already for the seven point function (\ref{rv,16}), there is one identity relating all the 26 appearing index configurations, namely (\ref{rv,18}). It can be traced back to $0 = \si^{[\mu} \bar{\si}^{ \nu} \si^{\la} \bar{\si}^{\rho} \si^{\tau]}$, i.e. to the vanishing of antisymmetric expressions in more than four vector indices. Eliminating one of the terms in (\ref{rv,16}) would certainly lead to a more complicated $z$ dependence.

The overcounting of the basis of index structures in (\ref{npt,7a}) and (\ref{npt,7}) grows with $n$. The expression for the eight point function $\om_{(4)}^{\mu_{1}...\mu_{6}} \, _{\al}\,^{\be}(z_{i})$ due to (\ref{npt,2}) contains 76 terms, but a group theoretic analysis determines the number of scalar representations in the tensor product to be 70. This difference is explained by the six independent reduction identities $0 = \si^{[\mu} \bar{\si}^{ \nu} \si^{\la} \bar{\si}^{\rho} \si^{\tau} \bar{\si}^{\xi ]}$ and $0 = \eta^{\mu [\nu} \vep^{\la \rho \tau \xi]}$
\footnote{Another way to write these equations is: \beq \eta_{\mu \nu} \, \vep^{\la \rho \tau \zeta} \eq \de_{(\mu}^{\la} \, \vep_{\nu)}\, \! ^{\rho \tau \zeta} \ + \ \de_{(\mu}^{\rho} \, \vep^{\la } \, \! _{\nu)}\, \! ^{\tau \zeta} \ + \ \de_{(\mu}^{\tau} \, \vep^{\la \rho} \, \! _{\nu)}\, \! ^{\zeta} \ + \ \de_{(\mu}^{\zeta} \, \vep^{\la \rho \tau} \, \! _{\nu)}\,.  \label{dep} \eeq}. Similarly, for higher point examples $\Om_{(n \geq 4)}$ and $\om_{(n \geq 5)}$, one can find relations of both types.

\section{Manifest NS antisymmetry in correlation functions}
\label{sec:MakingTheNSAntisymmetryManifest}

The vanishing of antisymmetrized products of more than four $\si$ matrices leads to further relations between the various Lorentz structures predicted by (\ref{npt,1}), (\ref{npt,2}) and (\ref{npt,3}). As we have indicated in the previous Section, this information is of limited help to simplify the correlation functions $\Om_{(n)}$ and $\om_{(n)}$ in the basis of index structure used before because plugging them into (\ref{npt,1}), (\ref{npt,2}) and (\ref{npt,3}) renders the $z$ dependence much more complicated.

It will turn out, however, that there is an alternative representation for these correlators in terms of antisymmetric $\si$ products $\si^{[\mu_{1}} ... \si^{\mu_{2k-1}]}$ and $\si^{[\mu_{1}} ... \bar{\si}^{\mu_{2k}]}$ where the $z$ dependence still follows some manageable rules. In this case, the vanishing of the antisymmetrized products of length $\geq 5$ simply removes some redundant index terms without any interference with the shorter $\si$ chains as in (\ref{rv,18}). This mechanisms truncates the $\ell$ sum in (\ref{npt,1}), (\ref{npt,2}) and (\ref{npt,3}) which is particularly helpful for large $n$. Even correlators $\om_{(n)}$ will only receive contributions from $\ell =0,1,2$ whereas the odd counterparts $\Om_{(n)}$ already terminate after two values $\ell=0,1$.

Another advantage of the different  representation of the correlators is their manifest antisymmetry under the exchange of $\psi^{\mu_i}(z_i) \leftrightarrow \psi^{\mu_j} (z_j)$ as required by the OPE (\ref{rv,1a}). 
Furthermore, let us point out that we are using the following normalization convention for the Lorentz generators:
\beq
\si^{\mu \nu} \ \ := \ \ \si^{[\mu} \, \bar{\si}^{\nu]} \eq \frac{1}{2} \; \bigl( \si^{\mu} \, \bar{\si}^{\nu} \ - \ \si^{\nu} \, \bar{\si}^{\mu} \bigr) \ .
\label{as,0}
\eeq

\subsection{Lower order examples}
\label{sec:LowerOrderExamples}

As a first example, we convert the four point functions (\ref{rv,5a}), (\ref{rv,5b}) into a representation with manifest symmetries:
\begin{subequations}
\begin{align}
\langle \psi^{\mu}(z_{1}) \, \psi^{\nu}(z_{2}) \, S_{\al}(z_{3}) \, S^{\be}(z_{4}) \rangle \ \ &= \ \ \frac{- \, 1}{z_{34}^{1/2} \, (z_{13} \, z_{14} \, z_{23} \, z_{24})^{1/2}} \; \left( \eta^{\mu \nu} \, \de_{\al}^{\be} \; \frac{z_{13} \, z_{24}}{z_{12}} \ + \ (\si^{\mu} \, \bar{\si}^{ \nu})_{\al} \, ^{\be} \; \frac{z_{34}}{2} \right) \notag \\
&= \ \ \frac{- \, 1}{z_{34}^{1/2} \, (z_{13} \, z_{14} \, z_{23} \, z_{24})^{1/2}} \; \left( \eta^{\mu \nu} \, \de_{\al}^{\be} \; \frac{z_{13} \, z_{24} \, + \, z_{14} \, z_{23}}{2 \, z_{12}} \ + \ \si^{\mu \nu}\,_{\al} \, ^{\be} \; \frac{z_{34}}{2} \right)\,, \label{as,1} \\
\langle \psi^{\mu}(z_{1}) \, \psi^{\nu}(z_{2}) \, S^{\dal}(z_{3}) \, S_{\dbe}(z_{4}) \rangle \ \ &= \ \ \frac{1}{z_{34}^{1/2} \, (z_{13} \, z_{14} \, z_{23} \, z_{24})^{1/2}} \; \left( \eta^{\mu \nu} \, \de^{\dal}_{\dbe} \; \frac{z_{13} \, z_{24}}{z_{12}} \ + \ (\bar{\si}^{\mu} \, \si^{ \nu})^{\dal} \, _{\dbe} \; \frac{z_{34}}{2} \right) \notag \\
&= \ \ \frac{1}{z_{34}^{1/2} \, (z_{13} \, z_{14} \, z_{23} \, z_{24})^{1/2}} \; \left( \eta^{\mu \nu} \, \de^{\dal}_{\dbe} \; \frac{z_{13} \, z_{24} \, + \, z_{14} \, z_{23}}{2 \, z_{12}} \ + \ \bar{\si}^{\mu \nu \dal} \, _{\dbe} \; \frac{z_{34}}{2} \right)\,.
\label{as,2}
\end{align}
\end{subequations}
Here we made use of the corollary $\si^{\mu}\,\bar{\si}^{\nu}=-\eta^{\mu \nu}+\si^{\mu \nu}$.

Similarly, the five--point function (\ref{rv,5}) can be written in terms of the antisymmetric product of three $\si$
matrices, i.e.\ $i\vep^{\mu \nu \la}\, \!_{\xi} \si^{\xi}_{ \al \dbe} = (\si^{[\mu} \bar{\si}^{\nu} \si^{\la]})_{\al
  \dbe}$ as \cite{6G}:
\begin{align}
\langle\psi^{\mu}(z_{1}) \, \psi^{\nu} (z_{2})  &\, \psi^{\la}(z_{3}) \, S_{\al}(z_{4}) \, S_{\dbe}(z_{5}) \rangle \ \ = \ \ \frac{1}{2\sqrt{2} \, (z_{14} \, z_{15} \, z_{24} \, z_{25} \, z_{34} \, z_{35})^{1/2}} \ 
\left(  z_{45}  \, i\vep^{\mu \nu \la}\, \!_{\xi} \, \si^{\xi}_{ \al \dbe} \ \ri.\notag \\
& \lf.+ \ \eta^{\mu \nu} \, \si^{\la}_{\al \dbe} \; \frac{z_{14} \, z_{25} \, + \, z_{15} \, z_{24}}{z_{12}}  - \ \eta^{\mu \la} \, \si^{\nu}_{\al \dbe} \; \frac{z_{14} \, z_{35} \, + \, z_{15} \, z_{34}}{z_{13}} \ + \ \eta^{\nu \la} \, \si^{\mu}_{\al \dbe} \; \frac{z_{24} \, z_{35} \, + \, z_{25} \, z_{34}}{z_{23}} \right).
\label{rv,8b}
\end{align}
The price to be paid for the manifest antisymmetry in the Lorentz indices is the more cumbersome dependence on $z_{ij}$. Indeed, the same effect occurs for the six point function $\om_{(3)}^{\mu \nu \la \rho} \,_{\al} \,^{\be}$:
\begin{align}
\om_{(3)}^{\mu \nu \la \rho} \,_{\al} \,^{\be} &(z_{1},...,z_{6}) \ \ = \ \ \frac{- \, 1}{4 \, z_{56}^{1/2} \, (z_{15} \, z_{16} \, z_{25} \, z_{26} \, z_{35} \, z_{36} \, z_{45} \, z_{46})^{1/2}} \notag \\
&\times \Biggl\{ \eta^{\mu \nu} \, \eta^{\la \rho} \, \de_{\al}^{\be} \; \frac{z_{15} \, z_{26} \, + \, z_{16} \, z_{25}}{z_{12}} \; \frac{z_{35} \, z_{46} \, + \, z_{36} \, z_{45}}{z_{34}}  \Biggr.\notag\\
&- \ \eta^{\mu \la} \, \eta^{\nu \rho} \, \de_{\al}^{\be} \; \frac{z_{15} \, z_{36} \, + \, z_{16} \, z_{35}}{z_{13}} \; \frac{z_{25} \, z_{46} \, + \, z_{26} \, z_{45}}{z_{24}} \notag \\
&  \ \ \ \ + \ \eta^{\mu \rho} \, \eta^{\nu \la} \, \de_{\al}^{\be} \; \frac{z_{15} \, z_{46} \, + \, z_{16} \, z_{45}}{z_{14}} \; \frac{z_{25} \, z_{36} \, + \, z_{26} \, z_{35}}{z_{23}} \ - \ i \vep^{\mu \nu \la \rho} \, \de_{\al}^{\be} \, z_{56}^{2} \notag \\
& \ \ \ \ + \ z_{56} \; \left( + \ \eta^{\mu \nu} \, \si^{\la \rho} \, _{\al} \, ^{\be} \; \frac{z_{15} \, z_{26} \, + \, z_{16} \, z_{25}}{z_{12}} \ + \ \eta^{\la \rho} \, \si^{\mu \nu} \, _{\al} \, ^{\be} \; \frac{z_{35} \, z_{46} \, + \, z_{36} \, z_{45}}{z_{34}} \right. \notag \\
& \ \ \ \ \ \ \ \ \ \ \ \ \ \ \ - \ \eta^{\mu \la} \, \si^{\nu \rho} \, _{\al} \, ^{\be} \; \frac{z_{15} \, z_{36} \, + \, z_{16} \, z_{35}}{z_{13}} \ - \ \eta^{\nu \rho} \, \si^{\mu \la} \, _{\al} \, ^{\be} \; \frac{z_{25} \, z_{46} \, + \, z_{26} \, z_{45}}{z_{24}} \notag \\
& \ \ \ \ \ \ \ \ \ \ \ \ \ \ \ \Biggl. \left. + \ \, \eta^{\mu \rho} \, \si^{\nu \la} \, _{\al} \, ^{\be} \; \frac{z_{15} \, z_{46} \, + \, z_{16} \, z_{45}}{z_{14}} \ + \ \eta^{\nu \la} \, \si^{\mu \rho} \, _{\al} \, ^{\be} \; \frac{z_{25} \, z_{36} \, + \, z_{26} \, z_{35}}{z_{23}} \right)  \Biggr\}\,.
\label{rv,12}
\end{align}
One can also express the seven point function $\Om_{(3)}^{\mu \nu \la \rho \tau} \,_{\al \dbe}$ in a form similar to (\ref{rv,12}) with a more involved $z_{ij}$ dependence but manifest antisymmetry in the NS fields. Moreover, the $z_{67}^{2} \si^{\mu}  \bar{\si}^{\nu}  \si^{\la}  \bar{\si}^{\rho}  \si^{\tau}$ term is absent by virtue of equation (\ref{rv,18}):
\begin{align}
&\Om_{(3)}^{\mu \nu \la \rho \tau}\,_{\al \dbe}(z_{1},...,z_{7}) \ \ = \ \ \frac{1}{4\sqrt{2} \, (z_{16} \, z_{17} \, z_{26} \, z_{27} \, z_{36} \, z_{37} \, z_{46} \, z_{47} \, z_{56} \, z_{57} )^{1/2}} \notag \\
&\times \Biggl\{ + \ \si^{\mu}_{\al \dbe} \, \left( \frac{\eta^{\nu \la} \, \eta^{\rho \tau}}{z_{23} \, z_{45}} \; (z_{26} \, z_{37} \, + \, z_{27} \, z_{36}) \, ( z_{46} \, z_{57} \, + \, z_{47} \, z_{56})   \right. \Biggr. \notag \\
&\left. + \, \frac{\eta^{\nu \tau} \, \eta^{\la \rho}}{z_{25} \, z_{34}} \; (z_{26} \, z_{57} \, + \, z_{27} \, z_{56}) \, ( z_{36} \, z_{47} \, + \, z_{37} \, z_{46}) \,  - \,  \frac{\eta^{\nu \rho} \, \eta^{\la \tau}}{z_{24} \, z_{35}} \; (z_{26} \, z_{47} \, + \, z_{27} \, z_{46}) \, ( z_{36} \, z_{57} \, + \, z_{27} \, z_{56}) \! \right) \notag \\
& \ \ \ \ \ - \ \si^{\nu}_{\al \dbe} \, \left( \frac{\eta^{\mu \la} \, \eta^{\rho \tau}}{z_{13} \, z_{45}} \; (z_{16} \, z_{37} \, + \, z_{17} \, z_{36}) \, ( z_{46} \, z_{57} \, + \, z_{47} \, z_{56}) \right. \notag \\
&\left. + \, \frac{\eta^{\mu \tau} \, \eta^{\la \rho}}{z_{15} \, z_{34}} \; (z_{16} \, z_{57} \, + \, z_{17} \, z_{56}) \, (z_{36} \, z_{47} \, + \, z_{37} \, z_{46}) \, - \,  \frac{\eta^{\mu \rho} \, \eta^{\la \tau}}{z_{14} \, z_{35}} \; (z_{16} \, z_{47} \, + \, z_{17} \, z_{46}) \, ( z_{36} \, z_{57} \, + \, z_{37} \, z_{56}) \! \right) \notag \\
& \ \ \ \ \ + \ \si^{\la}_{\al \dbe} \, \left( \frac{\eta^{\mu \nu} \, \eta^{\rho \tau}}{z_{12} \, z_{45}} \; (z_{16} \, z_{27} \, + \, z_{17} \, z_{26}) \, (z_{46} \, z_{57} \, + \, z_{47} \, z_{56}) \right. \notag \\
&\left. + \, \frac{\eta^{\mu \tau} \, \eta^{\nu \rho}}{z_{15} \, z_{24}} \; (z_{16} \, z_{57} \, + \, z_{17} \, z_{56}) \, (z_{26} \, z_{47} \, + \, z_{27} \, z_{46}) \, - \,  \frac{\eta^{\mu \rho} \, \eta^{\nu \tau}}{z_{14} \, z_{25}} \; (z_{16} \, z_{47} \, + \, z_{17} \, z_{46}) \, (z_{26} \, z_{57} \, + \, z_{27} \, z_{56}) \! \right) \notag \\
& \ \ \ \ \ - \ \si^{\rho}_{\al \dbe} \, \left( \frac{\eta^{\mu \nu} \, \eta^{\la \tau}}{ z_{12} \, z_{35}} \; (z_{16} \, z_{27} \, + \, z_{17} \, z_{26}) \, ( z_{36} \, z_{57} \, + \, z_{37} \, z_{56}) \right. \notag \\
&\left. + \, \frac{\eta^{\mu \tau} \, \eta^{\nu \la}}{z_{15} \, z_{23}} \; (z_{16} \, z_{57} \, + \, z_{17} \, z_{56}) \, ( z_{26} \, z_{37} \, + \, z_{27} \, z_{36}) \, - \,  \frac{\eta^{\mu \la} \, \eta^{\nu \tau}}{z_{13} \, z_{25}} \; (z_{16} \, z_{37} \, + \, z_{17} \, z_{36}) \, ( z_{26} \, z_{57} \, + \, z_{27} \, z_{56}) \! \right) \notag \\
& \ \ \ \ \ + \ \si^{\tau}_{\al \dbe} \, \left( \frac{\eta^{\mu \nu} \, \eta^{\la \rho}}{z_{12} \, z_{34}} \; (z_{16} \, z_{27} \, + \, z_{17} \, z_{26}) \, ( z_{36} \, z_{47} \, + \, z_{37} \, z_{46}) \right. \notag \\
&\left. + \, \frac{\eta^{\mu \rho} \, \eta^{\nu \la}}{z_{14} \, z_{23}} \; (z_{16} \, z_{47} \, + \, z_{17} \, z_{46}) \, ( z_{26} \, z_{37} \, + \, z_{27} \, z_{36}) \, - \,  \frac{\eta^{\mu \la} \, \eta^{\nu \rho}}{z_{13} \, z_{24}} \; (z_{16} \, z_{37} \, + \, z_{17} \, z_{36}) \, ( z_{26} \, z_{47} \, + \, z_{27} \, z_{46}) \! \right) \notag \\
& \ \ \  + \ z_{67}  \, \biggl( + \ \frac{\eta^{\mu \nu}}{z_{12}} \, i\vep^{ \la \rho \tau}\, \!_{\zeta} \, \si^{\zeta}_{ \al \dbe} \; (z_{16} \, z_{27} \, + \, z_{17} \, z_{26}) \ - \ \frac{\eta^{\mu \la}}{z_{13}} \, i\vep^{\nu \rho \tau}\, \!_{\zeta} \, \si^{\zeta}_{ \al \dbe} \; (z_{16} \, z_{37} \, + \, z_{17} \, z_{36}) \biggr. \notag \\
& \ \ \ \ \ \ \ \ \ \ \ \ \ \ + \ \frac{\eta^{\mu \rho}}{z_{14}} \, i\vep^{\nu \la \tau}\, \!_{\xi} \, \si^{\xi}_{ \al \dbe} \; (z_{16} \, z_{47} \, + \, z_{17} \, z_{46}) \ - \ \frac{\eta^{\mu \tau}}{z_{15}} \, i\vep^{\nu \la \rho}\, \!_{\xi} \, \si^{\xi}_{ \al \dbe} \; (z_{16} \, z_{57} \, + \, z_{17} \, z_{56}) \notag \\
& \ \ \ \ \ \ \ \ \ \ \ \ \ \ + \ \frac{\eta^{\nu \la}}{z_{23}} \, i\vep^{\mu \rho \tau}\, \!_{\xi} \, \si^{\xi}_{ \al \dbe} \; (z_{26} \, z_{37} \, + \, z_{27} \, z_{36}) \ - \ \frac{\eta^{\nu \rho}}{z_{24}} \, i\vep^{\mu \la \tau}\, \!_{\xi} \, \si^{\xi}_{ \al \dbe} \; (z_{26} \, z_{47} \, + \, z_{27} \, z_{46}) \notag \\
& \ \ \ \ \ \ \ \ \ \ \ \ \ \ + \ \frac{\eta^{\nu \tau}}{z_{25}} \, i\vep^{\mu \la \rho}\, \!_{\xi} \, \si^{\xi}_{ \al \dbe} \; (z_{26} \, z_{57} \, + \, z_{27} \, z_{56}) \ + \ \frac{\eta^{\la \rho}}{z_{34}} \, i\vep^{\mu \nu \tau}\, \!_{\xi} \, \si^{\xi}_{ \al \dbe} \; (z_{36} \, z_{47} \, + \, z_{37} \, z_{46}) \notag \\
& \ \ \ \ \ \ \ \ \ \ \ \ \ \ \Biggl. \biggl. \, - \, \ \frac{\eta^{\la \tau}}{z_{35}} \, i\vep^{\mu \nu \rho}\, \!_{\xi} \, \si^{\xi}_{ \al \dbe} \; (z_{36} \, z_{57} \, + \, z_{37} \, z_{56}) \ + \ \frac{\eta^{\rho \tau}}{z_{45}} \, i\vep^{\mu \nu \la}\, \!_{\xi} \, \si^{\xi}_{ \al \dbe} \; (z_{46} \, z_{57} \, + \, z_{47} \, z_{56}) \biggr)  \Biggr\}\,.
\label{rv,17}
\end{align}

\subsection[Generalization to $n$ NS fields]{Generalization to $\bm{n}$ NS fields}
\label{sec:GeneralizationToNNSFields}

Let us write down the generalization of the antisymmetrization prescription (\ref{as,3}), (\ref{rv,8a}), (\ref{si}) and
(\ref{rv,18}) to $n$ matrices:
\begin{subequations}
\begin{align}
\si^{[\mu_{1}}  \, \bar{\si}^{\mu_2} \, ... \, \bar{\si}^{\mu_{2k}]} \ \ &= \ \ \sum_{\ell = 0}^{k} \sum_{\rho \in S_{2k}/ {\cal Q}_{k+1,\ell}} \! \! \! \! \! \! \te{sgn}(\rho) \, \si^{\mu_{\rho(1)}} \, \bar{\si}^{\mu_{\rho(2)}} \, ... \, \bar{\si}^{\mu_{\rho(2\ell)}} \prod_{j=1}^{k-\ell} \eta^{\mu_{\rho(2\ell+2j-1)} \mu_{\rho(2\ell+2j)}}\,, \label{AS,1} \\
\si^{[\mu_{1}}  \, \bar{\si}^{\mu_2} \, ... \, \si^{\mu_{2k+1}]} \ \ &= \ \ \sum_{\ell = 0}^{k} \sum_{\rho \in S_{2k+1}/ {\cal P}_{k+1,\ell}} \! \! \! \! \! \! \! \! \! \te{sgn}(\rho) \, \si^{\mu_{\rho(1)}} \, \bar{\si}^{\mu_{\rho(2)}} \, ... \, \si^{\mu_{\rho(2\ell+1)}} \prod_{j=1}^{k-\ell} \eta^{\mu_{\rho(2\ell+2j)} \mu_{\rho(2\ell+2j+1)}\,.} \label{AS,2}
\end{align}
\end{subequations}
The proof is based upon induction and can be found in appendix
\ref{sec:AntisymmetrizedVersusOrderedProductsOfSigmaMatrices}. With equations (\ref{AS,1}), (\ref{AS,2}) in mind, we
claim that the $n$ point functions $\Om_{(n)}$, $\om_{(n)}$ and $\bar{\om}_{(n)}$ can alternatively be written as:
\begin{subequations}
\begin{align}
\Om&_{(n)}^{\mu_{1} ... \mu_{2n-1}}\,_{\al \dbe}(z_{i}) \ \ := \ \ \langle \psi^{\mu_{1}}(z_{1}) \, \psi^{\mu_{2}}(z_{2}) \, ... \, \psi^{\mu_{2n-1}}(z_{2n-1}) \, S_{\al}(z_{A}) \, S_{\dbe}(z_{B}) \rangle \notag \\
& = \ \ \frac{1}{2^{n-1} \, \sqrt{2}} \prod_{i=1}^{2n-1} (z_{iA} \, z_{iB})^{-1/2} \! \! \! \sum_{\ell = 0}^{\te{min}(1,n-1)} \! \! \!  z_{AB}^{\ell} \! \! \! \! \sum_{\rho \in S_{2n-1}/{\cal P}_{n,\ell}} \! \! \!  \te{sgn}(\rho) \, \bigl(\si^{[\mu_{\rho(1)}} \, \bar{\si}^{\mu_{\rho(2)}} \, ... \, \bar{\si}^{\mu_{\rho(2\ell)}} \, \si^{\mu_{\rho(2\ell+1)}]} \bigr)_{\al \dbe}  \notag \\
& \ \ \ \ \  \ \ \times \prod_{j=1}^{n-\ell-1} \frac{\eta^{\mu_{\rho(2\ell+2j)} \mu_{\rho(2\ell+2j+1)}}}{z_{\rho(2\ell+2j),\rho(2\ell+2j+1)} } \; \bigl( z_{\rho(2\ell+2j),A} \, z_{\rho(2\ell+2j+1),B} \, + \, z_{\rho(2\ell+2j),B} \, z_{\rho(2\ell+2j+1),A} \bigr)\,,
\label{AS,8a} \\
\om&_{(n)}^{\mu_{1} ... \mu_{2n-2}}\,_{\al} \, ^{ \be}(z_{i}) \ \ := \ \ \langle \psi^{\mu_{1}}(z_{1}) \, \psi^{\mu_{2}}(z_{2}) \, ... \, \psi^{\mu_{2n-2}}(z_{2n-2}) \, S_{\al}(z_{A}) \, S^{\be}(z_{B}) \rangle \notag \\
&= \ \ \frac{- \, 1}{2^{n-1} \, z_{AB}^{1/2}} \; \prod_{i=1}^{2n-2} (z_{iA} \, z_{iB})^{-1/2} \! \! \! \sum_{\ell = 0}^{\te{min}(2,n-1)} \! \! \!  z_{AB}^{\ell} \! \! \! \! \sum_{\rho \in S_{2n-2}/{\cal Q}_{n,\ell}} \! \! \!  \te{sgn}(\rho) \, \bigl(\si^{[\mu_{\rho(1)}} \, \bar{\si}^{\mu_{\rho(2)}} \, ... \, \si^{\mu_{\rho(2\ell-1)}} \, \bar{\si}^{\mu_{\rho(2\ell)}]} \bigr)_{\al} \, ^{\be} \notag \\
& \ \ \ \ \  \ \ \times \prod_{j=1}^{n-\ell-1} \frac{\eta^{\mu_{\rho(2\ell+2j-1)} \mu_{\rho(2\ell+2j)}}}{z_{\rho(2\ell+2j-1),\rho(2\ell+2j)} } \; \bigl( z_{\rho(2\ell+2j-1),A} \, z_{\rho(2\ell+2j),B} \, + \, z_{\rho(2\ell+2j-1),B} \, z_{\rho(2\ell+2j),A} \bigr)\,,
\label{AS,8b} \\
\bar{\om}&_{(n)}^{\mu_{1} ... \mu_{2n-2} \dal} \, _{ \dbe}(z_{i}) \ \ := \ \ \langle \psi^{\mu_{1}}(z_{1}) \, \psi^{\mu_{2}}(z_{2}) \, ... \, \psi^{\mu_{2n-2}}(z_{2n-2}) \, S^{\dal}(z_{A}) \,  S_{\dbe}(z_{B}) \rangle \notag \\
&= \ \ \frac{+ \, 1}{2^{n-1} \, z_{AB}^{1/2}} \; \prod_{i=1}^{2n-2} (z_{iA} \, z_{iB})^{-1/2} \! \! \! \sum_{\ell = 0}^{\te{min}(2,n-1)} \! \! \!   z_{AB}^{\ell} \! \! \! \! \sum_{\rho \in S_{2n-2}/{\cal Q}_{n,\ell}} \! \! \!  \te{sgn}(\rho) \, \bigl(\bar{\si}^{[\mu_{\rho(1)}} \, \si^{\mu_{\rho(2)}} \, ... \, \bar{\si}^{\mu_{\rho(2\ell-1)}} \, \si^{\mu_{\rho(2\ell)}]} \bigr)^{\dal} \,_{\dbe} \notag \\
& \ \ \ \ \ \ \ \times \prod_{j=1}^{n-\ell-1} \frac{\eta^{\mu_{\rho(2\ell+2j-1)} \mu_{\rho(2\ell+2j)}}}{z_{\rho(2\ell+2j-1),\rho(2\ell+2j)} } \; \bigl( z_{\rho(2\ell+2j-1),A} \, z_{\rho(2\ell+2j),B} \, + \, z_{\rho(2\ell+2j-1),B} \, z_{\rho(2\ell+2j),A} \bigr)\,.
\label{AS,8c}
\end{align}
\end{subequations}
For the cases $n=2,3$ these expressions reduce to (\ref{as,1}), (\ref{rv,8b}), (\ref{rv,12}) as well as
(\ref{rv,17}). The general proof is carried out in appendix \ref{sec:TheProofForTheAntisymmetricRepresentation}.

\section{Concluding remarks}
\label{sec:ConclusionAndOutlook}

In this work we have derived an algorithm to compute any tree level correlation 
function of the sort \req{BASIC} with $SO(1,3)$
Neveu--Schwarz vector fields $\psi^\mu$ and corresponding Ramond spin fields $S_\al, S_{\dbe}$. The construction of mixed correlators
\req{BASIC} with both vectors and spinors generally relies on the factorization \req{psiss} specific to four
dimensions. In addition, bosonization techniques play an essential role to decouple left-- and right--handed spin fields
eventually simplifying the spin field $2n+r+s$ point function  to a product of two pure spin field
correlators \req{BASICs} of opposite helicity.  With this description any correlator \req{BASIC} assumes the form
\req{noNS}.  Eq. \req{noNS} is one of the central results of this article and the starting point to determine various
correlators up to the eight--point level.  These results can be found in Section~6.  The factorization
property \req{noNS} allows to also write down a set of vanishing correlators
involving NS and R fermions, given in \req{vancor}. These identities play an important role in 
proving the vanishing of various superstring tree--level amplitudes.

Our findings can be used for both any superstring and heterotic string compactifications allowing for a CFT description.
Equipped with these results it is possible to compute superstring scattering amplitudes with many bosons and fermions,
in which correlators of the type \req{BASIC} enter as basic ingredients.  
The space--time Lorentz structure of the amplitudes is determined by these basic correlators. Such processes are both of considerable
theoretical and phenomenological interest.
E.g. the set of correlators \req{STARTover}, whose final result is given in Eq. 
\req{npt,1}, enters the computation of the class of $N$--point amplitudes \req{UNIVERSAL} with two gauginos or chiral fermions
and $N-2$ gluons. These $D=4$ amplitudes are especially interesting since they
do not depend on any compactification details and the class of amplitudes \req{UNIVERSAL}
is generic to any string compactification, cf. also \cite{LHC1,LHC2}.

Supersymmetric Ward identities valid at tree--level in field--theory can be extended to superstring theory valid 
to all orders in the string tension $\alpha'$ \cite{MHV}. Generically, these relations equate amplitudes with a different number
of fermions. With the techniques presented in this work it is tractable to 
compute any of these amplitudes explicitly to investigate further their structure and reveal the underlying symmetries in the corresponding Ward identities.

A natural question to ask is how these tree--level methods generalize to higher genus. Some correlators \req{BASIC} have
been constructed at one--loop by first principles by means of their short distance behaviour and their modular
properties on the one--loop Riemann surface \cite{ATICK}.  At one--loop, the generalization of the odd function
$z_{ij}=z_i - z_j$ is the unique odd theta function $\theta_1(z_{ij},\tau)$. However, this does not yet cover the spin
structure dependence. It turns out that loop correlators with left- and right handed spin fields no longer
factorize. The two chiralities couple through the spin structure dependent parts of the correlation
function. Bosonization does not work in straight forward fashion within the sector of definite spin structure, i.e. is
only valid after carrying out the spin sum \cite{LAG,OS}.

\vskip1cm
\goodbreak
\centerline{\noindent{\bf Acknowledgments} }\vskip 2mm

We wish to thank Alexander Dobrinevski, Olaf Lechtenfeld, Dieter L\"ust, and 
Dimitrios Tsimpis for fruitful discussions.

\section*{Appendix}
\appendix

\section{Sigma matrix identities}
\label{sec:SigmaMatrixTechnology}

This appendix collects the various sigma matrix identities which allow to reduce and compare different expressions for the
correlation functions. With increasing number of free spinor indices, the set of relations between various index
structures becomes particularly rich. The lists given here do not claim to be complete for all higher point correlators. In some cases, we rather give the tools and tricks to derive the $\si$--matrix equations from first principles and show representative examples.

The $\si$--matrix identities are organized according to the correlation functions they apply to. Therefore, after a brief
introduction to  conventions of $\si^{\mu \nu}$, the order in which they are displayed is in lines of  Section
\ref{sec:ExplicitExamplesWithSpinFields}.

\subsection{The Lorentz generators and symmetric bispinors}
\label{sec:TheLorentzGeneratorsAndSymmetricBispinors}

In section \ref{sec:ReviewOfLowerOrderCorrelators} it was already pointed out that we are using the normalization convention for the spinorial Lorentz generators:
\beq
(\si^{\mu \nu})_{\al}\,^{\be} \ \ := \ \ (\si^{[\mu} \, \bar{\si}^{\nu]})_{\al}\,^{\be} \eq \frac{1}{2} \; \bigl( \si^{\mu} \, \bar{\si}^{\nu} \ - \ \si^{\nu} \, \bar{\si}^{\mu} \bigr)_{\al}\,^{\be}\,.
\label{sig0}
\eeq
By lowering $\be$ one arrives at an object of definite symmetry property. Let us define
\beq
(\si^{\mu} \, \bar{\si}^{\nu} \, \vep)_{\al \be} \ \ := \ \ (\si^{\mu} \, \bar{\si}^{\nu})_{\al}\,^{\ga} \, \vep_{\ga \be}\,,
\label{sig0a}
\eeq
where the attachment of $\vep_{\ga \be}$ is treated here as a matrix multiplication. Under permutation of the Lorentz
and spinor indices this tensor catches a minus sign. This causes $\si^{\mu\nu}$ to be symmetric:
\beq
(\si^{\mu} \, \bar{\si}^{\nu} \, \vep)_{\al \be} \eq - \, (\si^{\nu} \, \bar{\si}^{\mu} \, \vep)_{\be \al} \ \ \ \Rightarrow \ \ \ (\si^{\mu \nu} \, \vep)_{\al \be} \eq (\si^{\mu \nu} \, \vep)_{(\al \be)}\,.
\label{sig0b}
\eeq
The analogous right handed expression is defined by
\beq
(\vep \, \bar{\si}^{\mu} \, \si^{\nu})_{\dal \dbe} \ \ := \ \ \vep_{\dal \dga} \, (\bar{\si}^{\mu} \, \si^{\nu})^{\dga} \, _{\dbe}\,,
\label{sig0c}
\eeq
from which we can form again a symmetric tensor:
\beq
(\vep \, \bar{\si}^{\mu} \, \si^{\nu})_{\dal \dbe} \eq - \, (\vep \, \bar{\si}^{\nu} \, \si^{\mu})_{\dbe \dal} \ \ \ \Rightarrow \ \ \ (\vep \, \bar{\si}^{\mu \nu})_{\dal \dbe} \eq (\vep \, \bar{\si}^{\mu \nu})_{(\dal \dbe)} \ .
\label{sig0d}
\eeq
The $\si$ matrices obey the well-known Dirac algebra
\begin{equation}
  \label{sigsymm}
  \si^\mu \, \sib^\nu \ + \ \si^\nu \, \sib^\mu \eq -2\,\eta^{\mu\nu} \ .
\end{equation}
Using this we can swap spinor indices in the following way:
\begin{equation}
  \label{sig7}
  (\si^\mu \, \sib^\nu \, \vep)_{\be\al} \eq - \, (\si^\nu \, \sib^\mu \, \vep)_{\al\be} \eq (\si^\mu \, \sib^\nu \, \vep)_{\al\be} \ + \ 2\,\eta^{\mu\nu} \ \vep_{\al\be}\,.
\end{equation}

\subsection{Spin field correlation functions and Fierz identities}
\label{sec:FourPointFunctionsFierzIdentities}

Four point functions $\langle S_\al S_\be S_\ga S_\de \rangle$ of spin fields with uniform chirality, clearly give rise
to a term proportional to $\vep_{\al \ga} \vep_{\be \de}$ in the limit $z_1 \mto z_3$ or $z_2 \mto z_4$. The reason for
these tensors not to appear in the results (\ref{4,1b}), (\ref{4,1c}) is the Fierz-identity:
\beq
\vep_{\al \ga} \, \vep_{\be \de} \eq \vep_{\al \be} \, \vep_{\ga \de} \ - \ \vep_{\al \de} \, \vep_{\ga \be}\,.
\label{sig1}
\eeq 
This equation can be derived in the following way: in four dimensions Weyl spinors can only take two values, i.e.\ $\al
= 1,2$. Hence, antisymmetric products of three or more indices, such as $\vep_{\al[\be} \vep_{\ga \de]}$, vanish
identically. The same reasoning applies to right handed spinors. In the following we make extensive use of this fact to
derive relations between index terms, like:
\begin{subequations}
\begin{align}
  \vep_{\al[\be} \, (\si^{\mu_1}\,\sib^{\mu_2}\dots\sib^{\mu_{2n}} \, \vep)_{\ga\de]}\eq 0\,,\label{sigreleven}\\
  \vep_{[\al\be} \, (\si^{\mu_1}\,\sib^{\mu_2}\dots\si^{\mu_{2n+1}})_{\ga]\dde} \eq 0\,.\label{sigrelodd}
\end{align}
\end{subequations}
A more involved identity can be derived from $\de^{\al}_{[\be} \de^{\ga}_{\de} \de^{\ep}_{\ze]} = 0$:
\begin{align}
  0 \eq &\vep_{\al \be} \,  \vep_{\ga \de} \, \vep_{\ep \ze} \ - \ \vep_{\al \be} \,  \vep_{\ga \ze} \, \vep_{\ep \de} \ - \ \vep_{\al \de}  \, \vep_{\ga \be} \,  \vep_{\ep \ze} \notag \\
   + \ &\vep_{\al \de} \, \vep_{\ga \ze} \, \vep_{\ep \be} \ + \ \vep_{\al \ze} \,  \vep_{\ga \be} \,  \vep_{\ep \de} \ - \ \vep_{\al \ze} \, \vep_{\ga \de} \, \vep_{\ep \be} \,.
\label{sigspin1}
\end{align}
This is helpful for the six point function $\langle S_\al S_\be S_\ga S_\de S_\ep S_{\zeta} \rangle$ and correlators derived from it. Similarly, antisymmetrizing even more spinor Kronecker deltas gives rise to analogous relations for handling spin field correlators $\langle S_{\al_1} S_{\al_2} ... S_{\al_{2M-1}} \, S_{\al_{2M}} \rangle$ of arbitrary size:
\beq
0 \eq \de^{\al_1}_{[\al_2} \, \de^{\al_3}_{\al_4} \, \ldots \, \de^{\al_{2M-1}}_{\al_{2M}]} \ \ \ \Rightarrow \ \ \ 0 \eq \sum_{\rho \in S_M} \te{sgn}(\rho) \, \prod_{m=1}^{M} \vep_{\al_{2m-1} \al_{\rho(2m)}} \ .
\label{2Mpt}
\eeq
The discussion is easily extended to the right handed sector with dotted indices, i.e.\ by complex conjugation of the
previous identities.

\subsection{Five- and six--point functions}

To reduce five--point functions like $\langle \psi^{\mu} S_{\al} S_{\be} S_{\ga} S_{\dde} \rangle$ the relation
\eqref{sigrelodd} for $n=0$
\beq
\si^{\mu}_{\be \dde} \, \vep_{\al \ga} \ \ = \ \ \si^{\mu}_{\al \dde} \, \vep_{\be \ga} \ + \ \si^{\mu}_{\ga \dde} \, \vep_{\al \be}
\label{sig4}
\eeq
proves to be useful. This is the starting point for several further relations involving two Lorentz- and four spinor
indices.

\bigskip
\noindent
\underline{{\bf Correlators $\langle \psi^{\mu} \psi^{\nu} S_{\al} S_{\dbe} S_{\ga} S_{\dde} \rangle$}}

\bigskip
\noindent
For this particular index configuration, it is worthwhile to multiply Eq. (\ref{sig4}) by $\vep_{\dbe \dal} \bar{\si}^{\mu
  \dal \be}$. Repeating this exercise with right handed representations, one obtains:
\begin{subequations}
\begin{align}
\si^{\mu}_{\al \dbe} \,  \si^{\nu}_{\ga \dde} \ - \ \si^{\mu}_{\al \dde}  \si^{\nu}_{\ga \dbe} \ \ &= \ \ (\si^{\mu}   \, \bar{\si}^{\nu} \, \vep)_{\al \ga} \, \vep_{\dbe \dde} \label{sig5a}\,, \\
\si^{\mu}_{\al \dbe} \,  \si^{\nu}_{\ga \dde} \ - \ \si^{\mu}_{\ga \dbe}  \si^{\nu}_{\al \dde} \ \ &= \ \ (\vep \, \bar{\si}^{\mu}   \, \si^{\nu} )_{\dbe \dde} \, \vep_{\al \ga} \label{sig5b} \, .
\end{align}
\end{subequations}
Finally, by symmetrizing in $(\mu \nu)$, we arrive at:
\beq
2 \, \eta^{\mu \nu} \, \vep_{\al \ga} \, \vep_{\dbe \dde} \eq \si^{\mu}_{\al \dde} \, \si^{\nu}_{\ga \dbe} \ + \ \si^{\mu}_{\ga \dbe} \, \si^{\nu}_{\al \dde} \ - \ \si^{\mu}_{\al \dbe} \, \si^{\nu}_{\ga \dde} \ - \ \si^{\mu}_{\ga \dde} \, \si^{\nu}_{\al \dbe} \,.
\label{sig6}
\eeq
These identities allows to reduce any tensor of type $T^{\mu \nu} _{\al \dbe \ga \dde}$ to a linear combination of the
four tensors $\si^{\mu}_{\al \dbe} \si^{\nu}_{\ga \dde}$, $\si^{\mu}_{\ga \dde}\si^{\nu}_{\al\dbe}$,
$\si^{\mu}_{\al\dde}\si^{\nu}_{\ga \dbe}$ and $\si^{\mu}_{\ga \dbe}\si^{\nu}_{\al \dde}$.  \bigskip
Besides, the following relations  prove to be useful:
\def\eps{\epsilon}\def\ov{\overline}
\beq\ba{rcl}
(\sigma\eps)_{\al\gamma}&=&(\sigma\eps)_{\gamma\al}\ ,\\
\si^{\mu}_{\al\dot\bet}\ \si^{\nu}_{\gamma\dot\delta}-
\si^{\mu}_{\al\dot\delta}\ \si^{\nu}_{\gamma\dot\beta}&=&
-\delta^{\mu\nu}\ 
\eps_{\al\gamma}\ \eps_{\dot\bet\dot\delta}+
(\sigma^{\mu\nu}\eps)_{\al\gamma}\ 
\eps_{\dot\bet\dot\delta}\ ,\cr
\sigma_{\al\dot\bet}^{\mu}\ \si^{\nu}_{\gamma\dot\delta}+
\sigma_{\al\dot\bet}^{\nu}\ \si^{\mu}_{\gamma\dot\delta}
&=&-\delta^{\mu\nu}\ \eps_{\al\gamma}\ \eps_{\dot\bet\dot\delta}+
(\si^{\lambda\nu}\eps)_{\al\gamma}\ 
(\eps\ov\sigma^{\lambda\mu})_{\dot\bet\dot\delta}\ .
\ea\eeq
These relations may be used to show, that the correlator
\req{spinii} has the correct properties under permutations.

\noindent
\underline{{\bf Correlators $\langle \psi^{\mu} \psi^{\nu} S_{\al} S_{\be} S_{\ga} S_{\de} \rangle$}}

\bigskip
\noindent
Organizing tensors of type $T^{\mu \nu}_{\al \be \ga \de}$ with uniform spinor indices requires a slightly different
reasoning. First of all, one can eliminate $\eta^{\mu \nu} \vep_{\al \ga} \vep_{\be \de}$ with the help of
(\ref{sig1}). The spinor indices at $(\si^{\mu} \bar{\si}^{\nu} \vep)\vep$ can be reordered using \eqref{sig7} so that
we only keep the index pairs $(\al \be)$, $(\al \de)$, $(\ga \de)$ and $(\ga \be)$. Then using \eqref{sigreleven} for
$n=1$ we can eliminate $(\si^{\mu} \bar{\si}^{\nu} \vep)_{\al \ga} \vep_{\be \de}$ and $(\si^{\mu} \bar{\si}^{\nu}
\vep)_{\be \de} \vep_{\al \ga}$:
\begin{subequations}
\begin{align}
(\si^{\mu} \, \bar{\si}^{\nu} \, \vep)_{\al \ga} \, \vep_{\be \de} \ \ &= \ \ (\si^{\mu} \, \bar{\si}^{\nu} \, \vep)_{\al \be} \, \vep_{\ga \de} \ - \ (\si^{\mu} \, \bar{\si}^{\nu} \, \vep)_{\al \de} \, \vep_{\ga \be}\,, \label{sig8b} \\
(\si^{\mu} \, \bar{\si}^{\nu} \, \vep)_{\be \de} \, \vep_{\al \ga} \ \ &= \ \ (\si^{\mu} \, \bar{\si}^{\nu} \, \vep)_{ \ga \de } \, \vep_{\al \be} \ - \ (\si^{\mu} \, \bar{\si}^{\nu} \, \vep)_{\al \de} \, \vep_{\ga \be}\,. \label{sig8c}
\end{align}
\end{subequations}
A further relation between the remaining $(\si \bar{\si} \vep) \vep$ terms is found in the same way:
\begin{align}
-2 \, \eta^{\mu \nu} \, \vep_{\al \ga} \, \vep_{\be \de} \eq (\si^{\mu} \, \bar{\si}^{\nu} \, \vep)_{\al \be} \, \vep_{\ga \de} \ - \ (\si^{\mu} \, \bar{\si}^{\nu} \, \vep)_{\al \de} \, \vep_{\ga \be} \  + \ (\si^{\mu} \, \bar{\si}^{\nu} \, \vep)_{ \ga \de } \, \vep_{\al \be} \ - \ (\si^{\mu} \, \bar{\si}^{\nu} \, \vep)_{\ga \be} \, \vep_{\al \de}\,.
\label{sig9}
\end{align}
Let us conclude this Subsection by the representation of (\ref{sig9}) in terms of antisymmetrized $\si$ products:
\beq
(\si^{\mu \nu} \, \vep)_{\al \be} \, \vep_{\ga \de} \ - \ (\si^{\mu \nu} \, \vep)_{\al \de} \, \vep_{\ga \be} \ + \ (\si^{\mu \nu} \, \vep)_{ \ga \de } \, \vep_{\al \be} \ - \ (\si^{\mu \nu} \, \vep)_{\ga \be} \, \vep_{\al \de} \eq 0\,.
\label{sig10} 
\eeq

\subsection{Seven--point functions}

\underline{{\bf Correlators $\langle \psi^{\mu} \psi^{\nu} \psi^{\la} S_{\al} S_{\be} S_{\ga} S_{\dde} \rangle$}}

\bigskip
\noindent
The antisymmetrization argument can be used for this correlator as well to derive relations between index terms, like:
\begin{subequations}
\begin{align}
\eta^{\nu \la} \, \si^{\mu}_{\be \dde} \, \vep_{\al \ga} \ \ &= \ \ \eta^{\nu \la} \,  \si^{\mu}_{\al \dde} \, \vep_{\be \ga} \ + \ \eta^{\nu \la} \, \si^{\mu}_{\ga \dde} \, \vep_{\al \be}\,, \label{sig11a} \\
(\si^{\mu} \, \bar{\si}^{\nu} \, \si^{\la})_{\be \dde} \, \vep_{\al \ga} \ \ &= \ \ (\si^{\mu} \, \bar{\si}^{\nu} \, \si^{\la})_{\al \dde} \, \vep_{\be \ga} \ + \ (\si^{\mu} \, \bar{\si}^{\nu} \, \si^{\la})_{\ga \dde} \, \vep_{\al \be} \,. \label{sig11b}
\end{align}
\end{subequations}
Furthermore, applying $(\si^{\mu} \bar{\si}^{\nu})_{\al} \,^{\ka} \vep_{[\ka \be} \si^{\la}_{\ga] \dde} = 0$ yields
identities that mix different $\si$ configurations:
\begin{subequations}
\begin{align}
\vep_{\be \ga} \, ( \si^{\mu} \, \bar{\si}^{\nu} \, \si^{\la})_{\al \dde} \ \ &= \ \ ( \si^{\mu} \, \bar{\si}^{\nu} \, \vep)_{\al \ga} \, \si^{\la}_{\be \dde} \ - \ ( \si^{\mu} \, \bar{\si}^{\nu} \, \vep)_{\al \be} \, \si^{\la}_{\ga \dde}\label{4,24a}\,,\\
\vep_{\al \be} \, ( \si^{\mu} \, \bar{\si}^{\nu} \, \si^{\la})_{\ga \dde} \ \ &= \ \ ( \si^{\mu} \, \bar{\si}^{\nu} \, \vep)_{\be \ga} \, \si^{\la}_{\al \dde} \ - \ ( \si^{\mu} \, \bar{\si}^{\nu} \, \vep)_{\al \ga} \, \si^{\la}_{\be \dde} \ - \ 2 \, \eta^{\mu \nu} \, \vep_{\al \be} \, \si^{\la}_{\ga \dde} \label{4,24b} \,.
\end{align}
\end{subequations}
Hence, triple products $\si^{\mu} \bar{\si}^{\nu} \si^{\la}$ can be completely eliminated.

By permuting the Lorentz indices in (\ref{4,24a}), one can derive four linearly independent relations:
\begin{subequations}
\begin{align}
2 \, \eta^{\nu \la} \, \si^{\mu}_{\al \dde} \, \vep_{\be \ga} \ \ &= \ \ (\si^{\mu} \, \bar{\si}^{\nu} \, \vep)_{\al \be} \, \si^{\la}_{\ga \dde} \ - \ (\si^{\mu} \, \bar{\si}^{\nu} \, \vep)_{\al \ga} \, \si^{\la}_{\be \dde} \notag \\
& \ \ \ \ \ \ \ \ + \ (\si^{\mu} \, \bar{\si}^{\la} \, \vep)_{\al \be} \, \si^{\nu}_{\ga \dde} \ - \ (\si^{\mu} \, \bar{\si}^{\la} \, \vep)_{\al \ga} \, \si^{\nu}_{\be \dde}
\label{4,25a}\,,\\
2 \, \eta^{\mu \nu} \, \si^{\la}_{\ga \dde} \, \vep_{\al \be} \ \ &= \ \ (\si^{\mu} \, \bar{\si}^{\la} \, \vep)_{\be \ga} \, \si^{\nu}_{\al \dde} \ - \ (\si^{\mu} \, \bar{\si}^{\la} \, \vep)_{\al \ga} \, \si^{\nu}_{\be \dde} \notag \\
& \ \ \ \ \ \ \ \ + \ (\si^{\nu} \, \bar{\si}^{\la} \, \vep)_{  \be \ga } \, \si^{\mu}_{\al \dde} \ - \ (\si^{\nu} \, \bar{\si}^{\la} \, \vep)_{\al \ga} \, \si^{\mu}_{\be \dde}
\label{4,25b}\,,\\
2 \, \eta^{\mu \la} \, \si^{\nu}_{\al \dde} \, \vep_{\be \ga} \ - \ 2 \, \eta^{\mu \nu} \, \si^{\la}_{\al \dde} \, \vep_{\be \ga} \ \ &= \ \ (\si^{\nu} \, \bar{\si}^{\la} \, \vep)_{\al \be} \, \si^{\mu}_{\ga \dde} \ - \ (\si^{\nu} \, \bar{\si}^{\la} \, \vep)_{\al \ga} \, \si^{\mu}_{\be \dde} \notag \\
& \ \ \ \ \ \ \ \ - \ (\si^{\mu} \, \bar{\si}^{\nu} \, \vep)_{\al \be} \, \si^{\la}_{\ga \dde} \ + \ (\si^{\mu} \, \bar{\si}^{\nu} \, \vep)_{\al \ga} \, \si^{\la}_{\be \dde}
\label{4,25c}\,,\\
2 \, \eta^{\mu \la} \, \si^{\nu}_{\ga \dde} \, \vep_{\al \be} \ - \ 2 \, \eta^{\nu \la} \, \si^{\mu}_{\ga \dde} \, \vep_{\al \be} \ \ &= \ \ (\si^{\nu} \, \bar{\si}^{\la} \, \vep)_{\al \ga} \, \si^{\mu}_{\be \dde} \ - \ (\si^{\nu} \, \bar{\si}^{\la} \, \vep)_{\be \ga} \, \si^{\mu}_{\al \dde} \notag \\
& \ \ \ \ \ \ \ \ + \ (\si^{\mu} \, \bar{\si}^{\nu} \, \vep)_{ \be \ga} \, \si^{\la}_{\al \dde} \ - \ (\si^{\mu} \, \bar{\si}^{\nu} \, \vep)_{\al \ga} \, \si^{\la}_{\be \dde} \,.
\label{4,25d}
\end{align}
\end{subequations}

These equations among the initially 21 Clebsch Gordan coefficients are not enough to obtain a minimal set of 10 tensors
$T^{\mu \nu \la}_{\al \be \ga \dde}$. We still have to perform one more elimination, which can be achieved by
multiplying (\ref{sig5a}) with $\bar{\si}^{\la \dbe \de} \vep_{\de \be}$. After a further permutation in the spinor
indices one finds:
\beq
0 \eq (\si^{\mu} \, \bar{\si}^{\nu} \, \vep)_{\al \be} \, \si^{\la}_{\ga \dde} \ - \ (\si^{\mu} \, \bar{\si}^{\la} \, \vep)_{\al \ga} \, \si^{\nu}_{\be \dde} \ + \ (\si^{\nu} \, \bar{\si}^{\la} \, \vep)_{\be \ga } \, \si^{\mu}_{\al \dde} \ .
\label{4,25f}
\eeq

\bigskip
\noindent
\underline{{\bf Correlators $\langle \psi^{\mu} S_{\al} S_{\be} S_{\ga} S_{\dde} S_{\dep} S_{\dze} \rangle$}}
\bigskip

\noindent
For this correlation function no relations are needed as it appears already in minimal form.

\goodbreak
\bigskip
\noindent
\underline{{\bf Correlators $\langle \psi^{\mu} S_{\al} S_{\be} S_{\ga} S_{\de} S_{\ep} S_{\dze} \rangle$}}
\bigskip

\noindent
The only relation necessary for putting this correlator into minimal form can be derived by contracting \eqref{sigspin1}
with $\vep^{\ze\ka}\,\si^\mu_{\ka\dze}$:
\begin{equation}
  \label{4.26}
  \si^\mu_{\ep\dze}\,\vep_{\al\be}\,\vep_{\ga\de} \, + \, \si^\mu_{\ep\dze}\,\vep_{\al\de}\,\vep_{\be\ga}\, + \,\si^\mu_{\al\dze}\,\vep_{\be\ep}\,\vep_{\ga\de} \, + \, \si^\mu_{\al\dze}\,\vep_{\be\ga}\,\vep_{\de\ep} \, + \, \si^\mu_{\ga\dze}\,\vep_{\al\be}\,\vep_{\de\ep}\, - \, \si^\mu_{\ga\dze}\,\vep_{\al\de}\,\vep_{\be\ep}\  = \ 0\,.
\end{equation}

\subsection{Eight--point functions}
\label{sec:EightPointFunctions}

In order to compute eight point functions with four NS fermions involved, the trace of (\ref{rv,11}) is needed:
\beq
\te{Tr} \Bigl\{ \si^{\mu} \, \bar{\si}^{\nu} \, \si^{\la} \, \bar{\si}^{\rho} \Bigr\} \eq 2 \, \bigl( \eta^{\mu \nu} \, \eta^{\la \rho} \ - \ \eta^{\mu \la} \, \eta^{\nu \rho} \ + \ \eta^{\mu \rho} \, \eta^{\nu \la} \bigr) \ - \ 2i \, \vep^{\mu \nu \la \rho} \ .
\label{trace}
\eeq

\bigskip
\noindent
\underline{{\bf Correlators $\langle \psi^{\mu} \psi^{\nu} \psi^{\la} \psi^{\rho} S_{\al} S_{\dbe} S_{\ga} S_{\dde} \rangle$}}

\bigskip
\noindent
This correlator can be expressed in terms of the index terms
\begin{equation}
  \label{8ptindexterms}
  \eta^{\mu\nu}\,\si^\la_{\al\dbe}\,\si^\rho_{\ga\dde} \ ,\quad
  (\si^\mu \, \sib^\nu\vep)_{\al\ga} \, (\vep \, \sib^\la \, \si^\rho)_{\dbe\dde} \ ,\quad
  (\si^\mu \, \sib^\nu \, \si^\la)_{\al\dbe}\,\si^\rho_{\ga\dde} \ ,\quad (\si^\mu \, \sib^\nu \, \si^\la \, \sib^\rho\vep)_{\al\ga} \, \vep_{\dbe\dde}
\end{equation}
and permutations in the Lorentz and spinors indices thereof. The two expressions consisting of a chain of four $\si$
matrices can be eliminated with the help of \eqref{sig5a} and \eqref{sig5b}:
\begin{subequations}
\begin{align}
  \label{els}
  (\si^\mu \, \sib^\nu \, \si^\la \, \sib^\rho \, \vep)_{\al\ga}\,\vep_{\dbe\dde} & \ =\ (\si^\mu \, \sib^\nu \, \si^\la)_{\al\dbe}\,\si^\rho_{\ga\dde} \, - \, (\si^\mu \, \sib^\nu\, \si^\la)_{\al\dde}\,\si^\rho_{\ga\dbe}\,,\\
  (\vep \, \sib^\mu \, \si^\nu \, \sib^\la \, \si^\rho)_{\dbe\dde}\,\vep_{\al\ga} & \ =\ (\si^\nu \, \sib^\la \, \si^\rho)_{\ga\dde}\,\si^\mu_{\al\dbe} \, - \, (\si^\nu \, \sib^\la\, \si^\rho)_{\al\dde}\,\si^\mu_{\ga\dbe}\,.
\end{align}
\end{subequations}
The relations \eqref{sig5a} and \eqref{sig5b} derived before can also be used to eliminate the terms of structure $(\si\sib\vep)(\vep\sib\si)$. In fact, they can be applied in two different ways, i.e.\ to the first or to the second
$\si$ chain
\beq  \label{elq}
  (\si^\mu \, \sib^\nu \, \vep)_{\al\ga} \, (\vep \, \sib^\la \, \si^\rho)_{\dbe\dde}  \ =\ \left\{ \begin{array}{ll} (\si^\mu \, \sib^\nu \, \si^\la)_{\al\dbe}\,\si^\rho_{\ga\dde} \ - \ (\si^\mu \, \sib^\nu\,\si^\rho)_{\al\dde}\,\si^\la_{\ga\dbe} &, \ \te{(\ref{sig5b})} \\
 (\si^\nu \, \sib^\la \, \si^\rho)_{\ga\dde}\,\si^\mu_{\al\dbe} \ - \ (\si^\mu \, \sib^\la \, \si^\rho)_{\al\dde}\,\si^\nu_{\ga\dbe} &, \ \te{(\ref{sig5a})} \end{array} \right. \,.
\eeq
Hence, it is not only possible to eliminate the terms $(\si\sib\vep)(\vep\sib\si)$ but furthermore to obtain relations between
the terms on the right hand side of \eqref{elq}:
\begin{subequations}
  \begin{align}
    \label{qmnrel}
    (\si^\mu \, \sib^\nu \, \si^\la)_{\al\dbe}\,\si^\rho_{\ga\dde}&\ =\
    (\si^\nu \, \sib^\la \, \si^\rho)_{\ga\dde}\,\si^\mu_{\al\dbe} \ - \ (\si^\mu \, \sib^\la\, \si^\rho)_{\al\dde}\,\si^\nu_{\ga\dbe} \ + \ (\si^\mu \, \sib^\nu \, \si^\rho)_{\al\dde}\,\si^\la_{\ga\dbe}\,,\\
    (\si^\mu \, \sib^\nu \, \si^\la)_{\al\dde}\,\si^\rho_{\ga\dbe}&\ =\
    (\si^\nu \, \sib^\la \, \si^\rho)_{\ga\dbe}\,\si^\mu_{\al\dde} \ - \ (\si^\mu \, \sib^\la\, \si^\rho)_{\al\dbe}\,\si^\nu_{\ga\dde} \ + \ (\si^\mu \, \sib^\nu \, \si^\rho)_{\al\dbe}\,\si^\la_{\ga\dde}\,,\\
    (\si^\mu \, \sib^\nu \, \si^\la)_{\ga\dde}\,\si^\rho_{\al\dbe}&\ =\
    (\si^\nu \, \sib^\la \, \si^\rho)_{\al\dbe}\,\si^\mu_{\ga\dde} \ - \ (\si^\mu \, \sib^\la\, \si^\rho)_{\ga\dbe}\,\si^\nu_{\al\dde} \ + \ (\si^\mu \, \sib^\nu \, \si^\rho)_{\ga\dbe}\,\si^\la_{\al\dde}\,,\\
    (\si^\mu \, \sib^\nu \, \si^\la)_{\ga\dbe}\,\si^\rho_{\al\dde}&\ =\
    (\si^\nu \, \sib^\la \, \si^\rho)_{\al\dde}\,\si^\mu_{\ga\dbe} \ - \ (\si^\mu \, \sib^\la \, \si^\rho)_{\ga\dde}\,\si^\nu_{\al\dbe} \ + \ (\si^\mu \, \sib^\nu \, \si^\rho)_{\ga\dde}\,\si^\la_{\al\dbe}\,.
  \end{align}
\end{subequations}
The last three equations were found by permuting the spinor indices in \eqref{elq}. However, one can also perform
permutations in the Lorentz indices. This yields
\begin{subequations}
  \label{qmrrel}
  \begin{align}
    2\,\eta^{\mu\nu} \, (\si^\la_{\al\dbe} \, \si^\rho_{\ga\dde} \ - \ \si^\la_{\ga\dbe}\, \si^\rho_{\al\dde})&\ =\
    (\si^\nu \, \sib^\la \, \si^\rho)_{\al\dde}\,\si^\mu_{\ga\dbe} \ - \ (\si^\nu \, \sib^\la\, \si^\rho)_{\ga\dde}\,\si^\mu_{\al\dbe}\notag\\
    & \hspace{.27cm}+ \ (\si^\mu \, \sib^\la \, \si^\rho)_{\al\dde}\,\si^\nu_{\ga\dbe} \ -\ (\si^\mu \, \sib^\la \, \si^\rho)_{\ga\dde}\,\si^\nu_{\al\dbe}\,,\\
    2\,\eta^{\mu\nu} \, (\si^\la_{\al\dde} \, \si^\rho_{\ga\dbe} \ - \ \si^\la_{\ga\dde} \,\si^\rho_{\al\dbe})&\ =\
    (\si^\nu \, \sib^\la \, \si^\rho)_{\al\dbe}\,\si^\mu_{\ga\dde} \ - \ (\si^\nu \, \sib^\la\, \si^\rho)_{\ga\dbe}\,\si^\mu_{\al\dde}\notag\\
    & \hspace{.27cm}+ \ (\si^\mu \, \sib^\la \, \si^\rho)_{\al\dbe}\,\si^\nu_{\ga\dde} \ - \ (\si^\mu \, \sib^\la \, \si^\rho)_{\ga\dbe}\,\si^\nu_{\al\dde}\,,\\
    -2\,\eta^{\la\rho} \, (\si^\mu_{\al\dbe} \, \si^\nu_{\ga\dde} \ - \ \si^\mu_{\al\dde}\,\si^\nu_{\ga\dbe})&\ =\
    (\si^\nu \, \sib^\la \, \si^\rho)_{\ga\dde}\,\si^\mu_{\al\dbe} \ - \ (\si^\nu \, \sib^\la\, \si^\rho)_{\ga\dbe}\,\si^\mu_{\al\dde}\notag\\ 
    & \hspace{.27cm}+ \ (\si^\mu \, \sib^\la \, \si^\rho)_{\al\dbe}\,\si^\nu_{\ga\dde} \ - \ (\si^\mu \, \sib^\la \, \si^\rho)_{\al\dde}\,\si^\nu_{\ga\dbe}\,,\\
    -2\,\eta^{\la\rho} \, (\si^\mu_{\ga\dde} \, \si^\nu_{\al\dbe} \ - \ \si^\mu_{\ga\dbe}\, \si^\nu_{\al\dde})&\ =\
    (\si^\nu \, \sib^\la \, \si^\rho)_{\al\dbe}\,\si^\mu_{\ga\dde} \ - \ (\si^\nu \, \sib^\la \, \si^\rho)_{\al\dde}\,\si^\mu_{\ga\dbe}\notag\\ 
    & \hspace{.27cm}+ \ (\si^\mu \, \sib^\la \, \si^\rho)_{\ga\dde}\,\si^\nu_{\al\dbe} \ -\ (\si^\mu \, \sib^\la \, \si^\rho)_{\ga\dbe}\,\si^\nu_{\al\dde}\,,
  \end{align}
\end{subequations}
and
\begin{subequations}
  \label{qlrrel}
  \begin{align}
    2\,(&\eta^{\mu\la} \, \si^\nu_{\al\dbe} \, \si^\rho_{\ga\dde} \ - \ \eta^{\mu\la} \, \si^\nu_{\ga\dbe} \, \si^\rho_{\al\dde} \
    - \ \eta^{\nu\la} \, \si^\mu_{\al\dbe} \, \si^\rho_{\ga\dde} \ + \ \eta^{\nu\la} \, \si^\mu_{\ga\dbe} \, \si^\rho_{\al\dde})\notag\\
    &= \ -(\si^\nu \, \sib^\la \, \si^\rho)_{\al\dde}\,\si^\mu_{\ga\dbe} \ + \ (\si^\nu \, \sib^\la \, \si^\rho)_{\ga\dde}\,\si^\mu_{\al\dbe}
    + \ (\si^\mu \, \sib^\nu \, \si^\rho)_{\al\dde}\,\si^\la_{\ga\dbe} \ - \ (\si^\mu \, \sib^\nu \, \si^\rho)_{\ga\dde}\,\si^\la_{\al\dbe}\,,\\
    2\,(&\eta^{\mu\la} \, \si^\nu_{\al\dde} \, \si^\rho_{\ga\dbe} \ - \ \eta^{\mu\la} \, \si^\nu_{\ga\dde} \, \si^\rho_{\al\dbe} \
    - \ \eta^{\nu\la} \, \si^\mu_{\al\dde} \, \si^\rho_{\ga\dbe} \ + \ \eta^{\nu\la} \, \si^\mu_{\ga\dde} \, \si^\rho_{\al\dbe})\notag\\
    &= \ -(\si^\nu \, \sib^\la \, \si^\rho)_{\al\dbe}\,\si^\mu_{\ga\dde} \ + \ (\si^\nu \, \sib^\la \, \si^\rho)_{\ga\dbe}\,\si^\mu_{\al\dde} \
    + \ (\si^\mu \, \sib^\nu \, \si^\rho)_{\al\dbe}\,\si^\la_{\ga\dde} \ - \ (\si^\mu \, \sib^\nu \, \si^\rho)_{\ga\dbe}\,\si^\la_{\al\dde}\,,\\
    2\,(&\eta^{\nu\la} \, \si^\mu_{\al\dbe} \, \si^\rho_{\ga\dde} \ - \ \eta^{\nu\la} \, \si^\mu_{\al\dde} \, \si^\rho_{\ga\dbe})
    \ - \ \eta^{\nu\rho} \, \si^\mu_{\al\dbe} \, \si^\la_{\ga\dde} \ + \ \eta^{\nu\rho} \, \si^\mu_{\al\dde} \, \si^\la_{\ga\dbe})\notag\\
    &= \ -(\si^\nu \, \sib^\la \, \si^\rho)_{\ga\dde}\,\si^\mu_{\al\dbe} \ + \ (\si^\nu \, \sib^\la \, \si^\rho)_{\ga\dbe}\,\si^\mu_{\al\dde} \
    + \ (\si^\mu \, \sib^\nu \, \si^\rho)_{\al\dbe}\,\si^\la_{\ga\dde} \ - \ (\si^\mu \, \sib^\nu \, \si^\rho)_{\al\dde}\,\si^\la_{\ga\dbe}\,,\\
    2\,(&\eta^{\mu\rho} \, \si^\nu_{\ga\dde} \, \si^\la_{\al\dbe} \ - \ \eta^{\mu\rho} \, \si^\nu_{\ga\dbe} \, \si^\la_{\al\dde} \
    - \ \eta^{\nu\rho} \, \si^\mu_{\ga\dde} \, \si^\la_{\al\dbe} \ - \ \eta^{\nu\rho} \, \si^\mu_{\ga\dbe} \, \si^\la_{\al\dde})\notag\\
    &= \ -(\si^\nu \, \sib^\la \, \si^\rho)_{\al\dbe}\,\si^\mu_{\ga\dde} \ + \ (\si^\nu \, \sib^\la \, \si^\rho)_{\al\dde}\,\si^\mu_{\ga\dbe}
    \ + \ (\si^\mu \, \sib^\nu \, \si^\rho)_{\ga\dde}\,\si^\la_{\al\dbe} \ - \ (\si^\mu \, \sib^\nu \, \si^\rho)_{\ga\dbe}\,\si^\la_{\al\dde}\,.
  \end{align}
\end{subequations}
In addition the following relations hold,
\begin{subequations}
  \begin{align}
    \label{otherrel1}
    2\,(&\eta^{\mu\rho} \, \si^\nu_{\al\dbe} \, \si^\la_{\ga\dde} \ - \ \eta^{\mu\rho} \, \si^\nu_{\ga\dbe} \, \si^\la_{\al\dde} \
    - \ \eta^{\nu\rho} \, \si^\mu_{\al\dbe} \, \si^\la_{\ga\dde} \ + \ \eta^{\nu\rho} \, \si^\mu_{\ga\dbe} \, \si^\la_{\al\dde} \
    - \ \eta^{\la\rho} \, \si^\mu_{\al\dbe} \, \si^\nu_{\ga\dde} \ + \ \eta^{\la\rho} \, \si^\mu_{\ga\dbe} \, \si^\nu_{\al\dde})\notag\\
    &= \ -(\si^\nu \, \sib^\la \, \si^\rho)_{\al\dbe}\,\si^\mu_{\ga\dde} \ + \ (\si^\nu \, \sib^\la \, \si^\rho)_{\al\dde}\,\si^\mu_{\ga\dbe} \
    - \ (\si^\nu \, \sib^\la \, \si^\rho)_{\ga\dde}\,\si^\mu_{\al\dbe} \ + \ (\si^\nu \, \sib^\la \, \si^\rho)_{\ga\dbe}\,\si^\mu_{\al\dde}\notag\\
    &\phantom{=} \ -(\si^\nu \, \sib^\la \, \si^\rho)_{\al\dbe}\,\si^\nu_{\ga\dde} \ +\ (\si^\mu \, \sib^\la \, \si^\rho)_{\ga\dbe}\,\si^\nu_{\al\dde} \
    + \ (\si^\mu \, \sib^\nu \, \si^\rho)_{\al\dbe}\,\si^\la_{\ga\dde} \ - \ (\si^\mu \, \sib^\nu \, \si^\rho)_{\ga\dbe}\,\si^\la_{\al\dde}\,,\\
    2\,(&\eta^{\mu\rho} \, \si^\nu_{\al\dde} \, \si^\la_{\ga\dbe} \ - \ \eta^{\mu\rho} \,\si^\nu_{\ga\dde} \, \si^\la_{\al\dbe} \
    - \ \eta^{\nu\rho} \, \si^\mu_{\al\dde} \, \si^\la_{\ga\dbe} \ + \ \eta^{\nu\rho} \, \si^\mu_{\ga\dde} \, \si^\la_{\al\dbe} \
    + \ \eta^{\la\rho} \, \si^\mu_{\al\dde} \, \si^\nu_{\ga\dbe} \ - \ \eta^{\la\rho} \, \si^\mu_{\ga\dde} \, \si^\nu_{\al\dbe})\notag\\
    &= \ +(\si^\nu \, \sib^\la \, \si^\rho)_{\al\dbe}\,\si^\mu_{\ga\dde} \ - \ (\si^\nu \,\sib^\la \, \si^\rho)_{\al\dde}\,\si^\mu_{\ga\dbe} \
    + \ (\si^\nu \, \sib^\la \, \si^\rho)_{\ga\dde}\,\si^\mu_{\al\dbe} \ - \ (\si^\nu \, \sib^\la \, \si^\rho)_{\ga\dbe}\,\si^\mu_{\al\dde}\notag\\
    &\phantom{=} \ - \ (\si^\mu \, \sib^\la \, \si^\rho)_{\al\dde}\,\si^\nu_{\ga\dbe} \ + \ (\si^\mu\, \sib^\la \, \si^\rho)_{\ga\dde}\,\si^\nu_{\al\dbe} \
    + \ (\si^\mu \, \sib^\nu \, \si^\rho)_{\al\dde}\,\si^\la_{\ga\dbe} \ - \ (\si^\mu \, \sib^\nu \, \si^\rho)_{\ga\dde}\,\si^\la_{\al\dbe}\,,
  \end{align}
\end{subequations}
as well as
\begin{align}
  \label{otherrel2}
  2\,(&\eta^{\mu\nu} \, \si^\la_{\al\dbe} \, \si^\rho_{\ga\dde}\ - \ \eta^{\mu\nu} \, \si^\la_{\ga\dde} \, \si^\rho_{\al\dbe} \
  - \ \eta^{\mu\la} \, \si^\nu_{\al\dbe} \, \si^\rho_{\ga\dde} \ + \ \eta^{\mu\la} \, \si^\nu_{\al\dde} \, \si^\rho_{\ga\dbe} \
  + \ \eta^{\mu\rho} \, \si^\nu_{\al\dbe} \, \si^\la_{\ga\dde} \ - \ \eta^{\mu\rho} \, \si^\nu_{\al\dde} \, \si^\la_{\ga\dbe})\notag\\
  &= \ (\si^\nu \, \sib^\la \, \si^\rho)_{\al\dde}\,\si^\mu_{\ga\dbe} \ - \ (\si^\mu \, \sib^\la \, \si^\rho)_{\ga\dbe}\,\si^\nu_{\al\dde} \
  + \ (\si^\mu \, \sib^\nu \, \si^\rho)_{\al\dbe}\,\si^\la_{\ga\dde} \ - \ (\si^\mu \, \sib^\nu \, \si^\la)_{\al\dbe}\,\si^\rho_{\ga\dde}\,.
\end{align}
Using the identities stated above it is possible to arrive at a set of 25 independent index terms.

\bigskip
\noindent
\underline{{\bf Correlators $\langle \psi^{\mu} \psi^{\nu} \psi^{\la} \psi^{\rho} S_{\al} S_{\be} S_{\ga} S_{\de} \rangle$}}

\bigskip
\noindent
Here it is possible to apply a generalization of (\ref{sig9}), i.e.:
\begin{align}
-2 \, \eta^{\mu \nu} \, \eta^{\la \rho} \, \vep_{\al \ga} \, \vep_{\be \de} \ \ &= \ \ \eta^{\la \rho} \  \Bigl[ (\si^{\mu} \, \bar{\si}^{\nu} \, \vep)_{\al \be} \, \vep_{\ga \de} \ - \ (\si^{\mu} \, \bar{\si}^{\nu} \, \vep)_{\al \de} \, \vep_{\ga \be} \Bigr. \notag \\
& \ \ \ \ \ \ \ \Bigl. \  + \ (\si^{\mu} \, \bar{\si}^{\nu} \, \vep)_{ \ga \de } \, \vep_{\al \be} \ - \ (\si^{\mu} \, \bar{\si}^{\nu} \, \vep)_{\ga \be} \, \vep_{\al \de} \Bigr] \notag \\
&= \ \ \eta^{\mu \nu} \   \Bigl[ (\si^{\la} \, \bar{\si}^{\rho} \, \vep)_{\al \be} \, \vep_{\ga \de} \ - \ (\si^{\la} \, \bar{\si}^{\rho} \, \vep)_{\al \de} \, \vep_{\ga \be} \Bigr. \notag \\
& \ \ \ \ \ \ \ \Bigl. \  + \ (\si^{\la} \, \bar{\si}^{\rho} \, \vep)_{ \ga \de } \, \vep_{\al \be} \ - \ (\si^{\la} \, \bar{\si}^{\rho} \, \vep)_{\ga \be} \, \vep_{\al \de} \Bigr] \,.
\label{sig12}
\end{align}
Again, (\ref{sig8b}) and (\ref{sig8c}) allow to eliminate the spinor index combinations $(\si \bar{\si} \vep)_{\al \ga}$ and $(\si \bar{\si} \vep)_{\be \de}$.

New identities arise from terms containing four sigma matrices. One can form for instance one single chain $(\si
\bar{\si} \si \bar{\si})$ and gradually apply (\ref{sig7}):
\begin{align}
(\si^{\mu} \, &\bar{\si}^{\nu} \, \si^{\la} \, \bar{\si}^{\rho} \, \vep)_{\be \al} \eq - \ (\si^{\rho} \, \bar{\si}^{\la} \, \si^{\nu} \, \bar{\si}^{\mu} \, \vep)_{\al \be} \notag \\
&= \ \ - \ (\si^{\mu} \, \bar{\si}^{\nu} \, \si^{\la} \, \bar{\si}^{\rho} \, \vep)_{\al \be} \ - \ 2 \, \eta^{\mu \nu} \, (\si^{\la} \, \bar{\si}^{\rho} \, \vep)_{\al \be}  \ + \ 2 \, \eta^{\mu \la} \, (\si^{\nu} \, \bar{\si}^{\rho} \, \vep)_{\al \be} \ - \ 2 \, \eta^{\mu \rho} \, (\si^{\nu} \, \bar{\si}^{\la} \, \vep)_{\al \be} \notag \\
&\hspace{.67cm} - \ 2 \, \eta^{\la \rho} \, (\si^{\mu} \, \bar{\si}^{\nu} \, \vep)_{\al \be} \ + \ 2 \, \eta^{\nu \rho} \, (\si^{\mu} \, \bar{\si}^{\la} \, \vep)_{\al \be} \ - \ 2 \, \eta^{\nu \la} \, (\si^{\mu} \, \bar{\si}^{\rho} \, \vep)_{\al \be} \notag \\
&\hspace{.67cm} - \ 4 \, \eta^{\mu \nu} \, \eta^{\la \rho} \, \vep_{\al \be} \ + \  4 \, \eta^{\mu \la} \, \eta^{\nu \rho} \, \vep_{\al \be} \ - \  4 \, \eta^{\mu \rho} \, \eta^{\nu \la} \, \vep_{\al \be}\,.
\label{sig13}
\end{align}
Due to \eqref{sigreleven} with $n=2$ the following relations hold:
\begin{subequations}
\begin{align}
(\si^{\mu} \, \bar{\si}^{\nu} \, \si^{\la} \, \bar{\si}^{\rho} \, \vep)_{\al \ga} \, \vep_{\be \de} \ \ &= \ \ (\si^{\mu} \, \bar{\si}^{\nu} \, \si^{\la} \, \bar{\si}^{\rho} \,  \vep)_{\al \be} \, \vep_{\be \de} \ - \ (\si^{\mu} \, \bar{\si}^{\nu} \, \si^{\la} \, \bar{\si}^{\rho} \, \vep)_{\al \de} \, \vep_{\ga \be} \label{sig14a} \\
(\si^{\mu} \, \bar{\si}^{\nu} \, \si^{\la} \, \bar{\si}^{\rho} \,  \vep)_{\be \de} \, \vep_{\al \ga} \ \ &= \ \ (\si^{\mu} \, \bar{\si}^{\nu} \, \si^{\la} \, \bar{\si}^{\rho} \,  \vep)_{\ga \de} \, \vep_{\al \be} \ - \ (\si^{\mu} \, \bar{\si}^{\nu} \, \si^{\la} \, \bar{\si}^{\rho} \,  \vep)_{\ga \be} \, \vep_{\al \de} \,. \label{sig14b}
\end{align}
\end{subequations}
Further permutations of $(\si^{\mu} \bar{\si}^{\nu} \si^{\la}  \bar{\si}^{\rho} \vep)_{\al [\be} \vep_{\ga \de]} =0$ yield
 \begin{align}
(\si^{\mu} \, &\bar{\si}^{\nu} \, \si^{\la} \, \bar{\si}^{\rho} \, \vep)_{\al \be} \, \vep_{\ga \de} \ - \ (\si^{\mu} \, \bar{\si}^{\nu} \, \si^{\la} \, \bar{\si}^{\rho} \, \vep)_{\al \de} \, \vep_{\ga \be} \ + \ (\si^{\mu} \, \bar{\si}^{\nu} \, \si^{\la} \, \bar{\si}^{\rho} \, \vep)_{\ga \be} \, \vep_{\al \de} \ - \ (\si^{\mu} \, \bar{\si}^{\nu} \, \si^{\la} \, \bar{\si}^{\rho} \, \vep)_{\ga \de} \, \vep_{\al \be} \notag \\
&= \ \ - \,\vep_{\be \de} \, \Bigl[  2 \, \eta^{\mu \nu} \, (\si^{\la} \, \bar{\si}^{\rho} \, \vep)_{\al \ga}  \ - \ 2 \, \eta^{\mu \la} \, (\si^{\nu} \, \bar{\si}^{\rho} \, \vep)_{\al \ga} \ + \ 2 \, \eta^{\mu \rho} \, (\si^{\nu} \, \bar{\si}^{\la} \, \vep)_{\al \ga}  \Bigr. \notag \\
& \ \ \ \ \ \ \ \ \ \ \ \ \ \ \ \ \Bigl. + \ 2 \, \eta^{\la \rho} \, (\si^{\mu} \, \bar{\si}^{\nu} \, \vep)_{\al \ga} \ - \ 2 \, \eta^{\nu \rho} \, (\si^{\mu} \, \bar{\si}^{\la} \, \vep)_{\al \ga} \ + \ 2 \, \eta^{\nu \la} \, (\si^{\mu} \, \bar{\si}^{\rho} \, \vep)_{\al \ga} \Bigr. \notag \\
& \ \ \ \ \ \ \ \ \ \ \ \ \ \ \ \ \Bigl. + \ 4 \, \eta^{\mu \nu} \, \eta^{\la \rho} \, \vep_{\al \ga} \ - \  4 \, \eta^{\mu \la} \, \eta^{\nu \rho} \, \vep_{\al \ga} \ + \  4 \, \eta^{\mu \rho} \, \eta^{\nu \la} \, \vep_{\al \ga}  \Bigr] \ . \label{sig15} 
 \end{align}
For this correlator tensors made up from two $\si$ chains, i.e.\ $(\si \bar{\si} \vep) (\si \bar{\si} \vep)$
arise. These terms fulfill the the relations
\begin{subequations}
\begin{align}
(\si^{\mu} \, \bar{\si}^{\nu} \, \vep)_{\ga \be} \, (\si^{\la} \, \bar{\si}^{\rho} \, \vep)_{\al \de} \ - \ &(\si^{\mu} \, \bar{\si}^{\nu} \, \vep)_{\al \be} \, (\si^{\la} \, \bar{\si}^{\rho} \, \vep)_{\ga \de} \ \, = \, \  \vep_{\al \ga} \, \Bigl[ (\si^{\mu} \, \bar{\si}^{\nu} \, \si^{\la} \, \bar{\si}^{\rho} \,  \vep)_{\be \de} \ + \ 2 \, \eta^{\mu \nu} \, (\si^{\la} \, \bar{\si}^{\rho} \, \vep)_{\be \de}\,, \Bigr] \label{sig16a} \\
(\si^{\mu} \, \bar{\si}^{\nu} \, \vep)_{\al \de} \, (\si^{\la} \, \bar{\si}^{\rho} \, \vep)_{\ga \be} \ - \ &(\si^{\mu} \, \bar{\si}^{\nu} \, \vep)_{\ga \de} \, (\si^{\la} \, \bar{\si}^{\rho} \, \vep)_{\al \be} \ \, = \, \  \vep_{\al \ga} \, \Bigl[ (\si^{\mu} \, \bar{\si}^{\nu} \, \si^{\la} \, \bar{\si}^{\rho} \,  \vep)_{\be \de} \ - \ 2 \, \eta^{\mu \la} \, (\si^{\nu} \, \bar{\si}^{\rho} \, \vep)_{\be \de} \Bigr. \notag \\
& \Bigl. \ + \ 2 \, \eta^{\mu \rho} \, (\si^{\nu} \, \bar{\si}^{\la} \, \vep)_{\be \de} \ + \ 2 \, \eta^{\la \rho} \, (\si^{\mu} \, \bar{\si}^{\nu} \, \vep)_{\be \de} \ - \ 2 \, \eta^{\nu \rho} \, (\si^{\mu} \, \bar{\si}^{\la} \, \vep)_{\be \de}  \Bigr. \notag \\
& \Bigl. \ + \ 2 \, \eta^{\nu \la} \, (\si^{\mu} \, \bar{\si}^{\rho} \, \vep)_{\be \de} \ - \ 4 \, \eta^{\mu \la} \, \eta^{\nu \rho} \, \vep_{\be \de} \ + \ 4 \, \eta^{\mu \rho} \, \eta^{\nu \la} \, \vep_{\be \de} \Bigr]\,,  \label{sig16b}
\end{align}
\end{subequations}
which stem from $(\si^{\mu} \bar{\si}^{\nu})_{\al}\, ^{\ka} \vep_{[\ka \be} (\si^{\la} \bar{\si}^{\rho} \vep)_{\ga] \de} =0$.

To remove unwanted poles in $z_{13} z_{24}$ from the eight functions, 
the following identities prove to be  useful:
\begin{subequations}
\begin{align}
- \, 2 \, \eta^{\nu \rho} \, (\si^{\mu} \, \bar{\si}^{\la} \, \vep)_{\al \ga} \, \vep_{\be \de} \ \ &= \ \ (\si^{\mu} \, \bar{\si}^{\nu} \, \vep)_{\al \be} \, (\si^{\la} \, \bar{\si}^{\rho} \, \vep)_{\ga \de} \ - \ (\si^{\mu} \, \bar{\si}^{\nu} \, \vep)_{\al \de} \, (\si^{\la} \, \bar{\si}^{\rho} \, \vep)_{\ga \be} \notag \\
& \ \ \ \ \ + \ (\si^{\mu} \, \bar{\si}^{\rho} \, \vep)_{\al \be} \, (\si^{\la} \, \bar{\si}^{\nu} \, \vep)_{\ga \de} \ - \ (\si^{\mu} \, \bar{\si}^{\rho} \, \vep)_{\al \de} \, (\si^{\la} \, \bar{\si}^{\nu} \, \vep)_{\ga \be}\,, \label{sig18a} \\
- \, 2 \, \eta^{\mu \la} \, (\si^{\nu} \, \bar{\si}^{\rho} \, \vep)_{\be \de} \, \vep_{\al \ga} \ \ &= \ \ (\si^{\mu} \, \bar{\si}^{\nu} \, \vep)_{\al \be} \, (\si^{\la} \, \bar{\si}^{\rho} \, \vep)_{\ga \de} \ - \ (\si^{\mu} \, \bar{\si}^{\nu} \, \vep)_{\ga \be} \, (\si^{\la} \, \bar{\si}^{\rho} \, \vep)_{\al \de} \notag \\
& \ \ \ \ \ + \  (\si^{\la} \, \bar{\si}^{\nu} \, \vep)_{\al \be} \, (\si^{\mu} \, \bar{\si}^{\rho} \, \vep)_{\ga \de} \ - \ (\si^{\la} \, \bar{\si}^{\nu} \, \vep)_{\ga \be} \, (\si^{\mu} \, \bar{\si}^{\rho} \, \vep)_{\al \de}\,, \label{sig18b} \\
4 \, \eta^{\mu \la} \, \eta^{\nu \rho} \, \vep_{\al \ga} \, \vep_{\be \de} \ \ &= \ \ (\si^{\mu} \, \bar{\si}^{\nu} \, \vep)_{\al \be} \, (\si^{\la} \, \bar{\si}^{\rho} \, \vep)_{\ga \de} \ - \ (\si^{\mu} \, \bar{\si}^{\nu} \, \vep)_{\ga \be} \, (\si^{\la} \, \bar{\si}^{\rho} \, \vep)_{\al \de} \notag \\
& \ \ \ \ \ + \ (\si^{\mu} \, \bar{\si}^{\nu} \, \vep)_{\ga \de} \, (\si^{\la} \, \bar{\si}^{\rho} \, \vep)_{\al \be} \ - \ (\si^{\mu} \, \bar{\si}^{\nu} \, \vep)_{\al \de} \, (\si^{\la} \, \bar{\si}^{\rho} \, \vep)_{\be \ga} \notag \\
& \ \ \ \ \ + \  (\si^{\la} \, \bar{\si}^{\nu} \, \vep)_{\al \be} \, (\si^{\mu} \, \bar{\si}^{\rho} \, \vep)_{\ga \de} \ - \ (\si^{\la} \, \bar{\si}^{\nu} \, \vep)_{\ga \be} \, (\si^{\mu} \, \bar{\si}^{\rho} \, \vep)_{\al \de} \notag \\
& \ \ \ \ \ + \ (\si^{\la} \, \bar{\si}^{\nu} \, \vep)_{\ga \de} \, (\si^{\mu} \, \bar{\si}^{\rho} \, \vep)_{\al \be} \ - \ (\si^{\la} \, \bar{\si}^{\nu} \, \vep)_{\al \de} \, (\si^{\mu} \, \bar{\si}^{\rho} \, \vep)_{\ga \be}\ .  \label{sig18c}
\end{align}
\end{subequations}

\bigskip
\noindent
\underline{{\bf Correlators $\langle \psi^{\mu} \psi^{\nu} S_{\al} S_{\be} S_{\ga} S_{\de} S_{\ep} S_{\ze} \rangle$}}

\bigskip
\noindent
The index terms appearing in this correlator have either the form $\eta \vep \vep$ or $(\si\sib\vep)\vep\vep$. The former can be reduced to five terms by virtue of \eqref{sigspin1}. For the latter terms we use \eqref{sig8b}:
\begin{equation}
  \label{8psig}
  (\si^{\mu}\,\sib^{\nu}\,\vep)_{\al\ga}\,\vep_{\be \de}\,\vep_{\ep\ze} \eq (\si^{\mu}\,\sib^{\nu}\,\vep)_{\al\be}\,\vep_{\ga \de}\,\vep_{\ep\ze}
  \ - \ (\si^{\mu}\,\sib^{\nu}\,\vep)_{\al\de}\,\vep_{\ga\be}\,\vep_{\ep\ze}\,.
\end{equation}
Permuting the spinor indices in this identity yields, together with the Fierz identity \eqref{sig1}, enough relations to
reduce the index terms down to a minimal set of 14.

\goodbreak
\section{Ordering and antisymmetrizing sigma matrix chains}
\label{sec:ConversionToAntisymmetrizedSiProducts}

To check the consistency of certain correlation functions in the limit $z_i \mto z_j$, one is sometimes faced with the
challenge to reorder a string of $\si$ matrices. The calculation of the seven point function (\ref{rv,16}) requires for
instance:
\begin{subequations}
\begin{align}
\si^{\nu} \, \bar{\si}^{\la} \, \si^{\mu} \ \ &= \ \ \si^{\mu} \, \bar{\si}^{\nu} \, \si^{\la} \ - \ 2 \, \eta^{\la \mu} \, \si^{\nu} \ + \ 2 \, \eta^{\nu \mu} \, \si^{\la}\,, \label{rv,15a} \\
\si^{\nu} \, \bar{\si}^{\la} \, \si^{\rho} \, \bar{\si}^{\tau} \, \si^{\mu} \ \ &= \ \ \si^{\mu} \, \bar{\si}^{\nu} \, \si^{\la} \, \si^{\rho} \, \bar{\si}^{\tau} \ - \ 2 \, \eta^{\mu \tau} \, \si^{\nu} \, \bar{\si}^{\la} \, \si^{\rho} \ + \ 2 \, \eta^{\mu \rho} \, \si^{\nu} \, \bar{\si}^{\la} \, \si^{\tau} \notag \\
& \ \ \ \ \ - \ 2 \, \eta^{\mu \la} \, \si^{\nu} \, \bar{\si}^{\rho} \, \si^{\tau} \ + \ 2 \, \eta^{\mu \nu} \, \si^{\la} \bar{\si}^{\rho} \, \si^{\tau} \,.
\label{rv,15}
\end{align}
\end{subequations}
These equations are easily generalized for arbitrary $n$:
\begin{align}
\si^{\la_{1}} \, \bar{\si}^{\la_{2}} \, ... \, &\si^{\la_{2n-1}} \, \bar{\si}^{\la_{2n}} \, \si^{\rho} \ \ = \ \ + \ \si^{\rho} \, \bar{\si}^{\la_{1}} \, \si^{\la_{2}} \, ... \, \bar{\si}^{\la_{2n-1}} \, \si^{\la_{2n}} \notag \\
& - \ 2  \, \sum_{k=1}^{n} \, \eta^{\rho \la_{2k}} \, \si^{\la_{1}} \, ... \, \si^{\la_{2k-1}} \, \bar{\si}^{\la_{2k+1}} \, ... \, \si^{\la_{2n}} \notag \\
& + \ 2 \, \sum_{k=1}^{n} \, \eta^{\rho \la_{2k-1}} \, \si^{\la_{1}} \, ... \, \bar{\si}^{\la_{2k-2}} \, \si^{\la_{2k}} \, ... \, \si^{\la_{2n}}\,,
\label{roeven}
\end{align}
\begin{align}
\si^{\la_{1}} \, \bar{\si}^{\la_{2}} \, ... \, &\bar{\si}^{\la_{2n-2}} \, \si^{\la_{2n-1}} \, \bar{\si}^{\rho} \ \ = \ \ - \ \si^{\rho} \, \bar{\si}^{\la_{1}} \, \si^{\la_{2}} \, ... \, \si^{\la_{2n-2}} \, \bar{\si}^{\la_{2n-1}} \notag \\
& - \ 2 \, \sum_{k=1}^{n} \, \eta^{\rho \la_{2k-1}} \, \si^{\la_{1}} \, ... \, \bar{\si}^{\la_{2k-2}} \, \si^{\la_{2k}} \, ... \, \bar{\si}^{\la_{2n-1}} \notag \\
& + \ 2 \, \sum_{k=1}^{n-1} \, \eta^{\rho \la_{2k}} \, \si^{\la_{1}} \, ... \, \si^{\la_{2k-1}} \, \bar{\si}^{\la_{2k+1}} \, ... \, \bar{\si}^{\la_{2n-1}} \,.
\label{roodd}
\end{align}
Let us now give relations, which convert ordered $\si$ chains into antisymmetrized products and vice
versa. Basically, these follow from repeated application of the ``anticommutation relation'':
\beq
\si^{\mu} \, \bar{\si}^{\nu} \ + \ \si^{\nu} \, \bar{\si}^{\mu} \eq - \, 2 \, \eta^{\mu \nu} \,.
\label{sig31}
\eeq
For two $\si$ matrices the conversion relation easily follows from the definition (\ref{as,0}) of the spinorial Lorentz generators and the Dirac algebra:
\beq
\si^{\mu} \, \bar{\si}^{\nu} \eq - \, \eta^{\mu \nu} \ + \ \si^{\mu \nu} \co \bar{\si}^{\mu} \, \si^{\nu} \eq - \, \eta^{\mu \nu} \ + \ \bar{\si}^{\mu \nu} \,.
\label{as,3}
\eeq
In the case of three $\si$'s their antisymmetric product can be written in terms of the Levi Civita tensor $\vep^{\mu \nu \la \rho}$ (with sign convention $\vep^{0123} = +1$):
\begin{align}
i\vep^{\mu \nu \la}\, \!_{\xi} \, \si^{\xi}_{ \al \dbe} \ \ &= \ \ (\si^{[\mu} \, \bar{\si}^{\nu} \, \si^{\la]})_{\al \dbe} \notag \\
&= \ \ (\si^{\mu} \, \bar{\si}^{\nu} \, \si^{\la})_{\al \dbe} \ + \ \eta^{\mu \nu} \, \si^{\la}_{\al \dbe} \ - \
\eta^{\mu \la} \, \si^{\nu}_{\al \dbe} \ + \ \eta^{ \nu \la} \, \si^{\mu}_{\al \dbe}\,.
\label{rv,8a}
\end{align}
The antisymmetrized product of four $\si$ matrices decomposes as:
\begin{align}
-i \vep^{\mu \nu \la \rho} \ \ &= \ \ \si^{[\mu} \, \bar{\si}^{\nu} \, \si^{\la} \, \bar{\si}^{\rho]} \notag \\
&= \ \ \si^{\mu} \, \bar{\si}^{\nu} \, \si^{\la} \, \bar{\si}^{\rho} \ + \ \eta^{\mu \nu} \, \si^{\la} \, \bar{\si}^{\rho} \ + \ \eta^{\la \rho} \, \si^{\mu} \, \bar{\si}^{\nu} \ - \ \eta^{\mu \la } \, \si^{\nu} \, \bar{\si}^{\rho} \ - \eta^{\nu \rho} \, \si^{\mu} \, \bar{\si}^{\la} \notag \\
& \ \ + \ \ \eta^{\mu \rho} \, \si^{\nu} \, \bar{\si}^{\la} \ + \ \eta^{\nu \la} \, \si^{\mu} \, \bar{\si}^{\rho} \ + \ \eta^{\mu \nu} \, \eta^{\la \rho} \ - \ \eta^{\mu \la} \, \eta^{\nu \rho} \ + \ \eta^{\mu \rho} \, \eta^{\nu \la} \,.
\label{si}
\end{align}
Finally, the antisymmetrized five $\si$ product vanishes identically in four space-time dimensions:
\begin{align}
0 \ \ &= \ \ \si^{[\mu} \, \bar{\si}^{\nu} \, \si^{\la} \, \bar{\si}^{\rho} \, \si^{\tau]} \notag \\
&= \ \ \si^{\mu} \, \bar{\si}^{\nu} \, \si^{\la} \, \bar{\si}^{\rho} \, \si^{\tau} \ + \ \eta^{\mu \nu} \, \si^{\la} \, \bar{\si}^{\rho} \, \si^{\tau} \ - \ \eta^{\mu \la} \, \si^{\nu} \, \bar{\si}^{\rho} \, \si^{\tau} \ + \ \eta^{\mu \rho} \, \si^{\nu} \, \bar{\si}^{\la} \, \si^{\tau} \notag \\
& \ \ \ \ \ \ - \ \eta^{\mu \tau} \, \si^{\nu} \, \bar{\si}^{\la} \, \si^{\rho} \ + \ \eta^{\nu \la} \, \si^{\mu} \, \bar{\si}^{\rho} \, \si^{\tau} \ - \ \eta^{\nu \rho} \, \si^{\mu} \, \bar{\si}^{\la} \, \si^{\tau} \ + \ \eta^{\nu \tau} \, \si^{\mu} \, \bar{\si}^{\la} \, \si^{\rho} \notag \\
& \ \ \ \ \ \ + \ \eta^{\la \rho} \, \si^{\mu} \, \bar{\si}^{\nu} \, \si^{\tau} \ - \ \eta^{\la \tau} \, \si^{\mu} \, \bar{\si}^{\nu} \, \si^{\rho} \ + \ \eta^{\rho \tau} \, \si^{\mu} \, \bar{\si}^{\nu} \, \si^{\la} \notag \\
& \ \ \ \ \ \ + \ \si^{\mu} \, \bigl( \eta^{\nu \la} \, \eta^{\rho \tau} \ - \ \eta^{\nu \rho} \, \eta^{\la \tau} \ + \ \eta^{\nu \tau} \, \eta^{ \la \rho } \bigr)  \ - \ \si^{\nu} \, \bigl( \eta^{\mu \la} \, \eta^{\rho \tau} \ - \ \eta^{\mu \rho} \, \eta^{\la \tau} \ + \ \eta^{\mu \tau} \, \eta^{ \la \rho } \bigr) \notag \\
& \ \ \ \ \ \ + \ \si^{\la} \, \bigl( \eta^{\mu \nu} \, \eta^{\rho \tau} \ - \ \eta^{\mu \rho} \, \eta^{\nu \tau} \ + \ \eta^{\mu \tau} \, \eta^{ \nu \rho } \bigr) \  - \ \si^{\rho} \, \bigl( \eta^{\mu \nu} \, \eta^{\la \tau} \ - \ \eta^{\mu \la} \, \eta^{\nu \tau} \ + \ \eta^{\mu \tau} \, \eta^{ \nu \la } \bigr) \notag \\
& \ \ \ \ \ \ + \ \si^{\tau} \, \bigl( \eta^{\mu \nu} \, \eta^{\la \rho} \ - \ \eta^{\mu \la} \, \eta^{\nu \rho} \ + \ \eta^{\mu \rho} \, \eta^{ \nu \la} \bigr)\,.
\label{rv,18}
\end{align}
It is easy to invert the previous relations (\ref{rv,8a}), (\ref{si}) and (\ref{rv,18}), which can be used to decompose
an ordered product of $\si$'s into an antisymmetrized basis. The first equation simply needs to be solved for $\si^{\mu}
\bar{\si}^{\nu} \si^{\la}$:
\beq
(\si^{\mu} \, \bar{\si}^{\nu} \, \si^{\la})_{\al \dbe} \eq i\vep^{\mu \nu \la}\,
\!_{\xi} \, \si^{\xi}_{ \al \dbe} \ - \ \eta^{\mu \nu} \, \si^{\la}_{\al \dbe} \ + \ \eta^{\mu \la} \, \si^{\nu}_{\al
  \dbe} \ - \ \eta^{ \nu \la} \, \si^{\mu}_{\al \dbe} \,.
\eeq
The latter yield the following expressions:
\begin{align}
\si^{\mu} \, \bar{\si}^{\nu} \, \si^{\la} \, \bar{\si}^{\rho} \ \ &= \ \ - \, i \vep^{\mu \nu \la \rho}  \ - \ \eta^{\mu \nu} \, \si^{\la \rho} \ - \ \eta^{\la \rho} \, \si^{\mu \nu} \ + \ \eta^{\mu \la} \, \si^{\nu \rho} \ + \ \eta^{\nu \rho} \, \si^{\mu \la} \notag \\
& \hspace{.67cm}  - \eta^{\mu \rho} \, \si^{\nu \la} \ - \ \eta^{\nu \la} \, \si^{\mu \rho} \ + \ \eta^{\mu \nu} \, \eta^{\la \rho} \ - \ \eta^{\mu \la} \, \eta^{\nu \rho} \ + \ \eta^{\mu \rho} \, \eta^{\nu \la}\,,
\label{rv,11} \\
\si^{\mu} \, \bar{\si}^{\nu} \, \si^{\la} \, \bar{\si}^{\rho} \, \si^{\tau} \ \ &= \ \ - \, \eta^{\mu \nu} \, i\vep^{\la \rho \tau}\, \!_{\xi} \, \si^{\xi} \ + \ \eta^{\mu \la} \, i\vep^{ \nu \rho \tau }\, \!_{\xi} \, \si^{\xi} \ - \ \eta^{\mu \rho} \, i\vep^{\nu \la \tau}\, \!_{\xi} \, \si^{\xi} \ + \ \eta^{\mu \tau} \, i\vep^{\nu \la \rho}\, \!_{\xi} \, \si^{\xi} \notag \\
&  \hspace{.67cm} - \eta^{\nu \la} \, i\vep^{\mu \rho \tau}\, \!_{\xi} \, \si^{\xi} \  + \ \eta^{\nu \rho} \, i\vep^{\mu \la \tau}\, \!_{\xi} \, \si^{\xi} \ - \ \eta^{\nu \tau} \, i\vep^{\mu \la \rho}\, \!_{\xi} \, \si^{\xi} \ - \ \eta^{\la \rho} \, i\vep^{\mu \nu \tau}\, \!_{\xi} \, \si^{\xi}  \notag \\
&  \hspace{.67cm} + \eta^{\la \tau} \, i\vep^{\mu \nu \rho}\, \!_{\xi} \, \si^{\xi} \ - \ \eta^{\rho \tau} \, i\vep^{\mu \nu \la}\, \!_{\xi} \, \si^{\xi} \ + \ \si^{\mu} \, \bigl( \eta^{\nu \la} \, \eta^{\rho \tau} \ - \ \eta^{\nu \rho} \, \eta^{\la \tau} \ + \ \eta^{\nu \tau} \, \eta^{ \la \rho } \bigr)  \notag \\
& \hspace{.67cm} - \si^{\nu} \, \bigl( \eta^{\mu \la} \, \eta^{\rho \tau} \ - \ \eta^{\mu \rho} \, \eta^{\la \tau} \ + \ \eta^{\mu \tau} \, \eta^{ \la \rho } \bigr) \ + \ \si^{\la} \, \bigl( \eta^{\mu \nu} \, \eta^{\rho \tau} \ - \ \eta^{\mu \rho} \, \eta^{\nu \tau} \ + \ \eta^{\mu \tau} \, \eta^{ \nu \rho } \bigr) \notag \\
& \hspace{.67cm}- \si^{\rho} \, \bigl( \eta^{\mu \nu} \, \eta^{\la \tau} \ - \ \eta^{\mu \la} \, \eta^{\nu \tau} \ + \ \eta^{\mu \tau} \, \eta^{ \nu \la } \bigr) \ + \ \si^{\tau} \, \bigl( \eta^{\mu \nu} \, \eta^{\la \rho} \ - \ \eta^{\mu \la} \, \eta^{\nu \rho} \ + \ \eta^{\mu \rho} \, \eta^{ \nu \la} \bigr)\,.
\label{rv,18inv}
\end{align}

\goodbreak
\section{The $\bm{n}$--point correlators with two spin fields: proof by induction}
\label{sec:TheProofByInduction}

The proof of equations (\ref{npt,1}), (\ref{npt,2}) and (\ref{npt,3}) is carried out by induction in this Section. They hold in the simple cases $n=2,3$ because these lower order examples were the starting point for guessing the general formulas. The induction step is much more involved: we show that $\Om_{(n)}$, $\om_{(n)}$ and $\bar{\om}_{(n)}$ have the correct behaviour for $z_{i} \mto z_{j}$ on basis of the induction hypothesis that the general formulae hold for $n-1$.

\subsection[An auxiliary correlator: $2n$ NS fields]{An auxiliary correlator: $\bm{2n}$ NS fields}
\label{sec:ThePureNSCorrelator}

Due to the OPEs (\ref{rv,1b}), (\ref{rv,1d}), (\ref{rv,1e}) all spin fields vanish from the correlators in the limit $z_{A} \mto z_{B}$. Both $\Om_{(n)}$ and $\om_{(n+1)}$ leave a $2n$ point function with NS fields only,
\beq
\Psi_{(n)}^{\mu_{1} ... \mu_{2n}} (z_{1},...,z_{2n}) \ \ := \ \ \langle \psi^{\mu_{1}}(z_{1}) \, \psi^{\mu_{2}}(z_{2}) \, ... \, \psi^{\mu_{2n}}(z_{2n}) \rangle \ ,
\label{npt,9}
\eeq
which is important to know for general $n$ in the following. For $n=1,2$ we have:
\begin{subequations}
\begin{align}
\Psi_{(1)}^{\mu_{1} \mu_{2}}(z_{1},z_{2}) \ \ &= \ \ \langle \psi^{\mu_{1}}(z_{1}) \, \psi^{\mu_{2}}(z_{2}) \rangle \eq \frac{\eta^{\mu_{1} \mu_{2}}}{z_{12}}\,, \label{npt,10a} \\
\Psi_{(2)}^{\mu_{1} \mu_{2} \mu_{3} \mu_{4}}(z_{1},z_{2},z_{3},z_{4}) \ \ &= \ \ \langle \psi^{\mu_{1}}(z_{1}) \, \psi^{\mu_{2}}(z_{2}) \, \psi^{\mu_{3}}(z_{3}) \, \psi^{\mu_{4}}(z_{4}) \rangle \notag \\
&= \ \ \frac{\eta^{\mu_{1} \mu_{2}} \, \eta^{\mu_{3} \mu_{4}}}{z_{12} \, z_{34}} \ - \ \frac{\eta^{\mu_{1} \mu_{3}} \, \eta^{\mu_{2} \mu_{4}}}{z_{13} \, z_{24}} \ + \ \frac{\eta^{\mu_{1} \mu_{4}} \, \eta^{\mu_{2} \mu_{3}}}{z_{14} \, z_{23}} \ .
\label{npt,10}
\end{align}
\end{subequations}
In the absence of spin fields $\psi^\mu$ is a free field. Hence, we can use Wick's theorem to reduce any higher order
correlator $\Psi_{(n)}$ to propagators $\frac{\eta^{\mu_{i} \mu_{j}}}{z_{ij}}$, just like the four point function
(\ref{npt,10}).

The decomposition of the $(2n)$ point function into pairwise contractions to $\frac{\eta^{\mu_{i} \mu_{j}}}{z_{ij}}$'s
can be identified with the $\ell = 0$ term (i.e. the $\si$ free piece) of the corresponding $\om_{(n+1)}$ correlator, so
we can express $\Psi_{(n)}$ in the notation of Section \ref{sec:TheNPointGeneralization} as:
\beq
\Psi_{(n)}^{\mu_{1} ... \mu_{2n}} (z_{1},...,z_{2n}) \eq \! \! \! \sum_{\rho \in S_{2n} / {\cal Q}_{n+1,0}} \! \! \te{sgn}(\rho) \prod_{j=1}^{n} \frac{\eta^{\mu_{\rho(2j-1)} \mu_{\rho(2j)}}}{z_{\rho(2j-1)} \, - \, z_{\rho(2j)}} \ .
\label{npt,11}
\eeq

\subsection{The web of limits}
\label{sec:TheWebOfLimits}

We have just given a closed formula for $\Psi_{(n)}$ in terms of $S_{2n} / {\cal Q}_{n+1,0}$ which is relevant for the
limits $z_{A}\mto z_{B}$ of both $\Om_{(n)}$ and $\om_{(n)}, \bar{\om}_{(n)}$. The asymptotic behaviour of the
correlators for $z_{i} \mto z_{j}$ does not only relate $\Om_{(n)} \leftrightarrow \Om_{(n-1)}$ and $\om_{(n)}
\leftrightarrow \om_{(n-1)}$ but also yields connections between correlators of different type such as $\Om_{(n)}
\leftrightarrow \om_{(n)},\bar{\om}_{(n)}$ and $\om_{(n)},\bar{\om}_{(n)} \leftrightarrow \Om_{(n-1)}$. Due to
the OPEs \eqref{rv,2a}, \eqref{rv,2b} the limits $z_{2n-1} \mto z_{A}$ or $z_{1} \mto z_{B}$ intertwine the $\Om_{(n)}$ sequence
with the $\om_{(n)}$ and $\bar{\om}_{(n)}$. So one cannot prove one of the equations (\ref{npt,1}),
(\ref{npt,2}), (\ref{npt,3}) separately.

\begin{figure}[H]
    \centering
\includegraphics[width=0.35\textwidth]{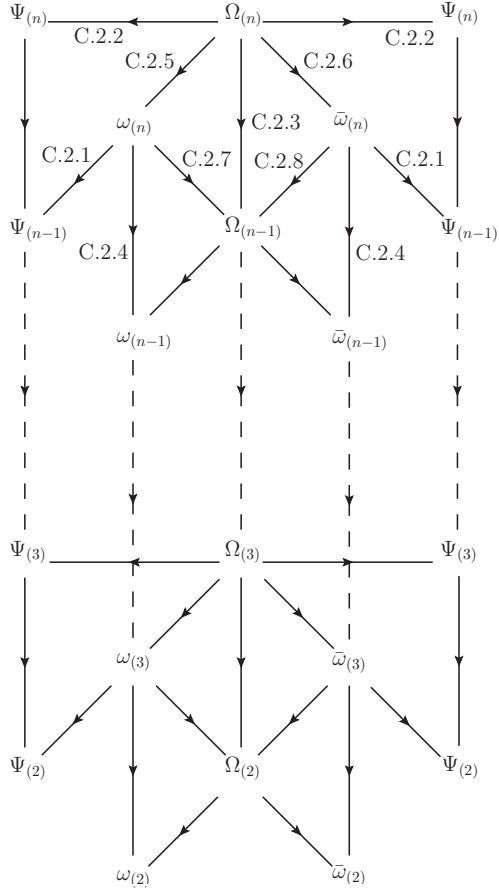}
\caption{Steps necessary for completing the proof by induction}
\label{ind}
\end{figure}

There is a web of limiting processes which we have to examine for a complete proof by induction. Figure \ref{ind} summarizes its structure.
\noindent
The following Subsections will verify the consistency of our central result (\ref{npt,1}) as well as (\ref{npt,2}),
(\ref{npt,3}) with every single relation between various $\Om_{(n)}$, $\om_{(n)}$, $\bar{\om}_{(n)}$, $\Psi_{(n)}$ due
to the OPEs given in Section \ref{sec:ReviewOfLowerOrderCorrelators}.

\subsubsection{$\om_{(n)}, \bar{\om}_{(n)}\ra\Psi_{(n-1)}$ by 
$z_{A} \mto z_{B}$ limit}
\label{sec:i}

Using the OPE \eqref{rv,1d} one finds the following $z_{A} \mto z_{B}$ behaviour for the $2n$ point correlators $\om_{(n)}$:
\begin{align}
\om_{(n)}^{\mu_{1} ... \mu_{2n-2}}\, _{\al} \, ^{\be} (z_{i}) \Bigl. \Bigr| \begin{smallmatrix} z_{A} \\ \downarrow \\ z_{B} \end{smallmatrix} \ \ &\sim \ \ \frac{- \, \de_{\al} ^{\be}}{(z_{A} \, - \, z_{B})^{1/2}} \; \langle \psi^{\mu_{1}}(z_{1}) \, ... \, \psi^{\mu_{2n-2}} (z_{2n-2}) \rangle \eq \frac{\de_{\al} ^{\be}}{z_{AB}^{1/2}} \; \Psi^{\mu_{1}...\mu_{2n-2}}_{(n-1)} (z_{i}) \notag \\
&= \ \ \frac{- \, \de_{\al} ^{\be}}{z_{AB}^{1/2}} \! \sum_{\rho \in S_{2n-2} / {\cal Q}_{n,0}} \! \! \te{sgn}(\rho) \prod_{j=1}^{n-1} \frac{\eta^{\mu_{\rho(2j-1)} \mu_{\rho(2j)}}}{z_{\rho(2j-1)} \, - \, z_{\rho(2j)}}\,.
\label{npt,14}
\end{align}
If we start from the expression (\ref{npt,2}) and isolate the highest $z_{AB}^{-1}$ powers, the sum over $\ell$ breaks down to the $\ell=0$ term where $\bigl(\si^{\mu_{\rho(1)}} ... \bar{\si}^{\mu_{\rho(2\ell)}} \bigr)_{\al} \, ^{\be} \bigl. \bigr|_{\ell =0} = \de^{\be}_{\al}$:
\begin{align}
\om_{(n)}^{\mu_{1} ... \mu_{2n-2}}&\,_{\al}\,^{\be} (z_{i}) \Bigl. \Bigr| \begin{smallmatrix} z_{A} \\ \downarrow \\ z_{B} \end{smallmatrix} \ \ \sim \ \ \frac{- \, 1}{z_{AB}^{1/2}} \; \prod_{i=1}^{2n-2} (z_{iA} \, z_{iB})^{-1/2} \, \sum_{\ell = 0}^{n-1} \, \biggl( \frac{z_{AB}}{2} \biggr)^{\ell} \, \de_{\ell,0} \! \! \! \! \sum_{\rho \in S_{2n-2}/{\cal Q}_{n,\ell}} \! \! \!  \te{sgn}(\rho) \notag \\
& \ \ \ \ \  \ \ \times \, \bigl(\si^{\mu_{\rho(1)}} \, \bar{\si}^{\mu_{\rho(2)}} \, ... \, \si^{\mu_{\rho(2\ell-1)}} \, \bar{\si}^{\mu_{\rho(2\ell)}} \bigr)_{\al} \, ^{\be} \, \prod_{j=1}^{n-\ell-1} \frac{\eta^{\mu_{\rho(2\ell+2j-1)} \mu_{\rho(2\ell+2j)}}}{z_{\rho(2\ell+2j-1),\rho(2\ell+2j)} } \; z_{\rho(2\ell +2j-1),A} \, z_{\rho(2\ell +2j),B} \notag \\
&= \ \ \frac{- \, \de_{\al} ^{\be}}{z_{AB}^{1/2}}  \! \sum_{\rho \in S_{2n-2}/{\cal Q}_{n,0}} \! \! \!  \te{sgn}(\rho) \, \prod_{j=1}^{n-1} \frac{\eta^{\mu_{\rho(2j-1)} \mu_{\rho(2j)}}}{z_{\rho(2j-1) ,\rho(2j)}}  \; \underbrace{\frac{z_{\rho(2j-1),A} \, z_{\rho(2j),B} }{\bigl( z_{\rho(2j-1),A} \, z_{\rho(2j-1),B} \, z_{\rho(2j),A} \, z_{\rho(2j),B} \bigr)^{1/2} }}_{\rightarrow 1 \ \te{as} \ z_{A} \mto z_{B}} \notag \\
&= \ \ \frac{- \, \de_{\al} ^{\be}}{z_{AB}^{1/2}} \! \sum_{\rho \in S_{2n-2} / {\cal Q}_{n,0}} \! \! \te{sgn}(\rho) \prod_{j=1}^{n} \frac{\eta^{\mu_{\rho(2j-1)} \mu_{\rho(2j)}}}{z_{\rho(2j-1)} \, - \, z_{\rho(2j)}} \ .
\label{npt,15}
\end{align}
Going to the third line we have rearranged the $\prod_{i=1}^{2n-2}(z_{iA}  z_{iB})^{-1/2}$ product such that the cancellation with the $z_{\rho(2j-1),A}  z_{\rho(2j),B}$ numerators in the $z_{A} \mto z_{B}$ limit becomes obvious. 

The argument for $\bar{\om}_{(n)}$ is completely analogous. Due to the OPE \eqref{rv,1e}, one arrives at:
\beq
\bar{\om}_{(n)}^{\mu_{1} ... \mu_{2n-2} \dal} \, _{\dbe} (z_{i}) \Bigl. \Bigr| \begin{smallmatrix} z_{A} \\ \downarrow \\ z_{B} \end{smallmatrix} \ \ \sim \ \ \frac{\de^{\dal} _{\dbe}}{z_{AB}^{1/2}} \! \sum_{\rho \in S_{2n-2} / {\cal Q}_{n,0}} \! \! \te{sgn}(\rho) \prod_{j=1}^{n} \frac{\eta^{\mu_{\rho(2j-1)} \mu_{\rho(2j)}}}{z_{\rho(2j-1)} \, - \, z_{\rho(2j)}} \ .
\label{npt,16}
\eeq 
This can also be obtained from (\ref{npt,3}) by virtue of the same truncation to the $\ell =0$ term and the identity
$\prod_{j=1}^{n-1} \frac{z_{\rho(2j-1),A} z_{\rho(2j),B} }{( z_{\rho(2j-1),A} z_{\rho(2j-1),B} z_{\rho(2j),A}
  z_{\rho(2j),B} )^{1/2} } \mto 1$ used above. The agreement (\ref{npt,14}) $\leftrightarrow$ (\ref{npt,15}) in two ways
of evaluating the $z_{A} \mto z_{B}$ asymptotics, which extends to $\bar{\om}_{(n)}$ by (\ref{npt,16}), allows to
proceed to the next arrow in figure \ref{ind}.

\subsubsection{$\Om_{(n)} \ra \Psi_{(n)}$ by $z_{A} \mto z_{B}$ limit}
\label{sec:ii}

The relevant OPE for this case is \eqref{rv,1b}. Let us plug this into $\Om_{(n)}$ for $z_{A} \mto z_{B}$:
\begin{align}
\Om_{(n)}&^{\mu_{1} ... \mu_{2n-1}}\,  _{\al \dbe} (z_{i}) \Bigl. \Bigr| \begin{smallmatrix} z_{A} \\ \downarrow \\ z_{B} \end{smallmatrix} \ \ \sim \ \ \frac{1}{\sqrt{2}} \; (\si_{\mu_{2n}})_{ \al \dbe} \, \langle \psi^{\mu_{1}}(z_{1}) \, ... \, \psi^{\mu_{2n-1}} (z_{2n-1}) \, \psi^{\mu_{2n}}(z_{B}) \rangle \notag \\
&= \ \ \frac{(\si_{\mu_{2n}})_{ \al \dbe}}{\sqrt{2}} \;  \Psi^{\mu_{1}...\mu_{2n}}_{(n)} (z_{i}) \Bigl. \Bigr|_{z_{2n}=z_{B}} \eq \frac{(\si_{\mu_{2n}})_{ \al \dbe}}{\sqrt{2}} \;  \! \sum_{\rho \in S_{2n} / {\cal Q}_{n+1,0}} \! \! \te{sgn}(\rho) \prod_{j=1}^{n} \frac{\eta^{\mu_{\rho(2j-1)} \mu_{\rho(2j)}}}{z_{\rho(2j-1)} \, - \, z_{\rho(2j)}} \Bigl. \Bigr|_{z_{2n}=z_{B}} \notag \\
&= \ \ \frac{(\si_{\mu_{2n}})_{ \al \dbe}}{\sqrt{2}} \;  \! \sum_{\rho \in S_{2n} / {\cal Q}_{n+1,0}} \! \! \te{sgn}(\rho) \; \frac{\eta^{\mu_{\rho(2j_{0}-1)} \mu_{2n}}}{z_{\rho(2j_{0}-1)} \, - \, z_{B}} \; \prod_{j=1 \atop {j\neq j_{0} \atop{ \rho(2j_{0}) = 2n}}}^{n}  \frac{\eta^{\mu_{\rho(2j-1)} \mu_{\rho(2j)}}}{z_{\rho(2j-1)} \, - \, z_{\rho(2j)}} \notag \\
&= \ \ \frac{1}{\sqrt{2}} \!  \sum_{\rho \in S_{2n} / {\cal Q}_{n+1,0}} \! \! \te{sgn}(\rho) \; \frac{\si^{\mu_{\rho(2j_{0}-1)}}_{\al \dbe}}{z_{\rho(2j_{0}-1)} \, - \, z_{B}} \; \prod_{j=1 \atop {j\neq j_{0} \atop{ \rho(2j_{0}) = 2n}}}^{n}  \frac{\eta^{\mu_{\rho(2j-1)} \mu_{\rho(2j)}}}{z_{\rho(2j-1)} \, - \, z_{\rho(2j)}} \notag \\
&= \ \ \frac{1}{\sqrt{2}} \!  \sum_{\bar{\rho} \in S_{2n-1} / {\cal P}_{n,0}} \! \! \te{sgn}(\bar{\rho}) \; \frac{\si^{\mu_{\bar{\rho}(1)}}_{\al \dbe}}{z_{\bar{\rho}(1)} \, - \, z_{B}} \; \prod_{j=1 }^{n-1}  \frac{\eta^{\mu_{\bar{\rho}(2j)} \mu_{\bar{\rho}(2j+1)}}}{z_{\bar{\rho}(2j)} \, - \, z_{\bar{\rho}(2j+1)}}\,.
\label{npt,17}
\end{align}
Since the index $\mu_{2n}$ is contracted from the third to the fourth line, one can sum over $S_{2n-1}$ subpermutations $\bar{\rho}$ acting on $(1,2,...,2n-1)$ instead of $\rho \in S_{2n}$. The number of terms is the same in both sums,
\begin{align}
\bigl| S_{2n} / {\cal Q}_{n+1,0} \bigr| \ \ = \ \ \frac{(2n)!}{n! \, 2^{n}} \ \ &= \ \ \frac{2n \, (2n \, - \, 1)!}{n \, (n \, - \, 1)! \,2 \  2^{n-1}} \notag \\
= \ \ \frac{(2n \, - \, 1)!}{(n \, - \, 1)! \, 2^{n-1}} \ \ &= \ \ \bigl| S_{2n-1} / {\cal P}_{n,0} \bigr| \ ,
\label{npt,18}
\end{align}
and one can check with the help of restrictions (\ref{npt,4}), (\ref{npt,5}) that the last equality of (\ref{npt,17}) holds exactly.

Can we reproduce (\ref{npt,17}) directly from $\Om_{(n)}$ as claimed in (\ref{npt,1}) by keeping the most singular $z_{AB}$ dependences only? Similar to (\ref{npt,15}), the $\ell$ sum will reduce to the $\ell =0$ contribution:
\begin{align}
\Om_{(n)}&^{\mu_{1} ... \mu_{2n-1}}\,  _{\al \dbe} (z_{i}) \Bigl. \Bigr| \begin{smallmatrix} z_{A} \\ \downarrow \\ z_{B} \end{smallmatrix} \ \ \sim \ \ \frac{1}{\sqrt{2}} \prod_{i=1}^{2n-1} (z_{iA} \, z_{iB})^{-1/2} \, \sum_{\ell = 0}^{n-1} \, \biggl( \frac{z_{AB}}{2} \biggr)^{\ell} \, \de_{\ell,0} \! \! \! \! \sum_{\rho \in S_{2n-1}/{\cal P}_{n,\ell}} \! \! \!  \te{sgn}(\rho) \notag \\
& \ \ \ \ \  \ \ \times \ \bigl(\si^{\mu_{\rho(1)}} \, \bar{\si}^{\mu_{\rho(2)}} \, ... \, \bar{\si}^{\mu_{\rho(2\ell)}} \, \si^{\mu_{\rho(2\ell+1)}} \bigr)_{\al \dbe} \prod_{j=1}^{n-\ell-1} \frac{\eta^{\mu_{\rho(2\ell+2j)} \mu_{\rho(2\ell+2j+1)}}}{z_{\rho(2\ell+2j),\rho(2\ell+2j+1)} } \;  z_{\rho(2\ell+2j),A} \,  z_{\rho(2\ell+2j+1),B} \notag \\
&= \ \ \frac{1}{\sqrt{2}} \! \sum_{\rho \in S_{2n-1}/{\cal P}_{n,0}} \! \! \!  \te{sgn}(\rho) \; \frac{1}{ z_{\rho(1),A}^{1/2} \, z_{\rho(1),B}^{1/2}} \; \underbrace{\left( \frac{z_{\rho(1),A}}{z_{\rho(1),B}} \right)^{1/2}}_{\rightarrow 1 \ \te{as} \ z_{A} \mto z_{B}} \, \si^{\mu_{\rho(1)}}_{\al \dbe} \notag \\
& \ \ \ \ \  \ \ \times \ \prod_{j=1}^{n-1} \frac{\eta^{\mu_{\rho(2j)} \mu_{\rho(2j+1)}}}{z_{\rho(2j) ,\rho(2j+1)}}  \; \underbrace{\frac{z_{\rho(2j),A} \, z_{\rho(2j+1),B} }{\bigl( z_{\rho(2j),A} \, z_{\rho(2j),B} \, z_{\rho(2j+1),A} \, z_{\rho(2j+1),B} \bigr)^{1/2} }}_{\rightarrow 1 \ \te{as} \ z_{A} \mto z_{B}} \notag \\
&= \ \ \frac{1}{\sqrt{2}} \! \sum_{\rho \in S_{2n-1}/{\cal P}_{n,0}} \! \! \!  \te{sgn}(\rho) \; \frac{\si^{\mu_{\rho(1)}}_{\al \dbe}}{ z_{\rho(1)} \, - \, z_{B}} \; \prod_{j=1}^{n-1} \frac{\eta^{\mu_{\rho(2j)} \mu_{\rho(2j+1)}}}{z_{\rho(2j)} \, - \, z_{\rho(2j+1)}}\,.
\label{npt,19}
\end{align}
Except for the insertion of $\left( \frac{z_{\rho(1),A}}{z_{\rho(1),B}} \right)^{1/2} \mto 1$, the mechanisms are the same as in (\ref{npt,15}).

\subsubsection{$\Om_{(n)} \ra \Om_{(n-1)}$ by $z_{2n-2} \mto z_{2n-1}$ limit}
\label{sec:iii}

This subsection treats the situation when the arguments of two NS field in $\Om_{(n)}$ approach each other. We are left with the correlator $\Om_{(n-1)}$ because of the OPE~\eqref{rv,1a}: 
\begin{align}
\Om_{(n)}&^{\mu_{1} ... \mu_{2n-1}}\,  _{\al \dbe} (z_{i}) \Bigl. \Bigr| \begin{smallmatrix} z_{2n-2} \\ \downarrow \\ z_{2n-1} \end{smallmatrix} \ \ \sim \ \ \frac{\eta^{\mu_{2n-2} \mu_{2n-1}}}{z_{2n-2} \, - \, z_{2n-1}} \; \langle \psi^{\mu_{1}}(z_{1}) \, ... \, \psi^{\mu_{2n-3}}(z_{2n-3}) \, S_{\al}(z_{A}) \, S_{\dbe}(z_{B}) \rangle \notag \\
&= \ \ \frac{\eta^{\mu_{2n-2} \mu_{2n-1}}}{z_{2n-2} \, - \, z_{2n-1}} \; \Om_{(n-1)}^{\mu_{1} ... \mu_{2n-3}}\,  _{\al \dbe} (z_{i}) \notag \\
&= \ \ \frac{\eta^{\mu_{2n-2} \mu_{2n-1}}}{z_{2n-2,2n-1}} \; \frac{1}{\sqrt{2}} \prod_{i=1}^{2n-3} (z_{iA} \, z_{iB})^{-1/2} \, \sum_{\ell = 0}^{n-2} \, \biggl( \frac{z_{AB}}{2} \biggr)^{\ell} \! \! \! \! \sum_{\rho \in S_{2n-3}/{\cal P}_{n-1,\ell}} \! \! \!  \te{sgn}(\rho) \, \bigl(\si^{\mu_{\rho(1)}} \, ... \,  \si^{\mu_{\rho(2\ell+1)}} \bigr)_{\al \dbe} \notag \\
& \ \ \ \ \  \ \ \times \ \prod_{j=1}^{n-\ell-2} \frac{\eta^{\mu_{\rho(2\ell+2j)} \mu_{\rho(2\ell+2j+1)}}}{z_{\rho(2\ell+2j),\rho(2\ell+2j+1)} } \;  z_{\rho(2\ell+2j),A}  \,  z_{\rho(2\ell+2j+1),B} \,.
\label{npt,20}
\end{align}
To bring the asymptotic behaviour of expression (\ref{npt,1}) for $z_{2n-2} \mto z_{2n-1}$ into the form (\ref{npt,20})
one has to isolate the permutations $S_{2n-1} / {\cal P}_{n,\ell}$ which provide a factor $\frac{\eta^{\mu_{2n-2}
    \mu_{2n-1}}}{z_{2n-2} -z_{2n-1}}$. A necessary condition for this to occur is $\ell \neq n-1$ because otherwise
there would be no $\eta$'s at all. So we include $(1-\de_{\ell,n-1})$.

Furthermore, the ordering of the $\eta$'s according to their second index (which is rephrased as $\rho(2\ell+3) <
\rho(2\ell+5) < ... < \rho(2n-1)$ in (\ref{npt,4})) makes sure that $\rho(2n-1) = 2n-1$, and then $\eta^{\mu_{2n-2} \mu_{2n-1}}$ appears whenever $\rho(2n-2) = 2n-2$. These arguments lead to
\begin{align}
\Om_{(n)}&^{\mu_{1} ... \mu_{2n-1}}\,  _{\al \dbe} (z_{i}) \Bigl. \Bigr| \begin{smallmatrix} z_{2n-2} \\ \downarrow \\ z_{2n-1} \end{smallmatrix} \ \ \sim \ \ \frac{1}{\sqrt{2}} \prod_{i=1}^{2n-1} (z_{iA} \, z_{iB})^{-1/2} \, \sum_{\ell = 0}^{n-1} \, \biggl( \frac{z_{AB}}{2} \biggr)^{\ell} \, \bigl(1 \, - \, \de_{\ell,n-1} \bigr) \! \! \! \!  \sum_{\rho \in S_{2n-1}/{\cal P}_{n,\ell}} \! \! \!  \te{sgn}(\rho) \notag \\
& \ \ \ \ \  \ \ \times \ \de_{\rho(2n-2),2n-2} \, \bigl(\si^{\mu_{\rho(1)}} \, ... \, \si^{\mu_{\rho(2\ell+1)}} \bigr)_{\al \dbe}  \, \prod_{j=1}^{n-\ell-1} \frac{\eta^{\mu_{\rho(2\ell+2j)} \mu_{\rho(2\ell+2j+1)}}}{z_{\rho(2\ell+2j),\rho(2\ell+2j+1)} } \; z_{\rho(2\ell+2j),A} \, z_{\rho(2\ell+2j+1),B} \notag \\
&= \ \ \frac{1}{\sqrt{2}} \prod_{i=1}^{2n-3} (z_{iA} \, z_{iB})^{-1/2} \, \sum_{\ell = 0}^{n-2} \, \biggl( \frac{z_{AB}}{2} \biggr)^{\ell} \! \! \! \!  \sum_{\rho \in S_{2n-1}/{\cal P}_{n,\ell}} \! \! \!  \te{sgn}(\rho) \, \de_{\rho(2n-2),2n-2}  \notag \\
& \ \ \ \ \  \ \ \times \ \bigl(\si^{\mu_{\rho(1)}} \, ... \, \si^{\mu_{\rho(2\ell+1)}} \bigr)_{\al \dbe} \; \frac{\eta^{\mu_{2n-2} \mu_{2n-1}}}{z_{2n-2,2n-1}} \; \underbrace{\frac{z_{2n-2,A} \, z_{2n-1,B}}{\bigl(z_{2n-2,A} \, z_{2n-2,B} \, z_{2n-1,A} \, z_{2n-1,B} \bigr)^{1/2} }}_{\mto 1 \ \te{as} \ z_{2n-2} \mto z_{2n-1}} \notag \\
& \ \ \ \ \  \ \ \times \ \prod_{j=1}^{n-\ell-2} \frac{\eta^{\mu_{\rho(2\ell+2j)} \mu_{\rho(2\ell+2j+1)}}}{z_{\rho(2\ell+2j),\rho(2\ell+2j+1)} } \; z_{\rho(2\ell+2j),A} \, z_{\rho(2\ell+2j+1),B} \notag \\
&= \ \ \frac{1}{\sqrt{2}} \; \frac{\eta^{\mu_{2n-2} \mu_{2n-1}}}{z_{2n-2,2n-1}} \prod_{i=1}^{2n-3} (z_{iA} \, z_{iB})^{-1/2} \, \sum_{\ell = 0}^{n-2} \, \biggl( \frac{z_{AB}}{2} \biggr)^{\ell} \! \! \! \!  \sum_{\bar{\rho} \in S_{2n-3}/{\cal P}_{n-1,\ell}} \! \! \!  \te{sgn}(\bar{\rho}) \notag \\
& \ \ \ \ \ \ \ \times \ \bigl(\si^{\mu_{\bar{\rho}(1)}} \, ... \, \si^{\mu_{\bar{\rho}(2\ell+1)}} \bigr)_{\al \dbe}  \, \prod_{j=1}^{n-\ell-2} \frac{\eta^{\mu_{\bar{\rho}(2\ell+2j)} \mu_{\bar{\rho}(2\ell+2j+1)}}}{z_{\bar{\rho}(2\ell+2j),\bar{\rho}(2\ell+2j+1)} } \; z_{\bar{\rho}(2\ell+2j),A} \, z_{\bar{\rho}(2\ell+2j+1),B} \ .
\label{npt,21}
\end{align}
In the last step we have used that $S_{2n-1}/{\cal P}_{n,\ell}$ with $\left( \begin{smallmatrix} \rho(2n-1) \\
    \rho(2n-2) \end{smallmatrix} \right) = \left( \begin{smallmatrix} 2n-1 \\ 2n-2 \end{smallmatrix} \right)$ is equivalent to $S_{2n-3}/{\cal P}_{n-1,\ell}$, where the subpermutation $\bar{\rho} \in S_{2n-3}$ is of the same sign as the corresponding $\rho \in S_{2n-1}$.

\subsubsection{$\om_{(n)}, \bar{\om}_{(n)} \ra \om_{(n-1)}, \bar{\om}_{(n-1)}$ by $z_{2n-3} \mto z_{2n-2}$ limit}
\label{sec:iv}

The singular behaviour of the $\om_{(n)}, \bar{\om}_{(n)}$ in two NS arguments can be studied in a similar fashion:
\begin{align}
\om_{(n)}&^{\mu_{1} ... \mu_{2n-2}}\,  _{\al} \, ^{ \be} (z_{i}) \Bigl. \Bigr| \begin{smallmatrix} z_{2n-3} \\ \downarrow \\ z_{2n-2} \end{smallmatrix} \ \ \sim \ \ \frac{\eta^{\mu_{2n-3} \mu_{2n-2}}}{z_{2n-3} \, - \, z_{2n-2}} \; \langle \psi^{\mu_{1}}(z_{1}) \, ... \, \psi^{\mu_{2n-4}}(z_{2n-4}) \, S_{\al}(z_{A}) \, S^{\be}(z_{B}) \rangle \notag \\
&= \ \ \frac{\eta^{\mu_{2n-3} \mu_{2n-2}}}{z_{2n-3} \, - \, z_{2n-2}} \; \om_{(n-1)}^{\mu_{1} ... \mu_{2n-4}}\, _{\al} \, ^{ \be} (z_{i}) \notag \\
&= \ \ - \, \frac{\eta^{\mu_{2n-3} \mu_{2n-2}}}{z_{2n-3,2n-2}} \; \left( \frac{1}{z_{AB}^{1/2}} \right) \prod_{i=1}^{2n-4} (z_{iA} \, z_{iB})^{-1/2} \, \sum_{\ell = 0}^{n-2} \, \biggl( \frac{z_{AB}}{2} \biggr)^{\ell} \! \! \! \! \sum_{\rho \in S_{2n-4}/{\cal Q}_{n-1,\ell}} \! \! \!  \te{sgn}(\rho) \, \bigl(\si^{\mu_{\rho(1)}} \, ... \,  \bar{\si}^{\mu_{\rho(2\ell)}} \bigr)_{\al} \, ^{\be} \notag \\
& \ \ \ \ \  \ \ \times \ \prod_{j=1}^{n-\ell-2} \frac{\eta^{\mu_{\rho(2\ell+2j-1)} \mu_{\rho(2\ell+2j)}}}{z_{\rho(2\ell+2j-1),\rho(2\ell+2j)} } \;  z_{\rho(2\ell+2j-1),A}  \,  z_{\rho(2\ell+2j),B}\,.
\label{npt,22}
\end{align}
When performing the limit $z_{2n-3} \mto z_{2n-2}$ in (\ref{npt,2}) we have to focus on the terms with
$\frac{\eta^{\mu_{2n-3} \mu_{2n-2}}}{z_{2n-3,2n-2}}$, i.e. firstly exclude $\ell = n-1$ (since this would distribute all
$^{\mu_{i}}$ indices among the $\si$ matrices) and secondly find the appropriate $S_{2n-2} / {\cal Q}_{n,\ell}$
permutations which really attach $\mu_{2n-3}$ and $\mu_{2n-2}$ to the same Kronecker symbol. Equation (\ref{npt,5})
requires $\rho(2\ell +2) < \rho(2\ell+4)<...< \rho(2n-2)$, so the projector to leading singular terms is simply
$\de_{\rho(2n-3),2n-3}$:
\begin{align}
\om_{(n)}&^{\mu_{1} ... \mu_{2n-2}}\,  _{\al} \,^{ \be} (z_{i}) \Bigl. \Bigr| \begin{smallmatrix} z_{2n-3} \\ \downarrow \\ z_{2n-2} \end{smallmatrix} \ \ \sim \ \ \frac{- \, 1}{z_{AB}^{1/2}} \prod_{i=1}^{2n-2} (z_{iA} \, z_{iB})^{-1/2} \, \sum_{\ell = 0}^{n-1} \, \biggl( \frac{z_{AB}}{2} \biggr)^{\ell} \, \bigl(1 \, - \, \de_{\ell,n-1} \bigr) \! \! \! \!  \sum_{\rho \in S_{2n-2}/{\cal Q}_{n,\ell}} \! \! \!  \te{sgn}(\rho) \notag \\
& \ \ \ \ \  \ \ \times \ \de_{\rho(2n-3),2n-3} \, \bigl(\si^{\mu_{\rho(1)}} \, ... \, \bar{\si}^{\mu_{\rho(2\ell)}} \bigr)_{\al}\, ^{\be}  \, \prod_{j=1}^{n-\ell-1} \frac{\eta^{\mu_{\rho(2\ell+2j-1)} \mu_{\rho(2\ell+2j)}}}{z_{\rho(2\ell+2j-1),\rho(2\ell+2j)} } \; z_{\rho(2\ell+2j-1),A} \, z_{\rho(2\ell+2j),B} \notag \\
&= \ \ \frac{- \, 1}{z_{AB}^{1/2}} \prod_{i=1}^{2n-4} (z_{iA} \, z_{iB})^{-1/2} \, \sum_{\ell = 0}^{n-2} \, \biggl( \frac{z_{AB}}{2} \biggr)^{\ell} \! \! \! \!  \sum_{\rho \in S_{2n-2}/{\cal Q}_{n,\ell}} \! \! \!  \te{sgn}(\rho) \, \de_{\rho(2n-3),2n-3}  \notag \\
& \ \ \ \ \  \ \ \times \ \bigl(\si^{\mu_{\rho(1)}} \, ... \, \bar{\si}^{\mu_{\rho(2\ell)}} \bigr)_{\al} \, ^{ \be} \; \frac{\eta^{\mu_{2n-3} \mu_{2n-2}}}{z_{2n-3,2n-2}} \; \underbrace{\frac{z_{2n-3,A} \, z_{2n-2,B}}{\bigl(z_{2n-3,A} \, z_{2n-3,B} \, z_{2n-2,A} \, z_{2n-2,B} \bigr)^{1/2} }}_{\mto 1 \ \te{as} \ z_{2n-3} \mto z_{2n-2}} \notag \\
& \ \ \ \ \  \ \ \times \ \prod_{j=1}^{n-\ell-2} \frac{\eta^{\mu_{\rho(2\ell+2j-1)} \mu_{\rho(2\ell+2j)}}}{z_{\rho(2\ell+2j-1),\rho(2\ell+2j)} } \; z_{\rho(2\ell+2j-1),A} \, z_{\rho(2\ell+2j),B} \notag \\
&= \ \ - \, \frac{ \eta^{\mu_{2n-3} \mu_{2n-2}}}{z_{2n-3,2n-2} \, z_{AB}^{1/2}} \prod_{i=1}^{2n-4} (z_{iA} \, z_{iB})^{-1/2} \, \sum_{\ell = 0}^{n-2} \, \biggl( \frac{z_{AB}}{2} \biggr)^{\ell} \! \! \! \!  \sum_{\bar{\rho} \in S_{2n-4}/{\cal Q}_{n-1,\ell}} \! \! \!  \te{sgn}(\bar{\rho}) \notag \\
& \ \ \ \ \ \ \ \times \ \bigl(\si^{\mu_{\bar{\rho}(1)}} \, ... \, \bar{\si}^{\mu_{\bar{\rho}(2\ell)}} \bigr)_{\al} \, ^{\be}  \, \prod_{j=1}^{n-\ell-2} \frac{\eta^{\mu_{\bar{\rho}(2\ell+2j-1)} \mu_{\bar{\rho}(2\ell+2j)}}}{z_{\bar{\rho}(2\ell+2j-1),\bar{\rho}(2\ell+2j)} } \; z_{\bar{\rho}(2\ell+2j-1),A} \, z_{\bar{\rho}(2\ell+2j),B}\,.
\label{npt,23}
\end{align}
Having read the arguments for $\Om_{(n)} \ra \Om_{(n-1)}$ in Subsection \ref{sec:iii}, the reader might not be
surprised about $S_{2n-2}/{\cal Q}_{n,\ell}$ with $\rho(2n-3)=2n-3,\,\rho(2n-2)=2n-2$ being equivalent to
$S_{2n-4}/{\cal Q}_{n-1,\ell}$.

This analysis is easily extended for $\bar{\om}$, so the analogue of (\ref{npt,22}) and (\ref{npt,23}) will not be displayed explicitly.

\subsubsection{$\Om_{(n)} \ra \om_{(n)}$ by $z_{1} \mto z_{B}$ limit}
\label{sec:v}

Let us now turn to a more sophisticated limiting process where a dotted spin field is converted into an undotted one via
OPE \eqref{rv,2b}:
\begin{align}
\Om_{(n)}&^{\mu_{1}... \mu_{2n-1}}\,_{\al \dbe}(z_{i}) \Bigl. \Bigr| \begin{smallmatrix} z_{1} \\ \downarrow \\ z_{B} \end{smallmatrix} \ \ \sim \ \ \frac{i}{\sqrt{2} \, (z_{1} \, - \, z_{B})^{1/2}} \; (\si^{\mu_{1}})_{\ga \dbe}  \, \langle \psi^{\mu_{2}}(z_{2}) \, ... \, \psi^{\mu_{2n-1}}(z_{2n-1}) \, S_{\al}(z_{A}) \, S^{\ga}(z_{B}) \rangle \notag \\
&= \ \ \frac{i}{\sqrt{2} \, (z_{1} \, - \, z_{B})^{1/2}} \; \om_{(n)}^{\mu_{2}...\mu_{2n-1}} \, _{\al} \, ^{\ga}(z_{i}) \, (\si^{\mu_{1}})_{\ga \dbe} \notag \\
&= \ \ \frac{1}{\sqrt{2} \, z_{1B}^{1/2}} \, \underbrace{\left( \frac{- \, i}{z_{AB}^{1/2}} \right)}_{\mto \, z_{1A}^{-1/2} \ \te{as} \ z_{1} \mto z_{B}} \, \prod_{i=2}^{2n-1} (z_{iA} \, z_{iB})^{-1/2} \, \sum_{\ell = 0}^{n-1} \, \biggl( \frac{z_{AB}}{2} \biggr)^{\ell} \! \! \! \! \sum_{\rho \in S_{2n-2}/{\cal Q}_{n,\ell}} \! \! \!  \te{sgn}(\rho)  \notag \\
& \ \ \ \ \ \times \, \bigl(\si^{\mu_{\rho(2)}} \,  ...  \, \bar{\si}^{\mu_{\rho(2\ell+1)}} \bigr)_{\al} \, ^{\ga} \, (\si^{\mu_{1}})_{\ga \dbe} \prod_{j=1}^{n-\ell-1} \frac{\eta^{\mu_{\rho(2\ell+2j)} \mu_{\rho(2\ell+2j+1)}}}{z_{\rho(2\ell+2j),\rho(2\ell+2j+1)} } \;  z_{\rho(2\ell+2j),A} \, z_{\rho(2\ell+2j+1),B}\,.
\label{npt,24}
\end{align}
The factor $i$ originates from moving the field $\psi^{\mu_{1}}(z_{1})$ across $S_{\al}(z_{A})$ before applying the OPE
of $\psi^{\mu_1}(z_1)$ with $S_{\dbe}(z_B)$. Using (\ref{roeven}) our result (\ref{npt,24}) becomes:
\begin{align}
\Om_{(n)}&^{\mu_{1}... \mu_{2n-1}}\,_{\al \dbe}(z_{i}) \Bigl. \Bigr| \begin{smallmatrix} z_{1} \\ \downarrow \\ z_{B} \end{smallmatrix} \ \ \sim \ \ \frac{1}{\sqrt{2} \, (z_{1A} \, z_{1B})^{1/2}} \, \prod_{i=2}^{2n-1} (z_{iA} \, z_{iB})^{-1/2} \, \sum_{\ell = 0}^{n-1} \, \biggl( \frac{z_{AB}}{2} \biggr)^{\ell} \! \! \! \! \sum_{\rho \in S_{2n-2}/{\cal Q}_{n,\ell}} \! \! \!  \te{sgn}(\rho)  \notag \\
& \ \ \ \ \ \ \ \times \, \biggl\{ \bigl(\si^{\mu_{1}} \, \bar{\si}^{\mu_{\rho(2)}} \, ... \, \si^{\mu_{\rho(2\ell+1)}} \bigr)_{\al \dbe} \biggr. \notag \\
& \ \ \ \ \ \ \ \ \ \  - \ \underline{2} \, \sum_{k=1}^{\ell} \eta^{\mu_{1} \mu_{\rho(2k+1)}} \; \underbrace{\frac{z_{1A} \, z_{\rho(2k+1),B}}{\underline{z_{AB}} \, z_{1,\rho(2k+1)}}}_{\mto 1 \ \te{as} \ z_{1} \mto z_{B}} \; \bigl( \si^{\mu_{\rho(2)}} \, ... \, \si^{\mu_{\rho(2k)}} \, \bar{\si}^{\mu_{\rho(2k+2)}} \, ... \, \si^{\mu_{\rho(2\ell+1)}} \bigr)_{\al \dbe} \biggr. \notag \\
& \ \ \ \ \ \ \ \ \ \ \ \biggl. + \ \underline{2} \, \sum_{k=1}^{\ell} \eta^{\mu_{1} \mu_{\rho(2k)}} \; \underbrace{\frac{z_{1A} \, z_{\rho(2k),B}}{\underline{z_{AB}} \, z_{1,\rho(2k)}}}_{\mto 1 \ \te{as} \ z_{1} \mto z_{B}} \; \bigl( \si^{\mu_{\rho(2)}} \, ... \, \bar{\si}^{\mu_{\rho(2k-1)}} \, \si^{\mu_{\rho(2k+1)}} \, ... \, \si^{\mu_{\rho(2\ell+1)}} \bigr)_{\al \dbe} \biggr\} \notag \\
& \ \ \ \ \ \ \ \times \, \prod_{j=1}^{n-\ell-1} \frac{\eta^{\mu_{\rho(2\ell+2j)} \mu_{\rho(2\ell+2j+1)}}}{z_{\rho(2\ell+2j),\rho(2\ell+2j+1)} } \;  z_{\rho(2\ell+2j),A} \, z_{\rho(2\ell+2j+1),B} \notag \\
&= \ \ \frac{1}{\sqrt{2} } \, \prod_{i=1}^{2n-1} (z_{iA} \, z_{iB})^{-1/2} \, \sum_{\ell' = 0}^{n-1} \, \biggl( \frac{z_{AB}}{2} \biggr)^{\ell'} \, \Biggl\{  \sum_{\rho \in S_{2n-2}/{\cal Q}_{n,\ell'}} \! \! \!  \te{sgn}(\rho)  \Biggr. \notag \\
& \ \ \ \ \ \ \ \ \ \ \times \ \Biggl. \bigl( \si^{\mu_{1}} \, \bar{\si}^{\mu_{\rho(2)}} \, ... \, \si^{\mu_{\rho(2\ell'+1)}} \bigr)_{\al \dbe} \, \prod_{j=1}^{n-\ell'-1} \frac{\eta^{\mu_{\rho(2\ell'+2j)} \mu_{\rho(2\ell'+2j+1)}}}{z_{\rho(2\ell'+2j),\rho(2\ell'+2j+1)} } \;  z_{\rho(2\ell'+2j),A} \, z_{\rho(2\ell'+2j+1),B} \notag \\
& \ \ \ \ \ + \ \! \! \! \! \! \! \! \! \sum_{\rho \in S_{2n-2}/{\cal Q}_{n,\ell'+1}} \! \! \! \! \! \! \!  \bigl( + \te{sgn}(\rho) \bigr) \, \sum_{k=1}^{\ell'+1} \bigl( \si^{\mu_{\rho(2)}} \, ... \, \bar{\si}^{\mu_{\rho(2k-1)}} \, \si^{\mu_{\rho(2k+1)}} \, ... \, \si^{\mu_{\rho(2\ell'+3)}} \bigr)_{\al \dbe} \; \frac{\eta^{\mu_{1} \mu_{\rho(2k)}}}{z_{1,\rho(2k)}} \; z_{1A} \, z_{\rho(2k),B} \notag \\
& \ \ \ \ \ \ \ \ \ \ \times \, \prod_{j=1}^{n-\ell'-2} \frac{\eta^{\mu_{\rho(2\ell'+2j+2)} \mu_{\rho(2\ell'+2j+3)}}}{z_{\rho(2\ell'+2j+2),\rho(2\ell'+2j+3)} } \;  z_{\rho(2\ell'+2j+2),A} \, z_{\rho(2\ell'+2j+3),B} \notag \\
& \ \ \ \ \ \ \Biggl. + \ \! \! \! \! \! \! \! \! \sum_{\rho \in S_{2n-2}/{\cal Q}_{n,\ell'+1}} \! \! \! \! \! \! \! \bigl(- \te{sgn}(\rho) \bigr) \, \sum_{k=1}^{\ell'+1} \bigl( \si^{\mu_{\rho(2)}} \, ... \, \si^{\mu_{\rho(2k)}} \, \bar{\si}^{\mu_{\rho(2k+2)}} \, ... \, \si^{\mu_{\rho(2\ell'+3)}} \bigr)_{\al \dbe} \; \frac{\eta^{\mu_{1} \mu_{\rho(2k+1)}}}{z_{1,\rho(2k+1)}} \; z_{1A} \, z_{\rho(2k+1),B} \notag \\
& \ \ \ \ \ \ \ \ \ \ \times \, \prod_{j=1}^{n-\ell'-2} \frac{\eta^{\mu_{\rho(2\ell'+2j+2)} \mu_{\rho(2\ell'+2j+3)}}}{z_{\rho(2\ell'+2j+2),\rho(2\ell'+2j+3)} } \;  z_{\rho(2\ell'+2j+2),A} \, z_{\rho(2\ell'+2j+3),B} \Biggr\} \notag \\
&= \ \ \frac{1}{\sqrt{2} } \, \prod_{i=1}^{2n-1} (z_{iA} \, z_{iB})^{-1/2} \, \sum_{\ell' = 0}^{n-1} \, \biggl( \frac{z_{AB}}{2} \biggr)^{\ell'} \! \! \! \! \sum_{\bar{\rho} \in S_{2n-1}/{\cal P}_{n,\ell'}} \! \! \!  \te{sgn}(\bar{\rho}) \notag \\
& \ \ \ \ \  \ \ \times \, \bigl(\si^{\mu_{\bar{\rho}(1)}} \,  ...  \, \si^{\mu_{\bar{\rho}(2\ell'+1)}} \bigr)_{\al \dbe} \, \prod_{j=0}^{n-\ell'-2} \frac{\eta^{\mu_{\bar{\rho}(2\ell'+2j+2)} \mu_{\bar{\rho}(2\ell'+2j+3)}}}{z_{\bar{\rho}(2\ell'+2j+2),\bar{\rho}(2\ell'+2j+3)} } \;  z_{\bar{\rho}(2\ell'+2j+2),A} \, z_{\bar{\rho}(2\ell'+2j+3),B} \ .
\label{npt,25}
\end{align}
Several steps might require some further explanation here: Firstly, in oder to see that the inserted terms $\frac{z_{1A}
  z_{\rho(2k+1),B}}{z_{AB} z_{1,\rho(2k+1)}}$ and $\frac{z_{1A} z_{\rho(2k),B}}{z_{AB} z_{1,\rho(2k)}}$ really tend to 1
in the limit $z_{1} \mto z_{B}$, one has to apply the crossing identity \eqref{zcrossing} to the numerators, i.e.\
$z_{1A} z_{\rho(2k),B} = z_{1B} z_{\rho(2k),A} + z_{1, \rho(2k)} z_{AB}$. Secondly, the underlined factors
$\frac{2}{z_{AB}}$ in the third and fourth line shift the summation variable $\ell$ by 1. Thirdly, we have replaced one
$S_{2n-2}/{\cal Q}_{n,\ell'}$ sum and $2\ell'+2$ sums over $S_{2n-2}/{\cal Q}_{n,\ell'+1}$ by the $S_{2n-1}/{\cal
  P}_{n,\ell'}$ sum over larger permutations $\bar{\rho}$ including $\mu_{1}$. The total number of terms is the same
since:
\begin{align}
\bigl| S_{2n-2}/{\cal Q}_{n,\ell} & \bigr| \ + \ (2\ell \, + \, 2) \, \bigl| S_{2n-2}/{\cal Q}_{n,\ell+1} \bigr| \ \ = \ \ \frac{(2n \, - \, 2)!}{(2\ell)! \, (n \, - \, \ell \, - \, 1)! \, 2^{n-\ell-1}} \notag \\
& \ \ \ \ \ \ + \ \frac{(2\ell \, + \, 2) \, (2n\, - \, 2)!}{(2\ell \, + \, 2)! \, (n \, - \, \ell \, - \, 2)! \, 2^{n-\ell-2}} \notag \\
&= \ \ \frac{(2n \, - \, 2)!}{(2\ell \, + \, 1)! \, (n \, - \, \ell \, - \, 1)! \, 2^{n-\ell-1}} \; \Bigl( 2\ell \, + \, 1 \, + \, 2 \  (n-\ell-1) \Bigr) \notag \\
&= \ \ \frac{(2n \, - \, 1)!}{(2\ell \, + \, 1)! \, (n \, - \, \ell \, - \, 1)! \, 2^{n-\ell-1}} \notag \\
&= \ \ \bigl| S_{2n-1}/{\cal P}_{n,\ell} \bigr| \ .
\label{npt,26}
\end{align}
Moreover, the index $\mu_{1}$ appears in all possible positions in (\ref{npt,25}), i.e.\ attached to $\si$'s as well as
to $\eta$'s. The relative sign between $\rho$ and $\bar{\rho}$ is taken into account.

Following the general principle of this proof, we should now compare the final line of (\ref{npt,25}) with the most
singular terms in the expression (\ref{npt,1}) for $\Om_{(n)}$. Due to our particular arrangement of $\si$ matrices
$\rho(1) < ... < \rho(2\ell+1)$, no factors of $z_{1}-z_{B}$ appear in any denominator. So equation (\ref{npt,1}) does
not simplify in the $z_{1} \mto z_{B}$ regime. However, it already matches with (\ref{npt,25}) up to a shift in the $j$
product, so we are done with the $\Om_{(n)} \ra \om_{(n)}$ reduction.

Actually, this is the reason we sent $z_{1}$ and not any other $z_{i}$, $i=2,3,...,2n-1$, to $z_{B}$. the more $z_{ij}$
dependences of different index structures are successfully reproduced, the more convincing becomes this part of the
proof.

\subsubsection{$\Om_{(n)} \ra \bar{\om}_{(n)}$ by $z_{2n-1} \mto z_{A}$ limit}
\label{sec:vi}

The ``dotted'' analogue of the preceding limit requires the use of the OPE \eqref{rv,2a}. For reasons given in the end
of the previous subsections the behaviour for $z_{2n-1} \mto z_{A}$  gives rise to the richest consistency checks as the
right hand side of (\ref{npt,1}) does not have any poles in $(z_{2n-1} - z_{A})$. In detail, the limiting
behaviour is given by:
\begin{align}
\Om&_{(n)}^{\mu_{1}... \mu_{2n-1}}\,_{\al \dbe}(z_{i}) \Bigl. \Bigr| \begin{smallmatrix} z_{2n-1} \\ \downarrow \\ z_{A} \end{smallmatrix} \ \ \sim \ \ \frac{(\si^{\mu_{2n-1}})_{\al \dga}}{\sqrt{2} \, (z_{2n-1} \, - \, z_{A})^{1/2}} \;   \langle \psi^{\mu_{1}}(z_{1}) \, ... \, \psi^{\mu_{2n-2}}(z_{2n-2}) \, S^{\dga}(z_{A}) \, S_{\dbe}(z_{B}) \rangle \notag \\
&= \ \ \frac{1}{\sqrt{2} \, (z_{2n-1} \, - \, z_{A})^{1/2}} \; (\si^{\mu_{2n-1}})_{\al \dga} \, \bar{\om}_{(n)}^{\mu_{1}...\mu_{2n-2}\dga} \, _{\dbe}(z_{i})  \notag \\
&= \ \ \frac{1}{\sqrt{2} \, z_{2n-1,A}^{1/2}} \, \underbrace{\left( \frac{1}{z_{AB}^{1/2}} \right)}_{\mto \, z_{2n-1,B}^{-1/2} \ \te{as} \ z_{2n-1} \mto z_{A}} \, \prod_{i=1}^{2n-2} (z_{iA} \, z_{iB})^{-1/2} \, \sum_{\ell = 0}^{n-1} \, \biggl( \frac{z_{AB}}{2} \biggr)^{\ell} \! \! \! \! \sum_{\rho \in S_{2n-2}/{\cal Q}_{n,\ell}} \! \! \!  \te{sgn}(\rho)  \notag \\
& \ \ \ \ \  \ \ \times \, (\si^{\mu_{2n-1}})_{\al \dga} \, \bigl(\bar{\si}^{\mu_{\rho(1)}} \,  ...  \, \si^{\mu_{\rho(2\ell)}} \bigr)^{\dga} \, _{\dbe} \, \prod_{j=1}^{n-\ell-1} \frac{\eta^{\mu_{\rho(2\ell+2j-1)} \mu_{\rho(2\ell+2j)}}}{z_{\rho(2\ell+2j-1),\rho(2\ell+2j)} } \;  z_{\rho(2\ell+2j-1),A} \, z_{\rho(2\ell+2j),B} \ .
\label{npt,27}
\end{align}
At this point we use the identity (\ref{roeven}):
\begin{align}
\Om&_{(n)}^{\mu_{1}... \mu_{2n-1}}\,_{\al \dbe}(z_{i}) \Bigl. \Bigr| \begin{smallmatrix} z_{2n-1} \\ \downarrow \\ z_{A} \end{smallmatrix} \ \ \sim \ \ \frac{1}{\sqrt{2} } \, \prod_{i=1}^{2n-1} (z_{iA} \, z_{iB})^{-1/2} \, \sum_{\ell = 0}^{n-1} \, \biggl( \frac{z_{AB}}{2} \biggr)^{\ell} \! \! \! \! \sum_{\rho \in S_{2n-2}/{\cal Q}_{n,\ell}} \! \! \!  \te{sgn}(\rho)  \notag \\
& \ \ \ \ \ \ \ \times \, \biggl\{ \bigl( \si^{\mu_{\rho(1)}} \,  ... \, \bar{\si}^{\mu_{\rho(2\ell)}} \, \si^{\mu_{2n-1}} \bigr)_{\al \dbe} \biggr. \notag \\
& \ \ \ \ \ \ \ \ \ \  - \ \underline{2} \, \sum_{k=1}^{\ell} \eta^{\mu_{\rho(2k-1)} \mu_{2n-1}} \; \underbrace{\frac{z_{\rho(2k-1),A} \, z_{2n-1,B}}{\underline{z_{AB}} \, z_{2n-1,\rho(2k-1)}}}_{\mto 1 \ \te{as} \ z_{2n-1} \mto z_{A}} \; \bigl( \si^{\mu_{\rho(1)}} \, ... \, \bar{\si}^{\mu_{\rho(2k-2)}} \, \si^{\mu_{\rho(2k)}} \, ... \, \si^{\mu_{\rho(2\ell)}} \bigr)_{\al \dbe} \biggr. \notag \\
& \ \ \ \ \ \ \ \ \ \ \ \biggl. + \ \underline{2} \, \sum_{k=1}^{\ell} \eta^{\mu_{\rho(2k)} \mu_{2n-1}} \; \underbrace{\frac{z_{\rho(2k),A} \, z_{2n-1,B}}{\underline{z_{AB}} \, z_{\rho(2k),2n-1}}}_{\mto 1 \ \te{as} \ z_{2n-1} \mto z_{A}} \; \bigl(\si^{\mu_{\rho(1)}} \, ... \, \si^{\mu_{\rho(2k-1)}} \, \bar{\si}^{\mu_{\rho(2k+1)}} \, ... \, \si^{\mu_{\rho(2\ell)}} \bigr)_{\al \dbe} \biggr\} \notag \\
& \ \ \ \ \ \ \ \times \, \prod_{j=1}^{n-\ell-1} \frac{\eta^{\mu_{\rho(2\ell+2j-1)} \mu_{\rho(2\ell+2j)}}}{z_{\rho(2\ell+2j-1),\rho(2\ell+2j)} } \;  z_{\rho(2\ell+2j-1),A} \, z_{\rho(2\ell+2j),B} \notag \\
&= \ \ \frac{1}{\sqrt{2} } \, \prod_{i=1}^{2n-1} (z_{iA} \, z_{iB})^{-1/2} \, \sum_{\ell' = 0}^{n-1} \, \biggl( \frac{z_{AB}}{2} \biggr)^{\ell'} \, \Biggl\{  \sum_{\rho \in S_{2n-2}/{\cal Q}_{n,\ell'}} \! \! \!  \te{sgn}(\rho)  \Biggr. \notag \\
& \ \ \ \ \ \ \ \ \ \ \times \ \Biggl. \bigl(\si^{\mu_{\rho(1)}} \,  ... \, \bar{\si}^{\mu_{\rho(2\ell')}} \, \si^{\mu_{2n-1}} \bigr)_{\al \dbe} \, \prod_{j=1}^{n-\ell'-1} \frac{\eta^{\mu_{\rho(2\ell'+2j-1)} \mu_{\rho(2\ell'+2j)}}}{z_{\rho(2\ell'+2j-1),\rho(2\ell'+2j)} } \;  z_{\rho(2\ell'+2j-1),A} \, z_{\rho(2\ell'+2j),B} \notag \\
& \ \ \ \ \ + \ \! \! \! \! \! \! \! \sum_{\rho \in S_{2n-2}/{\cal Q}_{n,\ell'+1}} \! \! \! \! \! \! \!  \bigl( + \te{sgn}(\rho) \bigr) \, \sum_{k=1}^{\ell'+1} \bigl( \si^{\mu_{\rho(1)}} \, ... \, \si^{\mu_{\rho(2k-1)}} \, \bar{\si}^{\mu_{\rho(2k+1)}} \, ... \si^{\mu_{\rho(2\ell'+2)}} \bigr)_{\al \dbe} \; \frac{\eta^{\mu_{\rho(2k)} \mu_{2n-1}}}{z_{\rho(2k),2n-1}} \notag \\
& \ \ \ \ \ \ \ \ \ \ \times \, z_{\rho(2k),A} \, z_{2n-1,B} \, \prod_{j=1}^{n-\ell'-2} \frac{\eta^{\mu_{\rho(2\ell'+2j+1)} \mu_{\rho(2\ell'+2j+2)}}}{z_{\rho(2\ell'+2j+1),\rho(2\ell'+2j+2)} } \;  z_{\rho(2\ell'+2j+1),A} \, z_{\rho(2\ell'+2j+2),B} \notag \\
& \ \ \ \ \ \ \Biggl. + \ \! \! \! \! \! \! \! \sum_{\rho \in S_{2n-2}/{\cal Q}_{n,\ell'+1}} \! \! \! \! \! \! \! \bigl(- \te{sgn}(\rho) \bigr) \, \sum_{k=1}^{\ell'+1} \bigl( \si^{\mu_{\rho(1)}} \, ... \, \bar{\si}^{\mu_{\rho(2k-2)}} \, \si^{\mu_{\rho(2k)}} \, ... \, \si^{\mu_{\rho(2\ell'+2)}} \bigr)_{\al \dbe} \; \frac{\eta^{ \mu_{\rho(2k-1)} \mu_{2n-1}}}{z_{\rho(2k-1),2n-1}} \notag \\
& \ \ \ \ \ \ \ \ \ \ \times \, z_{\rho(2k-1),A} \, z_{2n-1,B} \, \prod_{j=1}^{n-\ell'-2} \frac{\eta^{\mu_{\rho(2\ell'+2j+1)} \mu_{\rho(2\ell'+2j+2)}}}{z_{\rho(2\ell'+2j+1),\rho(2\ell'+2j+2)} } \;  z_{\rho(2\ell'+2j+1),A} \, z_{\rho(2\ell'+2j+2),B} \Biggr\} \notag \\
&= \ \ \frac{1}{\sqrt{2} } \, \prod_{i=1}^{2n-1} (z_{iA} \, z_{iB})^{-1/2} \, \sum_{\ell' = 0}^{n-1} \, \biggl( \frac{z_{AB}}{2} \biggr)^{\ell'} \! \! \! \! \sum_{\bar{\rho} \in S_{2n-1}/{\cal P}_{n,\ell'}} \! \! \!  \te{sgn}(\bar{\rho}) \notag \\
& \ \ \ \ \  \ \ \times \, \bigl(\si^{\mu_{\bar{\rho}(1)}} \,  ...  \, \si^{\mu_{\bar{\rho}(2\ell'+1)}} \bigr)_{\al \dbe} \, \prod_{j=0}^{n-\ell'-2} \frac{\eta^{\mu_{\bar{\rho}(2\ell'+2j+2)} \mu_{\bar{\rho}(2\ell'+2j+3)}}}{z_{\bar{\rho}(2\ell'+2j+2),\bar{\rho}(2\ell'+2j+3)} } \;  z_{\bar{\rho}(2\ell'+2j+2),A} \, z_{\bar{\rho}(2\ell'+2j+3),B}\,. 
\label{npt,28}
\end{align}
This is the exact expression for $\Om_{(n)}$. So we can move on to the next case $\om_{(n)} \ra \Om_{(n-1)}$.

\subsubsection{$\om_{(n)} \ra \Om_{(n-1)}$ by $z_{1} \mto z_{B}$ limit}
\label{sec:vii}

The reduction of $\om_{(n)}$ to correlators of $\Om_{(n-1)}$ type is based on the OPE \eqref{rv,2a}:
\begin{align}
\om&_{(n)}^{\mu_{1} ... \mu_{2n-2}}\,  _{\al} \, ^{ \be} (z_{i}) \Bigl. \Bigr| \begin{smallmatrix} z_{1} \\ \downarrow \\ z_{B} \end{smallmatrix} \ \ \sim \ \  \frac{i}{\sqrt{2} \, (z_{1} \, - \, z_{B})^{1/2}} \; (\bar{\si}^{\mu_{1}})^{\dga \be} \, \langle \psi^{\mu_{2}}(z_{2}) \, ... \, \psi^{\mu_{2n-2}}(z_{2n-2}) \, S_{\al}(z_{A}) \, S_{\dga}(z_{B}) \rangle \notag \\
&= \ \ \frac{i}{\sqrt{2} \, (z_{1} \, - \, z_{B})^{1/2}} \; \Om_{(n-1)}^{\mu_{2} ... \mu_{2n-2}}\, _{\al \dga} (z_{i}) \, (\bar{\si}^{\mu_{1}})^{\dga \be} \notag \\
&= \ \ \frac{i}{2 \, z_{1B}^{1/2}} \; \underbrace{\frac{z_{AB}}{z_{AB}^{1/2}  \, z_{1A}^{1/2}}}_{\mto \, - i \ \te{as} \ z_{1} \mto z_{B}} \, \prod_{i=2}^{2n-2} (z_{iA} \, z_{iB})^{-1/2} \, \sum_{\ell = 0}^{n-2} \, \biggl( \frac{z_{AB}}{2} \biggr)^{\ell} \! \! \! \! \! \sum_{\rho \in S_{2n-3}/{\cal P}_{n-1,\ell}} \! \! \! \! \te{sgn}(\rho) \notag \\
& \ \ \ \ \  \times  \, \bigl(\si^{\mu_{\rho(2)}} \,  ... \,  \si^{\mu_{\rho(2\ell+2)}} \bigr)_{\al \dga} \, (\bar{\si}^{\mu_{1}})^{\dga \be} \! \prod_{j=1}^{n-\ell-2} \frac{\eta^{\mu_{\rho(2\ell+2j+1)} \mu_{\rho(2\ell+2j+2)}}}{z_{\rho(2\ell+2j+1),\rho(2\ell+2j+2)} } \;  z_{\rho(2\ell+2j+1),A}  \, z_{\rho(2\ell+2j+2),B}\,.
\label{npt,29}
\end{align}
Here, it is necessary to move the $\bar{\si}^{\mu_{1}}$ to the left of $\si^{\mu_{\rho(2)}}
... \si^{\mu_{\rho(2\ell+2)}}$, i.e. across an odd number of $\si$ matrices. With the help of (\ref{roodd}) we obtain:
\begin{align}
\om_{(n)}&^{\mu_{1} ... \mu_{2n-2}}\,  _{\al} \, ^{ \be} (z_{i}) \Bigl. \Bigr| \begin{smallmatrix} z_{1} \\ \downarrow \\ z_{B} \end{smallmatrix} \ \ \sim \ \ \frac{\underline{ z_{AB}}}{\underline{2} \, z_{AB}^{1/2} \, z_{1A}^{1/2} \, z_{1B}^{1/2}} \; \prod_{i=2}^{2n-2} (z_{iA} \, z_{iB})^{-1/2} \, \sum_{\ell = 0}^{n-2} \, \biggl( \frac{z_{AB}}{2} \biggr)^{\ell}  \notag \\
& \ \ \ \ \  \ \ \times \! \! \! \! \sum_{\rho \in S_{2n-3}/{\cal P}_{n-1,\ell}} \! \! \! \! \! \te{sgn}(\rho) \, \Biggl\{ - \ \bigl( \si^{\mu_{1}} \, \bar{\si}^{\mu_{\rho(2)}} \,  ... \,  \bar{\si}^{\mu_{\rho(2\ell+2)}} \bigr)_{\al } \, ^{\be} \, \Biggr. \notag \\
&\ \ \ \ \ \ \ \ \ \  - \ \underline{2} \, \sum_{k=1}^{\ell+1} \eta^{\mu_{1} \mu_{\rho(2k)}} \; \underbrace{\frac{z_{1A} \, z_{\rho(2k),B}}{z_{1,\rho(2k)} \, \underline{z_{AB}}}}_{\mto 1 \ \te{as} \ z_{1} \mto z_{B}} \; \bigl( \si^{\mu_{\rho(2)}} \, ... \, \bar{\si}^{\mu_{\rho(2k-1)}} \,  \si^{\mu_{\rho(2k+1)}} \, ... \, \bar{\si}^{\mu_{\rho(2\ell+2)}} \bigr)_{\al } \, ^{\be} \notag \\
&\ \ \ \ \ \ \ \ \ \ \ \Biggl. + \ \underline{2} \, \sum_{k=1}^{\ell} \eta^{\mu_{1} \mu_{\rho(2k+1)}} \; \underbrace{\frac{z_{1A} \, z_{\rho(2k+1),B}}{z_{1,\rho(2k+1)} \, \underline{z_{AB}}}}_{\mto 1 \ \te{as} \ z_{1} \mto z_{B}} \; \bigl( \si^{\mu_{\rho(2)}} \, ... \, \si^{\mu_{\rho(2k)}} \,  \bar{\si}^{\mu_{\rho(2k+2)}} \, ... \, \bar{\si}^{\mu_{\rho(2\ell+2)}} \bigr)_{\al } \, ^{\be} \Biggr\} \notag \\
 & \ \ \ \ \  \ \ \times  \, \prod_{j=1}^{n-\ell-2} \frac{\eta^{\mu_{\rho(2\ell+2j+1)} \mu_{\rho(2\ell+2j+2)}}}{z_{\rho(2\ell+2j+1),\rho(2\ell+2j+2)} } \;  z_{\rho(2\ell+2j+1),A}  \, z_{\rho(2\ell+2j+2),B} \notag \\ 
  &= \ \ \frac{- \, 1}{z_{AB}^{1/2}} \, \prod_{i=1}^{2n-2} (z_{iA} \, z_{iB})^{-1/2} \, \sum_{\ell' = 0}^{n-1} \, \biggl( \frac{z_{AB}}{2} \biggr)^{\ell'}  \Biggl\{ \sum_{\rho \in S_{2n-3}/{\cal P}_{n-1,\ell'-1}} \! \! \! \! \! \te{sgn}(\rho) \Biggr. \notag \\
   & \ \ \ \ \  \ \ \times  \, \bigl( \si^{\mu_{1}} \, \bar{\si}^{\mu_{\rho(2)}} \,  ... \,  \bar{\si}^{\mu_{\rho(2\ell')}} \bigr)_{\al } \, ^{\be} \, \prod_{j=1}^{n-\ell'-1} \frac{\eta^{\mu_{\rho(2\ell'+2j-1)} \mu_{\rho(2\ell'+2j)}}}{z_{\rho(2\ell'+2j- 1),\rho(2\ell'+2j)} } \;  z_{\rho(2\ell'+2j-1),A}  \, z_{\rho(2\ell'+2j),B} \notag \\
     & \ \ \ \ + \! \! \! \! \sum_{\rho \in S_{2n-3}/{\cal P}_{n-1,\ell'}} \! \! \! \! \! \bigl( + \te{sgn}(\rho) \bigr) \, \sum_{k=1}^{\ell'+1} \bigl( \si^{\mu_{\rho(2)}} \, ... \, \bar{\si}^{\mu_{\rho(2k-1)}} \,  \si^{\mu_{\rho(2k+1)}} \, ... \, \bar{\si}^{\mu_{\rho(2\ell'+2)}} \bigr)_{\al } \, ^{\be} \; \frac{\eta^{\mu_{1} \mu_{\rho(2k)}}}{z_{1,\rho(2k)}} \notag \\ 
  & \ \ \ \ \ \ \ \ \times \ z_{1A} \, z_{\rho(2k),B} \prod_{j=1}^{n-\ell'-2} \frac{\eta^{\mu_{\rho(2\ell'+2j+1)} \mu_{\rho(2\ell'+2j+2)}}}{z_{\rho(2\ell'+2j+1),\rho(2\ell'+2j+2)} } \;  z_{\rho(2\ell'+2j+1),A}  \, z_{\rho(2\ell'+2j+2),B} \notag \\  
  & \ \ \ \ + \! \! \! \! \sum_{\rho \in S_{2n-3}/{\cal P}_{n-1,\ell'}} \! \! \! \! \! \bigl( - \te{sgn}(\rho) \bigr) \, \sum_{k=1}^{\ell'} \bigl( \si^{\mu_{\rho(2)}} \, ... \, \si^{\mu_{\rho(2k)}} \,  \bar{\si}^{\mu_{\rho(2k+2)}} \, ... \, \bar{\si}^{\mu_{\rho(2\ell'+2)}} \bigr)_{\al } \, ^{\be} \; \frac{\eta^{\mu_{1} \mu_{\rho(2k+1)}}}{z_{1,\rho(2k+1)}} \notag \\
  & \ \ \ \ \ \ \ \ \Biggl. \times \ z_{1A} \, z_{\rho(2k+1),B} \prod_{j=1}^{n-\ell'-2} \frac{\eta^{\mu_{\rho(2\ell'+2j+1)} \mu_{\rho(2\ell'+2j+2)}}}{z_{\rho(2\ell'+2j+1),\rho(2\ell'+2j+2)} } \;  z_{\rho(2\ell'+2j+1),A}  \, z_{\rho(2\ell'+2j+2),B} \Biggr\} \notag \\
  &= \ \ \frac{- \, 1}{z_{AB}^{1/2}} \, \prod_{i=1}^{2n-2} (z_{iA} \, z_{iB})^{-1/2} \, \sum_{\ell' = 0}^{n-1} \, \biggl( \frac{z_{AB}}{2} \biggr)^{\ell'}  \! \! \! \sum_{\bar{\rho} \in S_{2n-2}/{\cal Q}_{n,\ell'}} \! \! \! \! \! \te{sgn}(\bar{\rho}) \notag \\
  & \ \ \ \ \ \ \ \ \times \ \bigl( \si^{\mu_{\bar{\rho}(1)}} \, ... \, \bar{\si}^{\mu_{\bar{\rho}(2\ell')}} \bigr)_{\al} \,^{\be} \, \prod_{j=0}^{n-\ell'-2} \frac{\eta^{\mu_{\bar{\rho}(2\ell'+2j+1)} \mu_{\bar{\rho}(2\ell'+2j+2)}}}{z_{\bar{\rho}(2\ell'+2j+1),\bar{\rho}(2\ell'+2j+2)} } \;  z_{\bar{\rho}(2\ell'+2j+1),A}  \, z_{\bar{\rho}(2\ell'+2j+2),B}\,.
\label{npt,30}
\end{align}
The underlined factors of $\frac{2}{z_{AB}}$ in the third and fourth line cancel with the$ \frac{z_{AB}}{2}$ from the
prefactor, so only the $\si^{\mu_{1}} \bar{\si}^{\mu_{\rho(2)}}  ...  \bar{\si}^{\mu_{\rho(2\ell+2)}}$ term in the
second line gives rise to a resummation $\ell \mto \ell'=\ell+1$. In the last step we have regrouped the $S_{2n-3}$ permutations of total number:
\begin{align}
\bigl| S_{2n-3}/{\cal P}_{n-1,\ell-1} & \bigr| \ + \ (2\ell \, + \, 1) \, \bigl| S_{2n-3}/{\cal P}_{n-1,\ell} \bigr| \ \ = \ \ \frac{(2n \, - \, 3)!}{(2\ell \, - \, 1)! \, (n \, - \, \ell \, - \, 1)! \, 2^{n-\ell-1}} \notag \\
& \ \ \ \ \ \ + \ \frac{(2\ell \, + \, 1) \, (2n\, - \, 3)!}{(2\ell \, + \, 1)! \, (n \, - \, \ell \, - \, 2)! \, 2^{n-\ell-2}} \notag \\
&= \ \ \frac{(2n \, - \, 3)!}{(2\ell)! \, (n \, - \, \ell \, - \, 1)! \, 2^{n-\ell-1}} \; \Bigl( 2\ell \,  + \, 2 \  (n-\ell-1) \Bigr) \notag \\
&= \ \ \frac{(2n \, - \, 2)!}{(2\ell )! \, (n \, - \, \ell \, - \, 1)! \, 2^{n-\ell-1}} \notag \\
&= \ \ \bigl| S_{2n-2}/{\cal Q}_{n,\ell} \bigr| \ .
\label{npt,31}
\end{align}
They exhaust all the possible $S_{2n-2} / {\cal Q}_{n,\ell}$ elements $\bar{\rho}$ where $\mu_{1}$ is included. By carefully looking at the $\si$ strings, the reader can check that indeed $\te{sgn}(\rho) = \te{sgn}(\bar{\rho})$ in all cases. The result of (\ref{npt,30}) is exactly what was claimed in (\ref{npt,2}) for $\om_{(n)}$.

\subsubsection{$\bar{\om}_{(n)} \ra \Om_{(n-1)}$ 
by $z_{2n-2} \mto z_{A}$ limit}
\label{sec:viii}

This last Subsection completes the list of possible limits for our class of $n$ point correlators. It is analogous to
the preceding discussion with some minor differences: $(z_{1},z_{B})$ is replaced by $(z_{2n-2},z_{A})$ and here we need
the OPE \eqref{rv,2b}. As the OPE fields have to be permuted past $S_{\dbe}$ an additional minus sign appears:
\begin{align}
\bar{\om}&_{(n)}^{\mu_{1} ... \mu_{2n-2} \dal} \, _{ \dbe} (z_{i}) \Bigl. \Bigr| \begin{smallmatrix} z_{2n-2} \\ \downarrow \\ z_{A} \end{smallmatrix} \ \ \sim \ \ \frac{- \ (\bar{\si}^{\mu_{2n-2}})^{\dal \ga}}{\sqrt{2} \, (z_{2n-2} \, - \, z_{A})^{1/2}} \;   \langle \psi^{\mu_{1}}(z_{1}) \, ... \, \psi^{\mu_{2n-3}}(z_{2n-3}) \, S_{\ga}(z_{A}) \, S_{\dbe}(z_{B}) \rangle \notag \\
&= \ \ \frac{- \, 1}{\sqrt{2} \, (z_{2n-2} \, - \, z_{A})^{1/2}} \; (\bar{\si}^{\mu_{2n-2}})^{\dal \ga} \, \Om_{(n-1)}^{\mu_{1} ... \mu_{2n-3}}\, _{\ga \dbe} (z_{i})  \notag \\
&= \ \ \frac{- \, 1}{2 \, z_{2n-2,A}^{1/2}} \; \underbrace{\frac{z_{AB}}{z_{AB}^{1/2} \, z_{2n-2,B}^{1/2}}}_{\mto 1 \ \te{as} \ z_{2n-2} \mto z_{A}} \, \prod_{i=1}^{2n-3} (z_{iA} \, z_{iB})^{-1/2} \, \sum_{\ell = 0}^{n-2} \, \biggl( \frac{z_{AB}}{2} \biggr)^{\ell} \! \! \! \! \! \sum_{\rho \in S_{2n-3}/{\cal P}_{n-1,\ell}} \! \! \! \! \te{sgn}(\rho) \notag \\
& \ \ \ \ \times  \, (\bar{\si}^{\mu_{2n-2}})^{\dal \ga} \, \bigl(\si^{\mu_{\rho(1)}} \,  ... \,  \si^{\mu_{\rho(2\ell+1)}} \bigr)_{\ga \dbe}  \! \prod_{j=1}^{n-\ell-2} \frac{\eta^{\mu_{\rho(2\ell+2j)} \mu_{\rho(2\ell+2j+1)}}}{z_{\rho(2\ell+2j),\rho(2\ell+2j+1)} } \;  z_{\rho(2\ell+2j),A}  \, z_{\rho(2\ell+2j+1),B}\,.
\label{npt,32}
\end{align}
The appropriate order of $\si^{\mu_{i}}$ indices can be achieved with the help of (\ref{roodd}):
\begin{align}
\bar{\om}_{(n)}&^{\mu_{1} ... \mu_{2n-2} \dal} \, _{ \dbe} (z_{i}) \Bigl. \Bigr| \begin{smallmatrix} z_{2n-2} \\ \downarrow \\ z_{A} \end{smallmatrix} \ \ \sim \ \ \frac{- \, \underline{ z_{AB}}}{\underline{2} \, z_{AB}^{1/2} \, z_{2n-2,A}^{1/2} \, z_{2n-2,B}^{1/2}} \; \prod_{i=1}^{2n-3} (z_{iA} \, z_{iB})^{-1/2} \, \sum_{\ell = 0}^{n-2} \, \biggl( \frac{z_{AB}}{2} \biggr)^{\ell}  \notag \\
& \ \ \ \ \  \ \ \times  \! \! \! \! \sum_{\rho \in S_{2n-3}/{\cal P}_{n-1,\ell}} \! \! \! \! \!  \, \te{sgn}(\rho) \, \Biggl\{ - \ \bigl( \bar{\si}^{\mu_{\rho(1)}} \,  ... \,  \bar{\si}^{\mu_{\rho(2\ell+1)}} \, \si^{\mu_{2n-2}} \bigr)^{\dal } \, _{\dbe} \, \Biggr. \notag \\
&\ \ \ \ \ \ \ \ \ \  - \ \underline{2} \, \sum_{k=1}^{\ell+1} \eta^{\mu_{\rho(2k-1)} \mu_{2n-2}} \; \underbrace{\frac{z_{\rho(2k-1),A} \, z_{2n-2,B}}{z_{\rho(2k-1),2n-2} \, \underline{z_{AB}}}}_{\mto 1 \ \te{as} \ z_{2n-2} \mto z_{A}} \; \bigl( \bar{\si}^{\mu_{\rho(1)}} \, ... \, \si^{\mu_{\rho(2k-2)}} \,  \bar{\si}^{\mu_{\rho(2k)}} \, ... \, \si^{\mu_{\rho(2\ell+1)}} \bigr)^{\dal } \, _{\dbe} \notag \\
&\ \ \ \ \ \ \ \ \ \ \ \Biggl. + \ \underline{2} \, \sum_{k=1}^{\ell} \eta^{\mu_{\rho(2k)} \mu_{2n-2}} \; \underbrace{\frac{z_{\rho(2k),A} \, z_{2n-2,B}}{z_{\rho(2k),2n-2} \, \underline{z_{AB}}}}_{\mto 1 \ \te{as} \ z_{2n-2} \mto z_{A}} \; \bigl( \bar{\si}^{\mu_{\rho(1)}} \, ... \, \bar{\si}^{\mu_{\rho(2k-1)}} \,  \si^{\mu_{\rho(2k+1)}} \, ... \, \si^{\mu_{\rho(2\ell+1)}} \bigr)^{\dal } \, _{\dbe} \Biggr\} \notag \\
 & \ \ \ \ \  \ \ \times  \, \! \prod_{j=1}^{n-\ell-2} \frac{\eta^{\mu_{\rho(2\ell+2j)} \mu_{\rho(2\ell+2j+1)}}}{z_{\rho(2\ell+2j),\rho(2\ell+2j+1)} } \;  z_{\rho(2\ell+2j),A}  \, z_{\rho(2\ell+2j+1),B} \notag \\ 
  &= \ \ \frac{+1}{z_{AB}^{1/2}} \, \prod_{i=1}^{2n-2} (z_{iA} \, z_{iB})^{-1/2} \, \sum_{\ell' = 0}^{n-1} \, \biggl( \frac{z_{AB}}{2} \biggr)^{\ell'}  \Biggl\{ \sum_{\rho \in S_{2n-3}/{\cal P}_{n-1,\ell'-1}} \! \! \! \! \! \te{sgn}(\rho) \Biggr. \notag \\
   & \ \ \ \ \  \ \ \times  \, \bigl( \bar{\si}^{\mu_{\rho(1)}} \,  ... \,  \bar{\si}^{\mu_{\rho(2\ell'-1)}} \, \si^{\mu_{2n-2}} \bigr)^{\dal } \, _{\dbe} \, \prod_{j=1}^{n-\ell'-1} \frac{\eta^{\mu_{\rho(2\ell'+2j-2)} \mu_{\rho(2\ell'+2j-1)}}}{z_{\rho(2\ell'+2j-2),\rho(2\ell'+2j-1)} } \;  z_{\rho(2\ell'+2j-2),A}  \, z_{\rho(2\ell'+2j-1),B}
    \notag \\
     & \ \ \ \ + \! \! \! \! \sum_{\rho \in S_{2n-3}/{\cal P}_{n-1,\ell'}} \! \! \! \! \! \bigl( + \te{sgn}(\rho) \bigr) \, \sum_{k=1}^{\ell'+1} \bigl( \bar{\si}^{\mu_{\rho(1)}} \, ... \, \si^{\mu_{\rho(2k-2)}} \,  \bar{\si}^{\mu_{\rho(2k)}} \, ... \, \si^{\mu_{\rho(2\ell'+1)}} \bigr)^{\dal } \, _{\dbe} \; \frac{\eta^{\mu_{\rho(2k-1)} \mu_{2n-2}}}{z_{\rho(2k-1),2n-2}} \notag \\ 
  & \ \ \ \ \ \ \ \ \times \ z_{\rho(2k-1),A} \, z_{2n-2,B} \prod_{j=1}^{n-\ell'-2} \frac{\eta^{\mu_{\rho(2\ell'+2j)} \mu_{\rho(2\ell'+2j+1)}}}{z_{\rho(2\ell'+2j),\rho(2\ell'+2j+1)} } \;  z_{\rho(2\ell'+2j),A}  \, z_{\rho(2\ell'+2j+1),B} \notag \\  
  & \ \ \ \ + \! \! \! \! \sum_{\rho \in S_{2n-3}/{\cal P}_{n-1,\ell'}} \! \! \! \! \! \bigl( - \te{sgn}(\rho) \bigr) \, \sum_{k=1}^{\ell'} \bigl( \bar{\si}^{\mu_{\rho(1)}} \, ... \, \bar{\si}^{\mu_{\rho(2k-1)}} \,  \si^{\mu_{\rho(2k+1)}} \, ... \, \si^{\mu_{\rho(2\ell'+1)}} \bigr)^{\dal } \, _{\dbe} \; \frac{\eta^{\mu_{\rho(2k)} \mu_{2n-2}}}{z_{\rho(2k),2n-2}} \notag \\
  & \ \ \ \ \ \ \ \ \Biggl. \times \ z_{\rho(2k),A} \, z_{2n-2,B} \prod_{j=1}^{n-\ell'-2} \frac{\eta^{\mu_{\rho(2\ell'+2j)} \mu_{\rho(2\ell'+2j+1)}}}{z_{\rho(2\ell'+2j),\rho(2\ell'+2j+1)} } \;  z_{\rho(2\ell'+2j),A}  \, z_{\rho(2\ell'+2j+1),B} \Biggr\} \notag \\
  &= \ \ \frac{+1}{z_{AB}^{1/2}} \, \prod_{i=1}^{2n-2} (z_{iA} \, z_{iB})^{-1/2} \, \sum_{\ell' = 0}^{n-1} \, \biggl( \frac{z_{AB}}{2} \biggr)^{\ell'}  \! \! \! \sum_{\bar{\rho} \in S_{2n-2}/{\cal Q}_{n,\ell'}} \! \! \! \! \! \te{sgn}(\bar{\rho}) \notag \\
  & \ \ \ \ \ \ \ \ \times \ \bigl( \bar{\si}^{\mu_{\bar{\rho}(1)}} \, ... \, \si^{\mu_{\bar{\rho}(2\ell')}} \bigr)^{\dal} \, _{\dbe} \, \prod_{j=0}^{n-\ell'-2} \frac{\eta^{\mu_{\bar{\rho}(2\ell'+2j+1)} \mu_{\bar{\rho}(2\ell'+2j+2)}}}{z_{\bar{\rho}(2\ell'+2j+1),\bar{\rho}(2\ell'+2j+2)} } \;  z_{\bar{\rho}(2\ell'+2j+1),A}  \, z_{\bar{\rho}(2\ell'+2j+2),B}\,.
\label{npt,33}
\end{align}

\section{Manifest NS antisymmetry in $\bm{n}$--point correlators with two spin fields}
\label{sec:TheNPointCorrelator}

\subsection{Antisymmetrized versus ordered products of sigma matrices}
\label{sec:AntisymmetrizedVersusOrderedProductsOfSigmaMatrices}

The purpose of this appendix is to deliver the proof of the formulae (\ref{AS,1}), (\ref{AS,2}) claimed in Section~\ref{sec:MakingTheNSAntisymmetryManifest}. These equations relate an antisymmetrized chain of $\si$ matrices to an
ordered one. In order to avoid different notations for even and odd numbers of matrices, we will make use of the  notation
\beq
\ga^{\mu} \eq \ccb 0 &\si^{\mu} \\ \bar{\si}^{\mu} &0 \cce 
\co \ga^{\mu_1 ... \mu_n} \eq \left\{ \begin{array}{ll} \ccb 0 &\si^{[\mu_1} \, \bar{\si}^{\mu_2} \, ... \, \si^{\mu_n]} \\ \bar{\si}^{[\mu_1} \, \si^{\mu_2} \, ... \, \bar{\si}^{\mu_n]} &0 \cce &: \ n \ \te{odd} \\[.5cm]  \ccb \si^{[\mu_1} \, \bar{\si}^{\mu_2} \, ... \, \bar{\si}^{\mu_n]} &0 \\ 0 &\bar{\si}^{[\mu_1} \, \si^{\mu_2} \, ... \, \si^{\mu_n]} \cce &: \ n \ \te{even}\,, \end{array} \right.
\label{ga1}
\eeq
where as usual $\ga^{\mu_1 ... \mu_n} := \ga^{[\mu_1} \, ... \, \ga^{\mu_n]}$. Using this notation the claims (\ref{AS,1}), (\ref{AS,2})
become:
\begin{align}
\ga^{\mu_{1} ... \mu_{2k}} \ \ &= \ \ \sum_{\ell = 0}^{k} \sum_{\rho \in S_{2k}/ {\cal Q}_{k+1,\ell}} \! \! \! \! \! \! \te{sgn}(\rho) \, \ga^{\mu_{\rho(1)}} \, \ga^{\mu_{\rho(2)}} \, ... \, \ga^{\mu_{\rho(2\ell)}} \prod_{j=1}^{k-\ell} \eta^{\mu_{\rho(2\ell+2j-1)} \mu_{\rho(2\ell+2j)}}\,, \label{ga2a} \\
\ga^{\mu_{1} ... \mu_{2k+1}} \ \ &= \ \ \sum_{\ell = 0}^{k} \sum_{\rho \in S_{2k+1}/ {\cal P}_{k+1,\ell}} \! \! \! \! \! \! \! \! \! \te{sgn}(\rho) \, \ga^{\mu_{\rho(1)}} \, \ga^{\mu_{\rho(2)}} \, ... \, \ga^{\mu_{\rho(2\ell+1)}} \prod_{j=1}^{k-\ell} \eta^{\mu_{\rho(2\ell+2j)} \mu_{\rho(2\ell+2j+1)}}\,. \label{ga2b}
\end{align}
Under the assumption that this holds for some number $n$ of antisymmetrized $\ga$ matrices let us consider:
\beq
\ga^{\mu_{1} ... \mu_{n+2}} \eq \frac{1}{(n+2)!} \sum_{\rho \in S_{n+2}} \te{sgn}(\rho) \, \ga^{\mu_{\rho(1)}} \, ... \, \ga^{\mu_{\rho(n+2)}} \ .
\label{AS,3}
\eeq
We can bring each term of the right hand side to the ordered form $\ga^{\mu_{1}}  \ga^{\mu_2} ...  \ga^{\mu_{n+2}}$ by
virtue of $\{ \ga^{\mu} , \ga^{\nu} \} = -2\eta^{\mu \nu}$. The factor $\te{sgn}(\rho)$ makes sure that the ordered chain (of full length $n+2$) appears with a positive sign. Schematically this means:
\beq
\ga^{\mu_{1}... \mu_{n+2}} \eq \ga^{\mu_{1}}  \, \ga^{\mu_2} \, ... \, \ga^{\mu_{n+2}} \ + \ \frac{1}{(n+2)!}\sum_{\rho \in S_{n+2}} \te{anticommutators associated with $\rho$} \ .
\label{AS,4}
\eeq
To determine the structure of the $\eta$ terms note that a random permutation $\rho \in S_k$ of the sequence $(1,...,k)$ requires $\frac{k(k-1)}{4}$ transpositions in average to be ordered. In (\ref{AS,3}) we have a sum over $(n+2)!$ permutations of $S_{n+2}$ (with average transposition number $\frac{(n+1)(n+2)}{4}$), so we catch $\frac{(n+2)! \  (n+1)(n+2)}{4}$ anticommutators in total.

On the other hand there are $\left( \begin{smallmatrix} n+2 \\ 2 \end{smallmatrix} \right) = \frac{(n+2)(n+1)}{2}$
different $\eta^{\mu_i \mu_j} \ga^{\mu_{k_1} ...\mu_{k_n}}$ terms arising in the anticommutation process. The $\eta$ indices are taken as
$\mu_{\rho(1)} , \mu_{\rho(2)}$, and to capture each pair of $\mu_i, \mu_j$ exactly once we have to sum over the reduced
set
\begin{equation}
\label{permset}
  \tilde{S}_{n+2} := \Bigl\{ \rho \in S_{n+2} \, : \ \rho(1) < \rho(2) \Bigr\}
\end{equation}
of $S_{n+2}$ permutations. Each of the $\eta^{\mu_i \mu_j}$ appears $\frac{(n+2)! \ (n+1)(n+2)}{4} / \left(
  \frac{(n+2)(n+1)}{2} \right) = \frac{(n+2)!}{2}$ times, which cancels the denominator in front of the $S_{n+2}$ sum in
(\ref{AS,3}), (\ref{AS,4}). The shorter chain of $\ga$ matrices (with $n$ instead of $n+2$ factors) which come along with these $\eta^{\mu_{\rho(1)} \mu_{\rho(2)}}$'s naturally show up in antisymmetrized fashion $\ga^{[\mu_{\rho(3)} ...\mu_{\rho(n+2)}]}$,
\begin{align}
\ga^{\mu_{1} ... \mu_{n+2}} \ \ &= \ \ \ga^{\mu_{1}}   \, ... \, \ga^{\mu_{n+2}} \ - \  \sum_{\rho \in \tilde{S}_{n+2}} \te{sgn}(\rho) \; \frac{(n+2)!}{(n+2)! \, 2} \; \bigl\{ \ga^{\mu_{\rho(1)}} \, , \, \ga^{\mu_{\rho(2)}} \bigr\} \, \ga^{\mu_{\rho(3)}} \, ... \, \ga^{\mu_{\rho(n+2)}} \notag \\
&= \ \ \ga^{\mu_{1}}   \, ... \, \ga^{\mu_{n+2}} \ + \ \sum_{\rho \in \tilde{S}_{n+2}} \te{sgn}(\rho) \, \eta^{\mu_{\rho(1)} \mu_{\rho(2)}} \, n! \, \ga^{\mu_{\rho(3)}  ...  \mu_{\rho(n+2)}} \notag \\
&= \ \ \ga^{\mu_{1}}   \, ... \, \ga^{\mu_{n+2}} \ + \ \! \sum_{\rho \in \tilde{S}_{n+2} / S_n} \! \! \te{sgn}(\rho) \, \eta^{\mu_{\rho(1)} \mu_{\rho(2)}} \, \ga^{[\mu_{1}} \, ... \, \not{ \! \! \! \ga^{\mu_{\rho(1)}}} \, ... \, \not{ \! \! \! \ga^{\mu_{\rho(2)}}} \, ... \, \ga^{\mu_{ n+2}]}\,.
\label{AS,5}
\end{align}
The central idea in the last step is that the summation over the $S_n$ subgroup of $\tilde{S}_{n+2}$ permutations with fixed $\rho(1)$ and $\rho(2)$ is redundant, because every single terms $\ga^{\mu_{\rho(3)}  ...  \mu_{\rho(n+2)}}$ is already an antisymmetric expression with respect to $\rho(3)...\rho(n+2)$. We can now use the hypothesis of induction (\ref{AS,1}) or (\ref{AS,2}) to bring the chain $\ga^{[\mu_{1}} \, ... \, \not{ \! \! \! \ga^{\mu_{\rho(1)}}} \, ... \, \not{ \! \! \! \ga^{\mu_{\rho(2)}}} \, ... \, \ga^{\mu_{ n+2}]}$ of $n$ (instead of $n+2$) matrices into ordered form.

In order to show that (\ref{AS,5}) reproduces the $n+2$ version of (\ref{ga2a}) or (\ref{ga2b}) respectively, the crucial point is that $\tilde{S}_{n+2} / S_n$ exactly covers the $\ell = \left\lfloor n/2 \right\rfloor -1$ contribution of the $\rho \in S_{n}/ {\cal Q}_{\left\lfloor n/2 \right\rfloor+1,\ell}$ and $\rho \in S_{n}/ {\cal P}_{\left\lfloor n/2 \right\rfloor+1,\ell}$ sums. More concretely, in case of $n=2k$,
\beq
\tilde{S}_{2k+2} / S_{2k} \eq S_{2k+2} / {\cal Q}_{k+2,k} \ .
\label{S/Q}
\eeq
Therefore:
\begin{align}
\si^{[\mu_{1}} \, &\si^{\mu_2} \,  ... \bar{\si}^{\mu_{2k+2}]} \ \ = \ \  \si^{\mu_{1}}   \, ... \, \bar{\si}^{\mu_{2k+2}} \ + \ \! \sum_{\rho \in \tilde{S}_{2k+2}/ S_{2k}} \! \! \te{sgn}(\rho) \, \eta^{\mu_{\rho(1)} \mu_{\rho(2)}} \, \si^{[\mu_{1}} \, ... \, \not{ \! \! \! \si^{\mu_{\rho(1)}}} \, ... \, \not{ \! \! \! \si^{\mu_{\rho(2)}}} \, ... \, \bar{\si}^{\mu_{ 2k+2}]} \notag \\
&= \ \ \si^{\mu_{1}}   \, ... \, \bar{\si}^{\mu_{2k+2}} \ + \ \! \sum_{\rho \in S_{2k+2} / {\cal Q}_{k+2,k}} \! \! \te{sgn}(\rho) \, \eta^{\mu_{\rho(1)} \mu_{\rho(2)}} \sum_{\ell = 0}^{k} \sum_{\pi \in S_{2k}/ {\cal Q}_{k+1,\ell}} \! \! \! \! \! \! \te{sgn}(\pi) \, \si^{\mu_{\pi(\rho(3))}} \, \bar{\si}^{\mu_{\pi(\rho(4))}} \, ... \, \bar{\si}^{\mu_{\pi(\rho(2\ell+2))}} \notag \\
& \ \ \ \ \ \times \ \prod_{j=1}^{k-\ell} \eta^{\mu_{\pi(\rho(2\ell+2j+1))} \mu_{\pi(\rho(2\ell+2j+2))}}   \notag \\
&= \ \ \sum_{\ell = 0}^{k+1} \sum_{\bar{\rho} \in S_{2k+2}/ {\cal Q}_{k+2,\ell}} \! \! \! \! \! \! \te{sgn}(\bar{\rho}) \, \si^{\mu_{\bar{\rho}(1)}} \, \bar{\si}^{\mu_{\bar{\rho}(2)}} \, ... \, \bar{\si}^{\mu_{\bar{\rho}(2\ell)}} \prod_{j=1}^{k+1-\ell} \eta^{\mu_{\bar{\rho}(2\ell+2j-1)} \mu_{\bar{\rho}(2\ell+2j)}}\,.
\label{AS,6}
\end{align}
The argument for $n=2k+1$ is completely analogous. This completes our proof of (\ref{AS,1}) and (\ref{AS,2}).

\subsection{The proof for the antisymmetric representation of $\Om_{(n)}$, $\om_{(n)}$}
\label{sec:TheProofForTheAntisymmetricRepresentation}

The purpose of this Subsection is to deliver a proof of (\ref{AS,8a}) and (\ref{AS,8b}) being an equivalent representation of the $2n(+1)$ point correlation functions $\om_{(n)}$, $\Om_{(n)}$ with two spin fields. Basically, we will simply plug the conversion formulae (\ref{AS,1}) and (\ref{AS,2}) (for antisymmetrized $\si$ matrix chains in terms of ordered ones) into the claimed expression for $\om_{(n)}, \Om_{(n)}$ and see that we arrive at the well-established expressions (\ref{npt,1}) and (\ref{npt,2}):
\begin{align}
\Om&_{(n)}^{\mu_{1} ... \mu_{2n-1}}\,_{\al \dbe}(z_{i}) \eq  \underbrace{\frac{1}{2^{n-1} \, \sqrt{2}} \prod_{i=1}^{2n-1} (z_{iA} \, z_{iB})^{-1/2}}_{=: \ \Om_{0}(z_i)} \! \! \! \sum_{\ell = 0}^{\te{min}(1,n-1)} \! \! \!  z_{AB}^{\ell} \! \! \! \! \sum_{\rho \in S_{2n-1}/{\cal P}_{n,\ell}} \! \! \!  \te{sgn}(\rho) \notag \\
& \ \ \ \ \times \ \sum_{\bar{\ell}=0}^{\ell} \!  \sum_{\pi \in S_{2\ell+1} / {\cal P}_{\ell+1,\bar{\ell}}} \! \! \! \! \! \te{sgn}(\pi) \, \bigl(\si^{\mu_{\pi (\rho(1))}} \, \bar{\si}^{\mu_{\pi(\rho(2))}} \, ... \, \bar{\si}^{\mu_{\pi(\rho(2\ell))}} \, \si^{\mu_{\pi(\rho(2\bar{\ell}+1))}} \bigr)_{\al \dbe} \prod_{j=1}^{\ell-\bar{\ell}} \eta^{ \mu_{\pi(\rho(2\bar{\ell}+2j))} \mu_{\pi(\rho(2\bar{\ell}+2j+1))} } \notag \\
& \ \ \ \ \times \prod_{k=1}^{n-\ell-1} \frac{\eta^{\mu_{\rho(2\ell+2k)} \mu_{\rho(2\ell+2k+1)}}}{z_{\rho(2\ell+2k),\rho(2\ell+2k+1)} } \; \bigl( z_{\rho(2\ell+2k),A} \, z_{\rho(2\ell+2k+1),B} \, + \, z_{\rho(2\ell+2k),B} \, z_{\rho(2\ell+2k+1),A} \bigr) \notag \\
&= \ \ \Om_{0}(z_i) \  \sum_{\ell = 0}^{n-1} \sum_{\bar{\ell}=0}^{\ell} z_{AB}^{\bar{\ell}} \! \! \sum_{\rho \in S_{2n-1}/{\cal P}_{n,\ell}} \sum_{\pi \in S_{2\ell+1} / {\cal P}_{\ell+1,\bar{\ell}}} \! \! \te{sgn}(\rho) \, \te{sgn}(\pi) \notag \\
& \ \ \ \ \times \ \bigl(\si^{\mu_{\pi (\rho(1))}} \,  ... \,  \si^{\mu_{\pi(\rho(2\bar{\ell}+1))}} \bigr)_{\al \dbe} \, \prod_{j=1}^{\ell-\bar{\ell}} \frac{\eta^{ \mu_{\pi(\rho(2\bar{\ell}+2j))} \mu_{\pi(\rho(2\bar{\ell}+2j+1))} }}{z_{\pi(\rho(2\bar{\ell}+2j)) , \pi(\rho(2\bar{\ell}+2j+1)) }} \, \prod_{k=1}^{n-\ell-1} \frac{\eta^{\mu_{\rho(2\ell+2k)} \mu_{\rho(2\ell+2k+1)}}}{z_{\rho(2\ell+2k),\rho(2\ell+2k+1)} } \notag \\
&  \ \ \ \ \times \ z_{AB}^{\ell - \bar{\ell}} \, z_{\pi(\rho(2\bar{\ell}+2j)) , \pi(\rho(2\bar{\ell}+2j+1)) } \, \bigl( z_{\rho(2\ell+2k),A} \, z_{\rho(2\ell+2k+1),B} \, + \, z_{\rho(2\ell+2k),B} \, z_{\rho(2\ell+2k+1),A} \bigr) \notag \\
&= \ \ \Om_{0}(z_i) \  \sum_{\bar{\ell} = 0}^{n-1} z_{AB}^{\bar{\ell}} \sum_{\De \ell=0}^{n-\bar \ell-1}  \sum_{\rho \in S_{2n-1}/{\cal P}_{n,\bar \ell + \De \ell}} \ \sum_{\pi \in S_{2(\bar \ell + \De \ell)+1} / {\cal P}_{\bar \ell + \De \ell +1,\bar{\ell}}} \! \! \te{sgn}(\rho) \, \te{sgn}(\pi) \notag \\
& \ \ \ \ \times \ \bigl(\si^{\mu_{\pi (\rho(1))}} \,  ... \,  \si^{\mu_{\pi(\rho(2\bar{\ell}+1))}} \bigr)_{\al \dbe} \, \prod_{j=1}^{\De \ell} \frac{\eta^{ \mu_{\pi(\rho(2\bar{\ell}+2j))} \mu_{\pi(\rho(2\bar{\ell}+2j+1))} }}{z_{\pi(\rho(2\bar{\ell}+2j)) , \pi(\rho(2\bar{\ell}+2j+1)) }} \,\prod_{m=\De \ell+1}^{n-\bar{\ell}-1} \frac{\eta^{ \mu_{\rho(2\bar{\ell}+2m)} \mu_{\rho(2\bar{\ell}+2m+1)} }}{z_{\rho(2\bar{\ell}+2m) , \rho(2\bar{\ell}+2m+1) }} \notag \\
& \ \ \ \ \times \ \bigl( z_{\pi(\rho(2 \bar{\ell}+2j)),A} \, z_{\pi(\rho(2\bar{\ell}+2j+1)),B} \, - \, z_{\pi(\rho(2 \bar{\ell} +2j)),B} \, z_{\pi(\rho(2 \bar{\ell} +2j+1)),A} \bigr) \notag \\
& \ \ \ \ \times \ \bigl( z_{\rho(2 \bar{\ell}+2m),A} \, z_{\rho(2\bar{\ell}+2m+1),B} \, + \, z_{\rho(2 \bar{\ell} +2m),B} \, z_{\rho(2 \bar{\ell} +2m+1),A} \bigr) \notag \\
&= \ \ \frac{1}{2^{n-1} \, \sqrt{2}} \prod_{i=1}^{2n-1} (z_{iA} \, z_{iB})^{-1/2} \; \sum_{\bar{\ell} = 0}^{n-1} z_{AB}^{\bar{\ell}} \! \! \sum_{\wp \in S_{2n-1}/{\cal P}_{n,\bar{\ell}}} \! \! \! \te{sgn}(\wp) \, \bigl(\si^{\mu_{ \wp(1) }} \,  ... \,  \si^{\mu_{\wp(2\bar{\ell}+1)}} \bigr)_{\al \dbe} \notag \\
& \ \ \ \ \times \ \prod_{m=1}^{n-\bar{\ell}-1} \frac{\eta^{ \mu_{\wp(2\bar{\ell}+2m)} \mu_{\wp(2\bar{\ell}+2m+1)} }}{z_{\wp(2\bar{l}+2m) , \wp(2\bar{l}+2m+1) }} \; 2^{n-\bar{\ell}-1} \, \prod_{q=1}^{n-\bar{\ell}-1} z_{\wp(2\bar{\ell}+2q),A} \, z_{\wp(2\bar{\ell}+2q+1),B}\,.
\label{prfas1}
\end{align}
In the second step, we have changed summation variables from $(\ell,\bar{\ell})$ to $(\bar{\ell}, \De \ell = \ell - \bar{\ell})$ and used $z_{AB}^{ \De \ell }  z_{\pi(\rho(2\bar{\ell}+2j)) , \pi(\rho(2\bar{\ell}+2j+1)) } =  z_{\pi(\rho(2 \bar{\ell}+2j)),A}  z_{\pi(\rho(2\bar{\ell}+2j+1)),B} - z_{\pi(\rho(2 \bar{\ell} +2j)),B}  z_{\pi(\rho(2 \bar{\ell} +2j+1)),A}$ in context of the corresponding $j$ product. Let us further explain the emergence of the $2^{n-\bar{\ell}-1}$ term in the third step: 

Only those $z_{iA}$- and $z_{jB}$ combinations contribute where $i=\wp(2k)$ and $j = \wp(2k+1)$ for some $S_{2n-1} / {\cal P}_{n,\bar{\ell}}$ permutation $\wp$. (The other ones show up with a plus sign as often as with a minus sign in the $\rho$-, $\pi$- and $\De \ell$ sums and therefore drop out.) Their multiplicity at each fixed $\De \ell$ value can be found to be $\left( \begin{smallmatrix} n-\bar \ell-1 \\ \De \ell \end{smallmatrix} \right)$ by elementary combinatorics. We obtain overall multiplicity $2^{n-\bar \ell-1} = \sum_{\De \ell =0}^{n-\bar \ell-1} \left( \begin{smallmatrix} n-\bar \ell-1 \\ \De \ell \end{smallmatrix} \right)$.


\begin{thebibliography}{99}
\def\bi{\bibitem}


\bi{ST} S. Stieberger and T.R.~Taylor,
``Amplitude for N-gluon superstring scattering,''
  Phys.\ Rev.\ Lett.\  {\bf 97}, 211601 (2006)
  [arXiv:hep-th/0607184];
``Multi-gluon scattering in open superstring theory,''
  Phys.\ Rev.\  D {\bf 74}, 126007 (2006)
  [arXiv:hep-th/0609175].

\bibitem{MHV} S.~Stieberger and T.R.~Taylor,
``Supersymmetry Relations and MHV Amplitudes in Superstring Theory,''
  Nucl.\ Phys.\  B {\bf 793}, 83 (2008)
  [arXiv:0708.0574 [hep-th]].

\bibitem{6G} S.~Stieberger and T.R.~Taylor,
``Complete Six-Gluon Disk Amplitude in Superstring Theory,''
  Nucl.\ Phys.\  B {\bf 801}, 128 (2008)
  [arXiv:0711.4354 [hep-th]].


\bi{Medina} R.~Medina and L.A.~Barreiro,
``Higher N-point amplitudes in open superstring theory,''
  PoS {\bf IC2006}, 038 (2006)
  [arXiv:hep-th/0611349].


\bi{LHC1} D.~L\"ust, S.~Stieberger and T.R.~Taylor,
``The LHC String Hunter's Companion,''
  Nucl.\ Phys.\  B {\bf 808}, 1 (2009)
  [arXiv:0807.3333 [hep-th]].

\bi{LHC2}  D.~L\"ust, O.~Schlotterer, S.~Stieberger and T.R.~Taylor,
``The LHC String Hunter's Companion (II): Five-Particle Amplitudes and
Universal Properties,''
  Nucl.\ Phys.\  B {\bf 828} (2010) 139
  [arXiv:0908.0409 [hep-th]].

\bibitem{BD}
  T.~Banks and L.J.~Dixon,
 ``Constraints on String Vacua with Space-Time Supersymmetry,''
  Nucl.\ Phys.\  B {\bf 307} (1988) 93;\\
T.~Banks, L.J.~Dixon, D.~Friedan and E.J.~Martinec,
``Phenomenology and Conformal Field Theory Or Can String Theory Predict the
  Nucl.\ Phys.\  B {\bf 299} (1988) 613;\\
S.~Ferrara, D. L\"ust and S. Theisen,
``World Sheet Versus Spectrum Symmetries In Heterotic And Type II Superstrings.,''
  Nucl.\ Phys.\  B {\bf 325}, 501 (1989).




\bibitem{FMS}
  D.~Friedan, E.J.~Martinec and S.H.~Shenker,
``Conformal Invariance, Supersymmetry And String Theory,''
  Nucl.\ Phys.\  B {\bf 271} (1986) 93.

\bibitem{Cohn}
  J.~Cohn, D.~Friedan, Z.a.~Qiu and S.H.~Shenker,
`` Covariant Quantization Of Supersymmetric String Theories: The Spinor Field 
Of The Ramond-Neveu-Schwarz Model,''
  Nucl.\ Phys.\  B {\bf 278} (1986) 577.

\bibitem{KLLSW}
  V.A.~Kostelecky, O.~Lechtenfeld, W.~Lerche, S.~Samuel and S.~Watamura,
``Conformal Techniques, Bosonization and Tree Level String Amplitudes,''
  Nucl.\ Phys.\  B {\bf 288} (1987) 173.


\bibitem{GHMR} 
D.J.~Gross, J.A.~Harvey, E.J.~Martinec and R.~Rohm,
``Heterotic String Theory. 1. The Free Heterotic String,''
  Nucl.\ Phys.\  B {\bf 256}, 253 (1985);
``Heterotic String Theory. 2. The Interacting Heterotic String,''
  Nucl.\ Phys.\  B {\bf 267} (1986) 75.




\bibitem{Olaf}  V.A.~Kostelecky, O.~Lechtenfeld, W.~Lerche, S.~Samuel and S.~Watamura,
``A four-point amplitude for the $O(16)\times O(16)$ heterotic string,''
  Phys.\ Lett.\  B {\bf 182}, 331 (1986);

\bibitem{KLSVWS} 
V.A.~Kostelecky, O.~Lechtenfeld, S.~Samuel, D.~Verstegen, S.~Watamura and D.~Sahdev,
``The six-fermion amplitude in the superstring,''
  Phys.\ Lett.\  B {\bf 183}, 299 (1987).



\bibitem{Barreiro:2005hv} L.A.~Barreiro and R.~Medina,
``5-field terms in the open superstring effective action,''
  JHEP {\bf 0503}, 055 (2005)
  [arXiv:hep-th/0503182];\\
R.~Medina, F.T.~Brandt and F.R.~Machado,
``The open superstring 5-point amplitude revisited,''
  JHEP {\bf 0207}, 071 (2002)
  [arXiv:hep-th/0208121].


\bibitem{DAN}   D.~Oprisa and S.~Stieberger,
``Six gluon open superstring disk amplitude, multiple hypergeometric  series
 and Euler-Zagier sums,''
  arXiv:hep-th/0509042.
  
\bibitem{Hybrid} 
 N.~Berkovits, ``A new description of the superstring,''
  arXiv:hep-th/9604123.

\bibitem{KZ}
  V.G.~Knizhnik and A.B.~Zamolodchikov,
``Current algebra and Wess-Zumino model in two dimensions,''
  Nucl.\ Phys.\  B {\bf 247}, 83 (1984).



\bibitem{RW1} M. Bousquet-Melou, 
``Counting walks in the quarter plane,'' 
in: Trends Math., Birkh\"auser, Basel. pp. 49--67;\\
M. Bousquet-Melou and M. Mishna,
``Walks with small steps in the quarter plane,'' arXiv:0810.4387.



\bibitem{ATICK} J.J.~Atick and A.~Sen,
``Correlation Functions Of Spin Operators On A Torus,''
  Nucl.\ Phys.\  B {\bf 286}, 189 (1987);
``Spin Field Correlators On An Arbitrary Genus Riemann Surface And Nonrenormalization Theorems In String Theories,''
  Phys.\ Lett.\  B {\bf 186}, 339 (1987);
``Covariant One Loop Fermion Emission Amplitudes In Closed String Theories,''
  Nucl.\ Phys.\  B {\bf 293}, 317 (1987);\\
J.J.~Atick, L.J.~Dixon and A.~Sen,
``String Calculation Of Fayet-Iliopoulos D Terms In Arbitrary Supersymmetric
Compactifications,''
  Nucl.\ Phys.\  B {\bf 292}, 109 (1987).


\bibitem{LAG}
  L.~Alvarez-Gaum\'e, J.B.~Bost, G.W.~Moore, P.~C.~Nelson and C.~Vafa,
``Bosonization on higher genus Riemann surfaces,''
  Commun.\ Math.\ Phys.\  {\bf 112}, 503 (1987);\\
  L.~Alvarez-Gaum\'e, G.W.~Moore, P.C.~Nelson, C.~Vafa and J.B. Bost,
``Bosonization in arbitrary genus,''
  Phys.\ Lett.\  B {\bf 178}, 41 (1986).

\bibitem{OS} O.~Schlotterer,
``Higher Loop Spin Field Correlators in D=4 Superstring Theory,''
  arXiv:1001.3158 [hep-th].




\end{thebibliography}
\end{document}